\documentclass{aa}  
\usepackage{tablefootnote}
\usepackage{natbib}
\usepackage{graphicx}
\usepackage{txfonts}
\usepackage{array}
\usepackage{color}
\usepackage{rotating}
\usepackage{lscape}
\usepackage[colorlinks=true,linkcolor=blue,citecolor=blue,urlcolor=blue]{hyperref}
\bibpunct{(}{)}{;}{a}{}{,} 
\newcommand{\fado}{\textsc{Fado}}\newcommand{\starl}{\textsc{Starlight}}
\newcommand{\stec}{\textsc{Steckmap}}\newcommand{\reb}{\textsc{Rebetiko}}
\newcommand{\Con}{{\tt CONT}}\newcommand{\Tau}{{\tt TAU1}}
\newcommand{\logtL}{$\left<log(t_\ast)\right>_{L}$}\newcommand{\logtM}{$\left<log(t_\ast)\right>_{M}$}
\newcommand{\logZL}{$\left<log(Z_\ast)\right>_{L}$}\newcommand{\logZM}{$\left<log(Z_\ast)\right>_{M}$}
\begin{document} 

  \title{Self-consistent population spectral synthesis with \fado:}
  \subtitle{II. Star formation history of galaxies in spectral synthesis methods}
  \author{Ciro Pappalardo\inst{1,2}, Leandro S.M. Cardoso\inst{3,4}, Jean Michel Gomes\inst{4}, Polychronis Papaderos\inst{1,2,4}, Jos\'{e} Afonso\inst{1,2}, Iris Breda\inst{1}, Andrew Humphrey\inst{4}, Tom Scott\inst{4}, Stergios Amarantidis\inst{1,2}, Israel Matute\inst{1,2}, Rodrigo Carvajal\inst{1,2}, Silvio Lorenzoni\inst{1}, Patricio Lagos\inst{4}, Ana Paulino-Afonso\inst{5}, Henrique Miranda\inst{2}
          }
  \institute{Instituto de Astrof\'{i}sica e Ci\^{e}ncias do Espa\c{c}o, Universidade de Lisboa - OAL, Tapada da Ajuda, PT1349-018 Lisboa, Portugal
  \and
  Departamento de F\'{i}sica, Faculdade de Ci\^{e}ncias da Universidade de Lisboa, Edif\'{i}cio C8, Campo Grande, PT1749-016 Lisboa, Portugal
  \and
  Department of Mathematics, University of \'{E}vora, R. Rom\~{a}o Ramalho 59, 7000 \'{E}vora, Portugal
  \and
  Instituto de Astrof\'{i}sica e Ci\^{e}ncias do Espa\c{c}o, Universidade do Porto - CAUP, Rua das Estrelas, PT4150-762 Porto, Portugal
  \and
  CENTRA/COSTAR, Instituto Superior T\'{e}cnico, Universidade de Lisboa, Av. Rovisco Pais 1, 1049-001 Lisboa, Portugal
  }

  \abstract
  {The field of galaxy evolution will make a great leap forward in the next decade as a consequence of the huge effort by the scientific community in multi-object spectroscopic facilities. Various future surveys will enormously increase the number of available galaxy spectra, providing new insights into unexplored areas of research. To maximise the impact of such incoming data, the analysis methods must also step up, extracting reliable information from the available spectra. It is therefore urgent to refine and test reliable analysis tools that are able to infer the properties of a galaxy from medium- or high-resolution spectra.}
  {In this paper we aim to investigate the limits and the reliability of different spectral synthesis methods in the estimation of the mean stellar age and metallicity. These two quantities are fundamental to determine the assembly history of a galaxy by providing key insights into its star formation history. The main question this work aims to address is which signal-to-noise ratios (S/N) are needed to reliably determine the mean stellar age and metallicity from a galaxy spectrum and how this depends on the tool used to model the spectra.}
  {To address this question we built a set of realistic simulated spectra containing stellar and nebular emission, reproducing the evolution of a galaxy in two limiting cases: a constant star formation rate and an exponentially declining star formation with a single initial burst. We degraded the synthetic spectra built from these two star formation histories (SFHs) to different S/N and analysed with three widely used spectral synthesis codes, namely \fado, \stec, and \starl,\ assuming similar fitting set-ups and the same base of spectral templates.}
  {For S/N $\le5$ all the three tools show a large diversity in the results. The \fado\ and \starl\ tools find median differences in the light-weighted mean stellar age of $\sim$0.1 dex, while \stec\ shows a higher value of $\sim$0.2 dex. For S/N $> 50$ the median differences in \fado\ are $\sim$0.03 dex ($\sim7$\%), a factor 3 and 4 lower than the 0.08 dex ($\sim$20\%) and 0.11 dex ($\sim$30\%) obtained from \starl\ and \stec, respectively. Detailed investigations of the best-fit spectrum for galaxies with overestimated mass-weighted quantities point towards the inability of purely stellar models to fit the observed spectral energy distribution around the Balmer jump.}
  {Our results imply that when a galaxy enters a phase of high specific star formation rate (sSFR) the neglect of the nebular continuum emission in the fitting process has a strong impact on the estimation of its SFH when purely stellar fitting codes are used, even in presence of high S/N spectra. The median value of these differences are of the order of 7\% (\fado), 20\% (\starl), and 30\% (\stec) for light-weighted quantities, and 20\% (\fado), 60\% (\starl), and 20\% (\stec) for mass-weighted quantities. More specifically, for a continuous SFH both \stec\ and \starl\ overestimate the stellar age by $>$2 dex within the first $\sim$100 Myr even for high S/N spectra. This bias, which stems from the neglect of nebular continuum emission, obviously implies a severe overestimation of the mass-to-light ratio and stellar mass. But even in the presence of a mild contribution from nebular continuum, there is still the possibility to misinterpret the data as a consequence of the poor quality of the observations. Our work underlines once more the importance of a self-consistent treatment of nebular emission, as implemented in FADO, which, according to our analysis, is the only viable route towards a reliable determination of the assembly of any high-sSFR galaxy at high and low redshift.}

\titlerunning{Star formation histories of galaxies in spectral synthesis methods}
\authorrunning{C. Pappalardo et al.}
  \maketitle
%

\section{Introduction}
\label{intro}

 The study of galaxy evolution will see tremendous breakthroughs in the next decade as a consequence of the huge effort by the scientific community using multi-object spectroscopic facilities. Surveys carried out with instruments such as 4MOST \citep{4most}, WEAVE \citep{weave}, MOONS \citep{moons,moons2,mai}, and PFS \citep{pfs} will enormously increase the number of available galaxy spectra, opening new windows into unexplored areas of research. 

 These spectra will tackle the crucial problem of the assembly of galaxies, shedding light on the physical processes shaping galaxies throughout their evolution. From this perspective, it is of paramount importance to refine and test reliable analysis tools that are able to infer the properties of a galaxy from medium- or high-resolution spectra. To achieve this goal a widely used technique is so-called spectral synthesis, which comes in two flavours: evolutionary and population synthesis methods (see the seminal works of \citealt{stu} and \citealt{fab}).
 
 Evolutionary synthesis methods estimate the evolution of the spectral energy distribution (SED) of galaxies according to a specific set of star formation histories (SFH) and chemical evolution models (CEM). The observed spectrum is compared to different sets of models,  and the best-fit provides the physical parameters of interest. The main limitation of such methods is in the underlying assumption of a well-defined star formation law, which even with more complex parametrisations (e.g. delayed exponential, double exponential), or SFHs derived from cosmological simulations (e.g. \citealt{pac}), lacks the needed accuracy to reproduce the complexity of star formation processes within a galaxy (see a more complete discussion in \citealt{gom}).

 Population synthesis methods instead do not assume a fixed SFH and/or CEM, since by definition they attempt to approximate the SFH of a galaxy as a linear superposition of simple stellar populations (SSPs), each one corresponding to the SED of an instantaneously formed stellar population of a given age, metallicity, and initial mass function (IMF) \citep{tin,bru}. This approach is complementary to evolutionary methods and has a number of advantages. The most important advantage is that this approach is free of prior assumptions about the SFH and CEM. This allows the investigation of galaxies with complex SFHs: sudden bursts due to a merger, active galactic nuclei (AGNs) triggering, or the abrupt decline of star formation due to ram pressure stripping \citep[e.g.][]{car2,pap}. 
 
 The main limits of spectral synthesis codes are the following: a) degeneracies inherent to theoretical stellar physics; b) different mathematical recipes of the available codes, which have their own intrinsic limitations; c) the non-perfect coverage of the adopted SSP library, resulting in increasing uncertainties in modelling techniques; and, d) the neglect of important physical ingredients to the SED (e.g. nebular continuum emission), as pointed out in \cite{gom}.
 
 Different approaches have been proposed to overcome this limitation. An interesting example is to join semi-analytical libraries of stars to reproduce the spectra in missing regions of the parameter space \citep{coe}. This sparse coverage can introduce large errors; depending on the application, semi-empirical or fully theoretical models can be effective. With the development of more advanced computational techniques, analysis methods have become more sophisticated, and new approaches such as genetic algorithms \citep{gom}, principal component analysis (PCA, \citealt{mck}), mixed PCA-neural networks analysis \citep{als}, or Monte Carlo methods \citep{lej}  have been implemented. Despite the significant improvements provided by these approaches, the SFH and CEM remain complicated elements to disentangle, and different methods have been refined to break the well-known problem of age-metallicity-extinction degeneracy. 
 
 For example in the {\it STEllar Content and Kinematics via Maximum A Posteriori likelihood} tool (\stec, \citealt{ocv2,ocv}), the matrix solution is regularised through the introduction of a penalty function, while in \cite{iye} the star formation is modelled by adding a series of Gaussian bursts, reproducing various star formation episodes.

 A different approach is taken in \starl\ \citep{cid}, where an observed spectrum is fit with a linear combination of SSPs of several ages and metallicities, assuming a given extinction law and a Gaussian distribution for stellar velocities. The results are very accurate, recovering reliable stellar ages and metallicities but, as noted in \cite{cid}, these analyses are valid "as long as one does not attempt a very detailed description of the star formation and chemical histories". For this reason subsequent versions of the code have tackled different aspects of the issues mentioned in their paper, including photometric data from ultraviolet (UV, see e.g. \citealt{lop}). This approach is not free of caveats, one of which is that aperture effects and differences in the point spread function (PSF) of GALEX and optical spectrographs need to be accounted for.
 
 The problem of determining the SFH and CEM in a galaxy from spectra remains hard to solve, and the main outcome from these previous investigations is that only for high-resolution spectra with sufficient signal-to-noise ratio (S/N $\sim$ 10) it is possible to achieve reliable results. This is because it is only possible to disentangle two bursts with 0.8 dex difference in age at high S/N, as shown in \cite{ocv2}. However, this effect depends on the ages considered and could also be a consequence of the regularisation itself and thus specific to \stec. This is because the penalty function tends to avoid solutions with high derivative, excluding short bursts and favouring a more continuous SFH. Beyond the quality of the data, we should also consider the quality of the models used to fit the data, which has a crucial impact on the final outcome of the analysis \citep{kol2}.

 In this context, the {\it Fitting Analysis using Differential evolution Optimization} tool (\fado, \citealt{gom}) represents a conceptually novel approach because it is the first code to reproduce the nebular continuum in star-forming galaxies. This approach is of paramount importance to quantify the role of warm ionised gas in star-forming regions, particularly in starburst systems. Dealing self-consistently with nebular and stellar emission allows \fado\ to reliably reproduce emission fluxes and equivalent widths (EW) during starburst phases of galaxies and to identify the stellar continuum in UV emitting regions, thereby removing the bias introduced by neglecting the nebular gas \citep{gom}. A correct modelling of this component will become more relevant in the near future because upcoming surveys will provide us with many more galaxies at higher redshift, where we will analyse spectra that contain or are dominated by nebular emission. This is because the rest-frame blue part of the spectrum of a galaxy will be redshifted into optical spectral window and also because star-forming galaxies become more common as we move to higher redshifts.
 
 In light of these future surveys, it is necessary to understand and quantify the reliability of the results obtained with such methods by determining their ability to recover fundamental evolutionary parameters. Motivated by these problems, and having in mind the plethora of multi-object spectrograph instruments in the deployment phase, in this paper we aim to investigate the limits and reliability of different spectral synthesis methods in the estimation of two specific quantities being representatives of the evolutionary status of a galaxy: the mean stellar age and mean stellar metallicity. These two parameters are fundamental for determining the assembly history of a galaxy, by providing key insights into its SFH and stellar mass growth. The latter will be the subject of a forthcoming article of this series.
 
 The main question we want to address in this work is at which S/N values is it still possible to determine, from a galaxy spectrum, its mean stellar age and metallicity and how this depends on the tool used to model the spectra.
 
To address this question we built a set of simulated spectra reproducing the evolution of a galaxy considering two limiting cases: a continuous star formation and a single initial burst with an exponentially declining star formation law. We degraded the synthetic spectra built from these two SFHs to different S/N, simulating the different conditions encountered within the different surveys mentioned at the beginning of Sec. \ref{intro}.
 
 The obtained mock spectra were then analysed with three widely used spectral synthesis codes: \fado, \stec, and \starl\ assuming similar initial set-ups and spectral bases. The main goals are as follows: a) to determine at which S/N the results for the parameters considered are still reliable; b) to identify which stellar ages the fitting tools are more sensitive to; and c) to quantify the effect of nebular emission component on the physical quantities inferred for the stellar component in population synthesis codes.
 
 The paper is organised as follows: Sections \ref{rebetikoSection} and \ref{methods} describe the construction of synthetic spectra and the spectral fitting tools used for the analysis, respectively. Sections \ref{results} and \ref{discussion} report and discuss the results, while Section \ref{conclusions} presents the main conclusion of this work.
 
 \section{Synthetic spectra: Dancing \reb}
 \label{rebetikoSection}
 
 The synthetic spectra analysed in this paper were built with the evolutionary synthesis code {\it Reckoning galaxy Emission By means of Evolutionary Tasks with Input Key Observables} (\reb), applying the following equation:
 \begin{equation}\begin{split}
     F(\lambda,t)= \int_0^t \Phi(t-t') & F^{SSP}_{\lambda}(t',Z)dt'
     \\
     & +\frac{\gamma_{eff}(T_e)}{\alpha_B(T_e)}\int_0^t \Phi(t-t')q^{SSP}(t',Z)dt'
     \label{eq1}
 \end{split}.\end{equation}
 \noindent
 Eq. \ref{eq1} reproduces the evolution of the SED of a galaxy as a function of time. For each $t$ the flux $F(\lambda,t)$ at a specific wavelength $\lambda$ depends on two terms: The first term represents the emission due to the stellar component, which is the integral of the different SSPs weighted for the star formation rate (SFR; $\Phi$). The second term represents the flux contribution due to the nebular continuum, which depends on the number of hydrogen ionising photons $q^{SSP}$ weighted for the star formation (analogously to the stellar component). The only difference here is the addition of an extra term quantifying the efficiency of the process, which is estimated by the ratio of the photons continuous emission $\gamma_{eff}$ to the photons recombination rate, $\alpha_B$. These values depend on the gas temperature and the star formation-weighted nebular emission since both processes drive the number of photons emitted \citep{sch}. We assume all ionising photons produced by the stars are absorbed by the interstellar medium (ISM), and the nebular continuum is computed assuming case B recombination under typical physical conditions of HII regions, that is an electron density of $n_e$ = 100 cm$^{-3}$ and a temperature of $T_e = 10^4$ K \citep{ost,ost2}. 
 
 In the construction of the models, we did not take into account potential misalignments of stellar and gas kinematics. Decoupling of gas and stars in a galaxy can be indicative of gas accretion \citep{sar,che}, AGN or starburst-driven feedback \citep{flo,cic}, and/or gas loss due to environmental effects \citep{duc}. The misalignment affects mainly early-type galaxies and only a small fraction of star-forming discs \citep{jin}. For such cases, the integral field unit analysis of \citealt{mar} found a misalignment on the order of 2 km s$^{-1}$. Our mock galaxies are built assuming a closed-box model neglecting external processes, such as accretion, ram pressure stripping, and merging. Finally, emission lines are masked out in fits with \stec\ or \starl, thus a kinematical decoupling between gas and stars does not notably influence these fits. Nebular continuum emission (excluding its Balmer and Paschen jump) is featureless, so a kinematical decoupling does not play a role. As for \fado, it measures the H$\alpha$ and H$\beta$ (within $\pm$400 km/s from the stellar velocity) and uses their luminosity to self-consistently infer the SFH, so kinematical decoupling is not an issue.

 One of the goals of this paper is to quantify the effect of the inclusion of nebular continuum in the determination of evolutionary parameters inferred by the different codes and describing the evolution of a galaxy. To accomplish such an objective the strategy is to estimate the reliability of the results obtained with different analysis methods as a function of S/N. For these reasons we built our synthetic spectra assuming two SFHs. First, the {\tt Cont} model simulates a galaxy undergoing a continuous accretion of gas, which is therefore able to sustain star formation at a constant rate. Second, the \Tau\ model simulates a galaxy undergoing an initial instantaneous burst, on a timescale of $\sim10^8$ Myr, followed by an exponentially declining SFR ($\Phi$ = e$^{-t/\tau}$, with $\tau$ = 1 Myr).
 
 From the mathematical point of view extracting at the same time age and metallicity from a spectrum is an `ill-posed' problem because of the well-known age-metallicity degeneracy. The introduction of a complex chemical evolutionary history for our models would complicate the interpretation of the results, and for this reason, we assumed for our models a constant solar metallicity with $Z_\odot$ = 0.02. We note that  \cite{bc03} made use of the STELIB stellar library \citep{leb}, which is biased towards solar metallicity. The spectra of the simulated galaxies cover 3400-8900 \AA\ and have a spectral resolution of 2.3 \AA \ and a $\Delta\lambda$ = 1 \AA. The wavelengths coverage emulates the average range observed by the Sloan Digital Sky Survey (SDSS, \citealt{ahu,doi}), including the Balmer and Paschen jumps, which are relevant features at young ages ($<10^7$ yr) and are impossible to fit with purely stellar templates \citep{car}. The evolution of a galaxy is divided into 716 time-steps spanning $1 < t < 1500$ Myr, assuming as a spectral basis the 221 ages in \citealt{bc03}, a \citealt{cha} IMF, and Padova 1994 evolutionary tracks \citep{alo,bres,fag,fag2,gir}.

\begin{figure*}     \centering
     \includegraphics[clip=,width=0.89\textwidth]{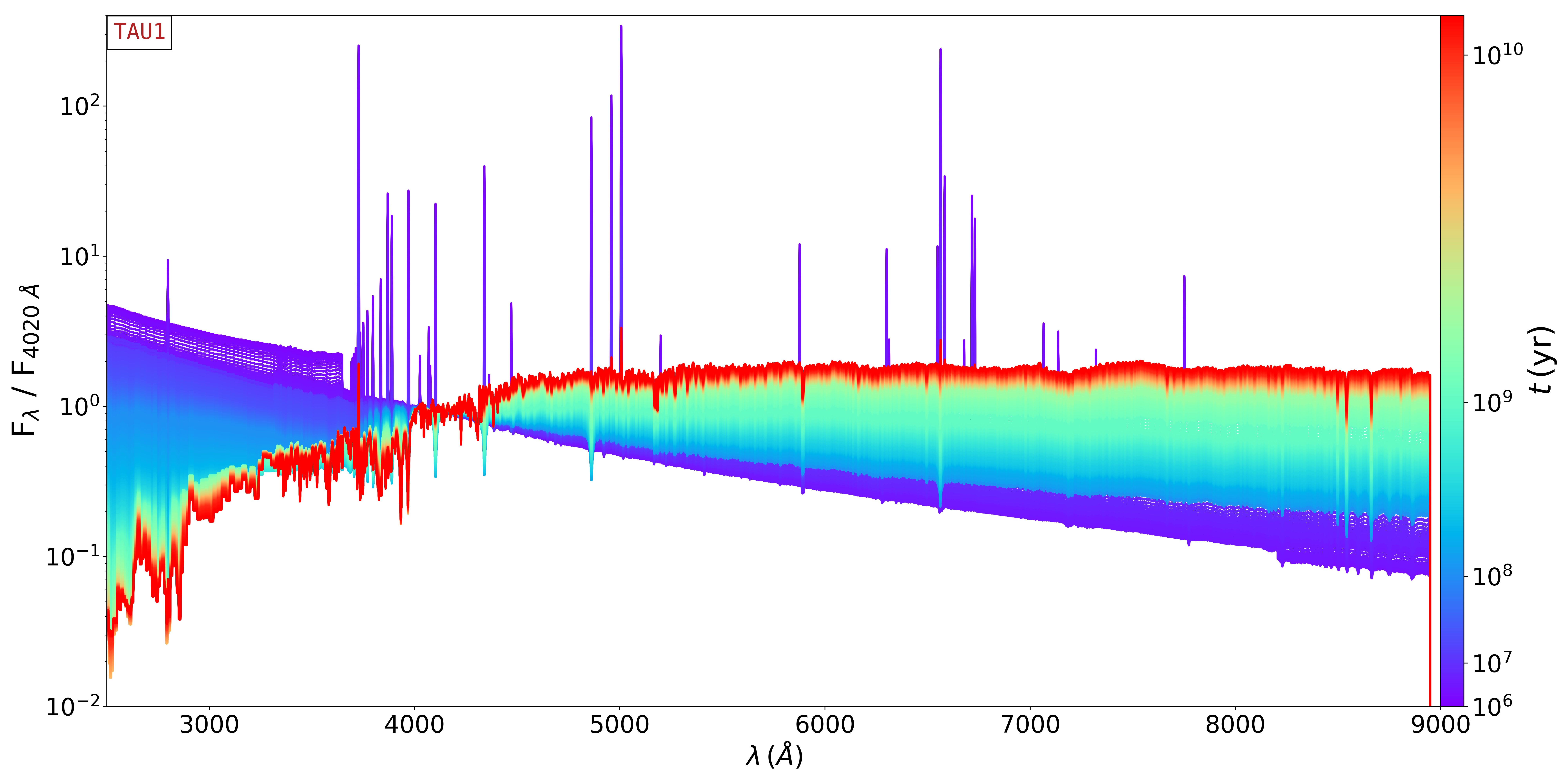}
     \includegraphics[clip=,width=0.89\textwidth]{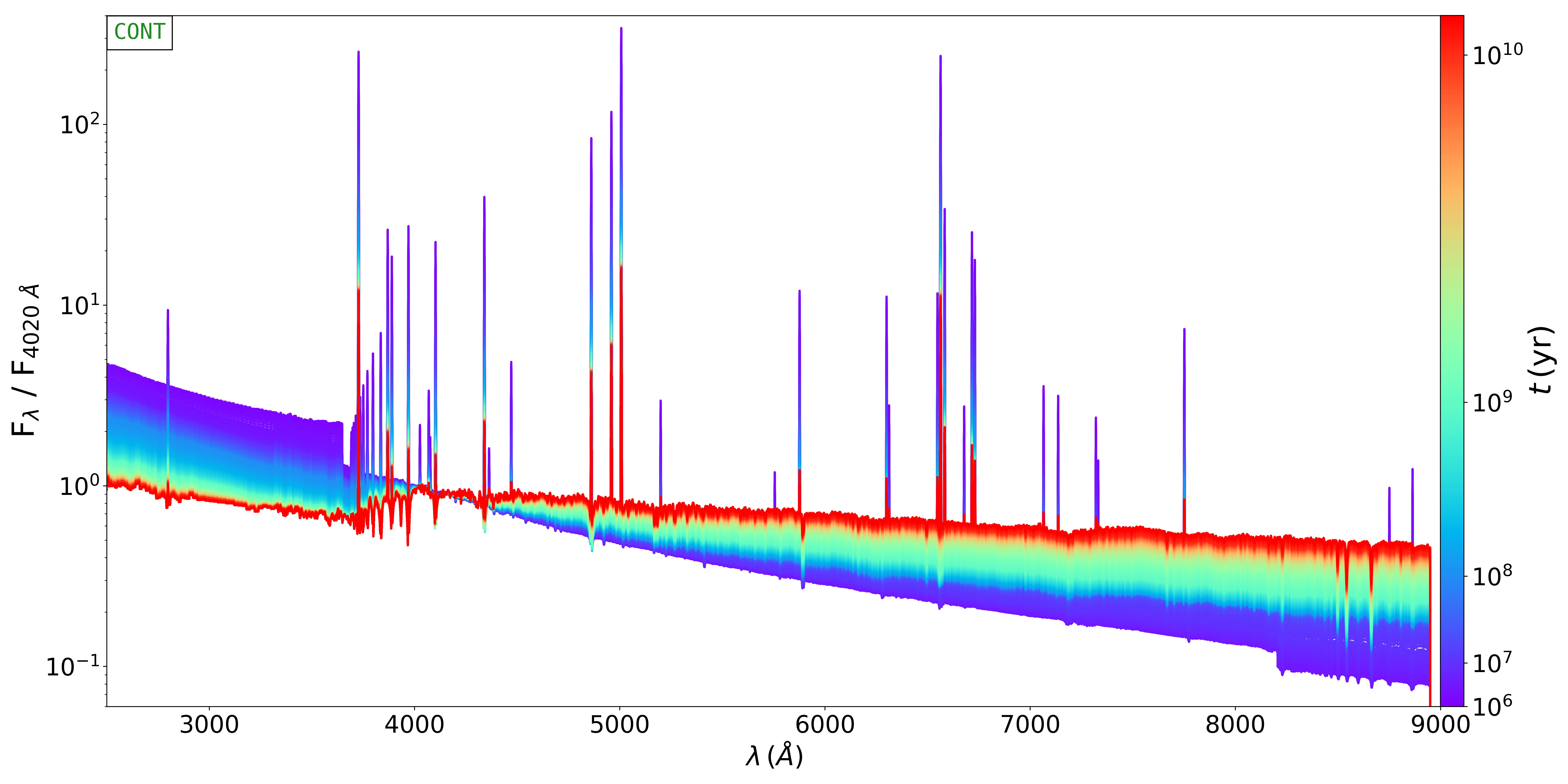}     
     \caption{Synthetic spectra normalised at $\lambda$ = 4020 \AA\ for \Tau\ (instantaneous burst, top panel) and \Con\ (continuous, bottom panel) SFH models. Figures adapted from Fig. 1 and Fig. 2 of \citealt{car}.}
     \label{rebetikospectra}
 \end{figure*}

 The models assume no extinction (V-band extinction of A$_V$ = 0 mag) and no extra line broadening from stellar kinematics to reduce the parameters space studied. The latter is a necessary step in spectral fitting methods, usually estimated convolving the models with a proper kinematical kernel, including physical and instrumental effects. This procedure allows us to improve the constraints on the galaxy kinematics, spot an anomalous PSF, and determine the redshift with higher accuracy when unknown. The \reb\ spectra for the \Con\ and \Tau\ models are shown in Fig. \ref{rebetikospectra} (adapted from \citealt{car}).
 
 The evolution of the synthetic SEDs simulated in \reb\ is characterised by two parameters: the mean stellar age and the mean stellar metallicity.  Each of these parameters is weighted by the light (mass) contribution to the total galaxy light (mass), considering logarithmic time-steps, as in \citealt{cid}:
 \begin{equation}\begin{split}
   \left<log(t_\ast)\right>_L = \sum_{j=1}^{N_\ast} x_j \cdot log (t_j),
  \\
   \left<log(t_\ast)\right>_M = \sum_{j=1}^{N_\ast} \mu_j \cdot log (t_j)
 \end{split}\label{sspAge}.\end{equation}
 
 Analogously for the metallicity,
 
 \begin{equation}\begin{split}
   \left<log(Z_\ast)\right>_L = \sum_{j=1}^{N_\ast} x_j \cdot log (Z_j),
  \\
   \left<log(Z_\ast)\right>_M = \sum_{j=1}^{N_\ast} \mu_j \cdot log (Z_j).
 \end{split}\label{sspZ2}\end{equation} 
 
\begin{figure}\centering
   \includegraphics[clip=,width=0.49\textwidth]{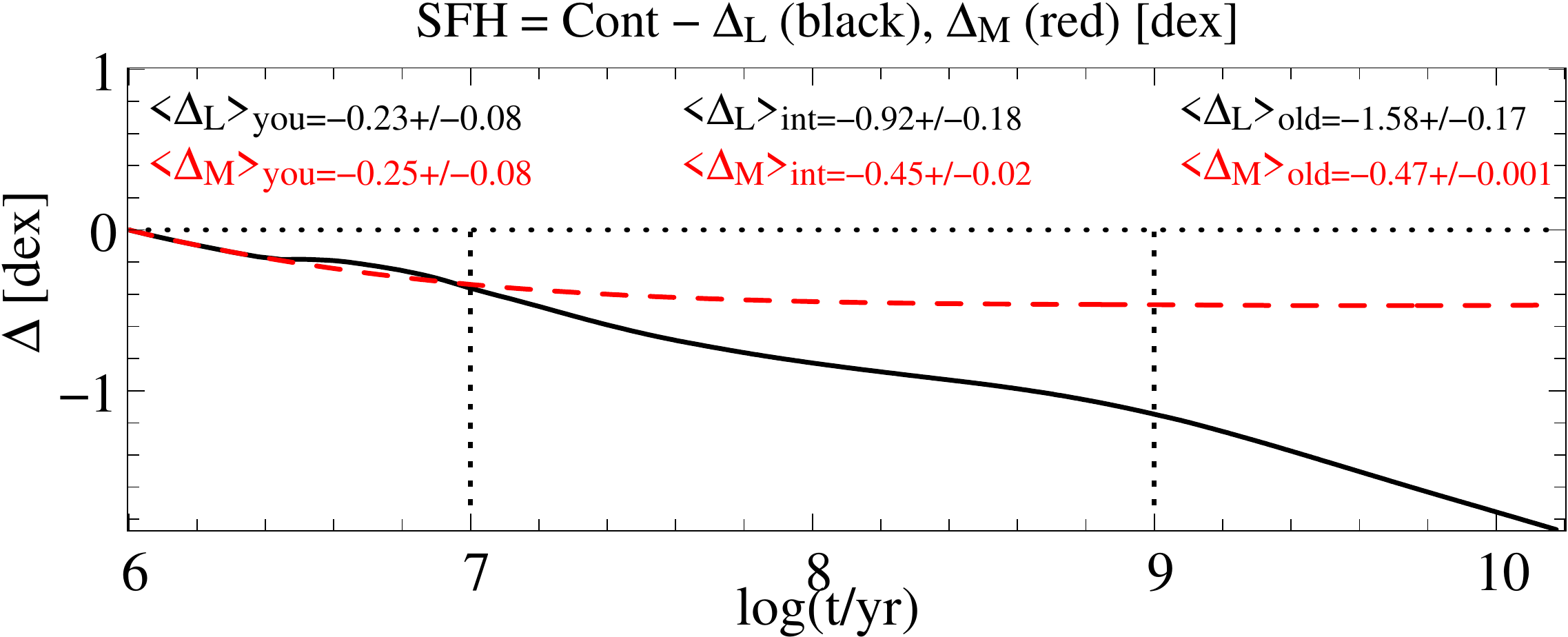}   
     \caption{Differences between luminosity-weighted (solid black line) and mass-weighted (dashed red line) mean stellar ages and the galaxy ages ($\Delta_L$ and $\Delta_M$ in Eq. \ref{del}) as a function of galaxy age for a continuous SFH (\Con). The horizontal dotted line reports 0 dex variations. On top of this line are reported the average $\left<\Delta\right>_L$ (black text) and $\left<\Delta\right>_M$ (red text) for young ($7<\log(t/\mathrm{yr})$), intermediate($7<\log(t/\mathrm{yr})<9$), and old ($\log(t/\mathrm{yr})>9$) ages, separated by vertical dotted lines.}\label{delta}
 \end{figure}

 \noindent
 In Eq. \ref{sspAge} and \ref{sspZ2}, $t_j$ ($Z_j$) represents the age (metallicity) of the j-th SSP element, where $x_j$ and $\mu_j$ are their associated light and mass fractions at 4020 \AA, respectively. Specifically, $\mu_j$ denotes the existent stellar mass $M_\ast$ corrected for the fraction returned to the ISM during stellar evolution. These parameters characterise the stellar population mixture of a galaxy, tracing the main contributors to the observed light and mass. To analyse more quantitatively the evolution of these parameters we defined
 
  \begin{equation}\begin{split}
    \Delta_L = \left<log(t_\ast)\right>_L-log(t)
  \\
    \Delta_M = \left<log(t_\ast)\right>_M-log(t),
 \end{split}\label{del}\end{equation} 
 
 \noindent
 which is the ratio of the mean light (mass) weighted stellar age to the galaxy age. The parameters in Eq. \ref{del} quantify the uncertainties related to the use of average quantities, as in Eq. \ref{sspAge}, to estimate the age of a galaxy. Fig. \ref{delta} shows the evolution of such parameters for the \Con\ model.
 
 Variations of light-weighted quantities with respect to the galaxy age are mainly due to the dominant luminosity contribution that stellar populations younger than 10$^8$ yr give to the total galaxy luminosity. This is a well-known selection effect \citep{mar,gom3}, in which young populations are more luminous than older populations, biasing the determination of the galaxy age. The new generation of stars systematically exceeds the luminosity of older populations, increasing over time the difference between the galaxy age and the luminosity-weighted mean stellar age \citep{con}, as shown in in Fig. \ref{delta}.
 
Mass- and light-weighted quantities underestimate the galaxy ages in the first 10$^7$ yr when star formation begins and the mass and luminosity of the stellar populations are growing. After 10$^7$ yr, the mass-weighted stellar ages evolve steadily, showing a constant offset with respect to the galaxy age of -0.45 dex. To better quantify this trend we divided our simulated galaxy into three different time-steps representing its main evolutionary phases: young (7 $<\log(t/\mathrm{yr})$, intermediate ($7 <\log(t/\mathrm{yr}) < 9$), and old ($\log(t/\mathrm{yr}) > 9$) ages (see also \citealt{wes}).
 
 Fig. \ref{delta} shows that after 10$^7$ yr the mass-weighted stellar ages consistently underestimate the galaxy ages by -0.45 dex (a factor of 3), while for luminosity-weighted quantities the underestimation grows constantly with time, reaching for ages above 1 Gyr an average value of -1.58 dex (a factor of 40). This implies that for galaxies undergoing recent bursts of star formation, because of mergers or self-induced starburst phases, the average stellar population estimated with luminosity-weighted quantities could underestimate the real galaxy age by more than one order of magnitude.
 
 A further consideration regards \logtM, which better reproduces the overall galaxy ages, and besides this, it also has a more physical meaning than luminosity-weighted quantities. As shown in the following sections, despite this theoretical advantage, the determination of \logtM\ remains very challenging because of the large variations of mass-to-light ratios as stellar populations evolve \citep{lei,cid,gom}.
 
\begin{figure}\centering
   \includegraphics[clip=,width=0.49\textwidth]{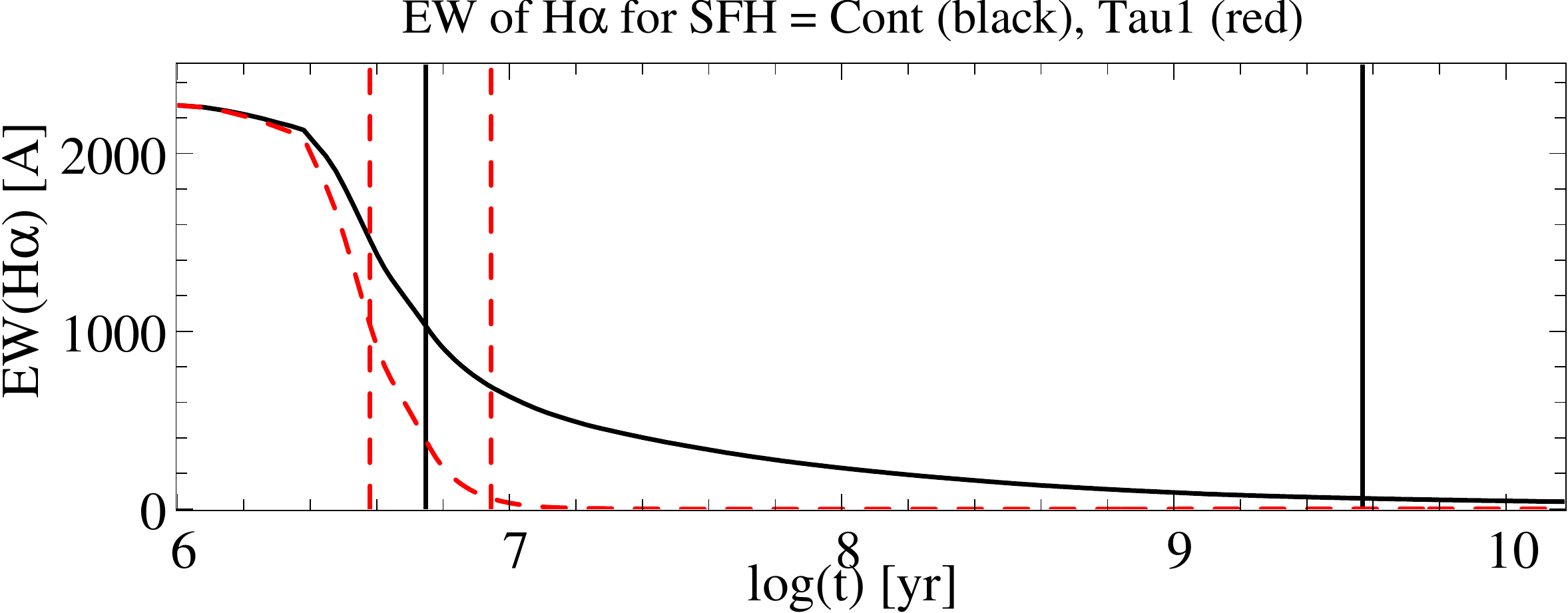}   
     \caption{Evolution of EW(H$\alpha$) as a function of galaxy age for the \Con\ (solid black line) and \Tau\ (dashed red line) models. The solid black and dashed vertical red lines show the ages at which, for each SFH model, the nebular emission contribution is dominant or negligible.}\label{Ha}
 \end{figure}

 To better characterise \reb\ galaxies, we use the H$\alpha$ line emission line EW, EW(H$\alpha$), which gives an indication about the contribution of the nebular emission to the total observed continuum. This contribution, produced by ionised gas, can be significant in specific phases of the  evolution of a galaxy, for example during starburst phases, producing an overall increase in the EW of the emitted H$\alpha$ line. Fig. \ref{Ha} reports the evolution of EW(H$\alpha$) as a function of galaxy age for \Con\ and \Tau\ SFHs. We are interested in two specific phases of the evolution of galaxies, indicated by the vertical dotted lines in the figure where the galaxy experiences a peak in the production of ionising radiation resulting from the starburst phase. 
 
 The Lyman continuum production rate, hence the EW(H$\alpha$) steeply decrease at these ages, albeit remaining high (200-1000 \AA\ for both star formation scenarios). This phase is identified in Fig. \ref{Ha} by the regions EW(H$\alpha$) $> 1000\ \AA$, corresponding to galaxy ages of 5.6 Myr and 3.8 Myr for the \Con\ and \Tau\ models, respectively. After the initial burst, the nebular emission decreases rapidly, depending on the SFH. This is an intermediate phase where nebular emission continues providing an appreciable fraction of the total emission to the case of a continuous SFH; but the intensity of ionising radiation slowly decreases, producing, in turn, weaker emission lines. There is still a rapid jump around 100 Myr, owing to the onset of the post-AGB phase, which produces further gas photo-ionisation \citep{bin,sta,cid2,gom2}. But overall the decline of the nebular emission finally becomes negligible for EW(H$\alpha$) $< 60\ \AA$, which is the typical value of late-type galaxies in nearby Universe (see e.g. \citealt{bred}). These phases are fairly different for the two SFHs chosen: in \Tau\ this drop in the nebular continuum is immediately after the burst, at 8.8 Myr, while in \Con\ model it occurs at later stages, around 3.7 Gyr. This implies for such models a presence of nebular contamination for most of the evolution of the galaxies  \citep{lei}. These timescales must be taken into account when comparing the different results of our analysis in the following sections.
 
 \section{Methods: \fado, \starl, and \stec}
 \label{methods}
 We add to the input \reb\ spectra random Gaussian noise with a varying standard deviation, scaled to a fixed value of the S/N. This procedure assures that each spectrum has a constant S/N throughout the entire range of wavelengths considered in the analysis. In this way we obtained a set of spectra with constant S/N values of 3, 5, 10, 20, 50, and 100, reproducing the range covered by SDSS DR7 (e.g. Fig. 2 in \citealt{bri})
 
 The spectral basis for the analysis is built from the {\tt BaseL} of \citealt{bc03}, considering 100 SSPs for 25 ages (t = 0.001, 0.00209, 0.00316, 0.00501, 0.00661, 0.00871, 0.01, 0.01445, 0.02512, 0.04, 0.055, 0.10152, 0.1609, 0.28612, 0.5088, 0.90479, 1.27805, 1.434, 2.5, 4.25, 6.25, 7.5, 10, 13, and 15 Gyr) and four metallicities ($Z$ = 0.004, 0.008, 0.02 and 0.05).

 We selected three widely used spectral fitting tools for the analysis, which are representative of different approaches to extract evolutionary parameters from a galaxy spectrum: \fado\ \citep{gom}, which includes in its analysis nebular emission; the purely stellar code \starl\ \citep{cid}; and \stec\ \citep{ocv2,ocv}, which in many aspects can be considered a `hybrid' approach (see Sec. \ref{Secste}). 
 In the following section we describe the most important aspects of each code, which are summarised in Table \ref{methodSummary}.
 
 \begin{table}     \centering
     \begin{tabular}{|c|c|c|c|}
     \hline
          &  \fado & \starl & \stec \\
    \hline
    Algorithm used & DEO\tablefootnote{Differential evolution optimisation (\citealt{sto,sto2})}  & SA+MS\tablefootnote{Simulated annealing plus Metropolis scheme \citep{mac}} & MAP\tablefootnote{Maximum a posteriori method}\\
    Nebular emission & Yes & No & No\\
    Automatic masking & Yes & No & No\\
    Predefined ext. law & Yes & Yes & Yes\\
    Gaussian kernel & Yes & Yes & Yes\\
    \hline
     \end{tabular}
     \caption{Main features of the spectral synthesis codes used in this paper.}\label{methodSummary}
 \end{table}
 
 \subsection{\fado}
 
 The \fado\ \citep{gom} population synthesis code allows for simultaneous modelling of stellar and nebular emission in a self-consistent framework. It is currently the only publicly available tool permitting determination of SFH and CEM that consistently accounts for the nebular characteristics (H$\alpha$, H$\beta$ luminosity and EW). This approach reduces the biases in the determination of mass and light-weighted stellar ages, as shown in \cite{car}, and improves the accuracy of the SFH estimation. The code employs a differential genetic optimisation approach called differential evolution optimisation (DEO, \citealt{sto,sto2}), which assures convergence at an affordable expense of computational time. 
 
 The nebular continuum is estimated on the fly, together with electron density and temperature. This separation between stellar and nebular continuum is also maintained for the determination of the intrinsic extinction, which is followed by an automated spectroscopic classification based on diagnostic emission-line ratios. Important features of the code are the following: a) the inclusion of the nebular continuum to the fit, which can have a significant effect on stellar mass determination \citep{gom,car,yua}; b) the automatic masking of emission lines and spurious features (e.g. residuals from the sky line subtraction); c) a pre-defined extinction law, which is set as input parameter; d) a Gaussian kernel considered for broadening effects, both instrumental and physical.

  \subsection{\starl}
  
  The \starl\ code \citep{cid} fits the observed spectrum with a linear combination of SSPs, modelling the extinction with a parametric law and assuming a Gaussian distribution for the line-of-sight stellar motions. Stellar continuum is normalised to an anchor point (usually set at 4020 \AA) varying its shape through a user-defined extinction law.
  
  The minimisation is done through simulated annealing plus the Metropolis scheme, masking regions around emission lines, and bad pixels \citep{mac}. The Metropolis exploration of the parameter space identifies regions of convergence within the solutions space and after an iteration, the step-size of each parameter is changed accordingly. At the same time, the number of steps of the parameters varies by an amount that scales with the step-size variation. The overall effect of such iterations is a gradual focus of the solutions towards the most likely region in the parameter space, thereby avoiding the problem of local minima.
  
  \subsection{\stec}
  \label{Secste}
 
  The \stec\ \citep{ocv2,ocv} is a tool for interpreting galaxy spectra in terms of combinations of different stellar populations. The high level of degeneracy for the solutions is tackled by regularising the problem through different penalty functions in a Bayesian framework. The main advantages of this approach are the following: a) the method is non-parametric and does not assume any specific shape for the SFH and CEM; b) stellar continuum is normalised to have an average of one, and a user defined extinction law is applied on the fly during the minimisation steps; c) the degeneracy problem is partially solved through a proper regularisation, calibrated with high-resolution spectra; and d) the minimal request for the solutions is the continuity and an overall low gradient within the solutions space.
  
  At the beginning of this section, we referred to \stec\ as an example of a hybrid method. This definition originates from the idea that \stec\ cannot be fully considered as a pure population synthesis code since the matrix of the solution is regularised through an ad hoc penalty function, which forces the solution towards a smooth shape, suppressing large gradients. The a priori exclusion of specific solutions is per se incompatible with the concept of population synthesis, which relies totally on the algorithm chosen for the fitting procedure. To be exact, all the methods considered in this work have some sort of a priori assumption in the sense that all the algorithms exclude negative contributions. This can be considered a `pseudo' penalisation, but it is more related to the mathematical formalism of the problem than the physical motivation behind it. For this reason, we consider \stec\ to be a hybrid method because it approaches the problem with a minimisation procedure similar to many spectral synthesis methods, but reserving the right to remove valid solutions on the basis of an underlying assumption about the continuity of the solution.
  
 \begin{figure*}     \centering
     \includegraphics[clip=,width=0.99\textwidth]{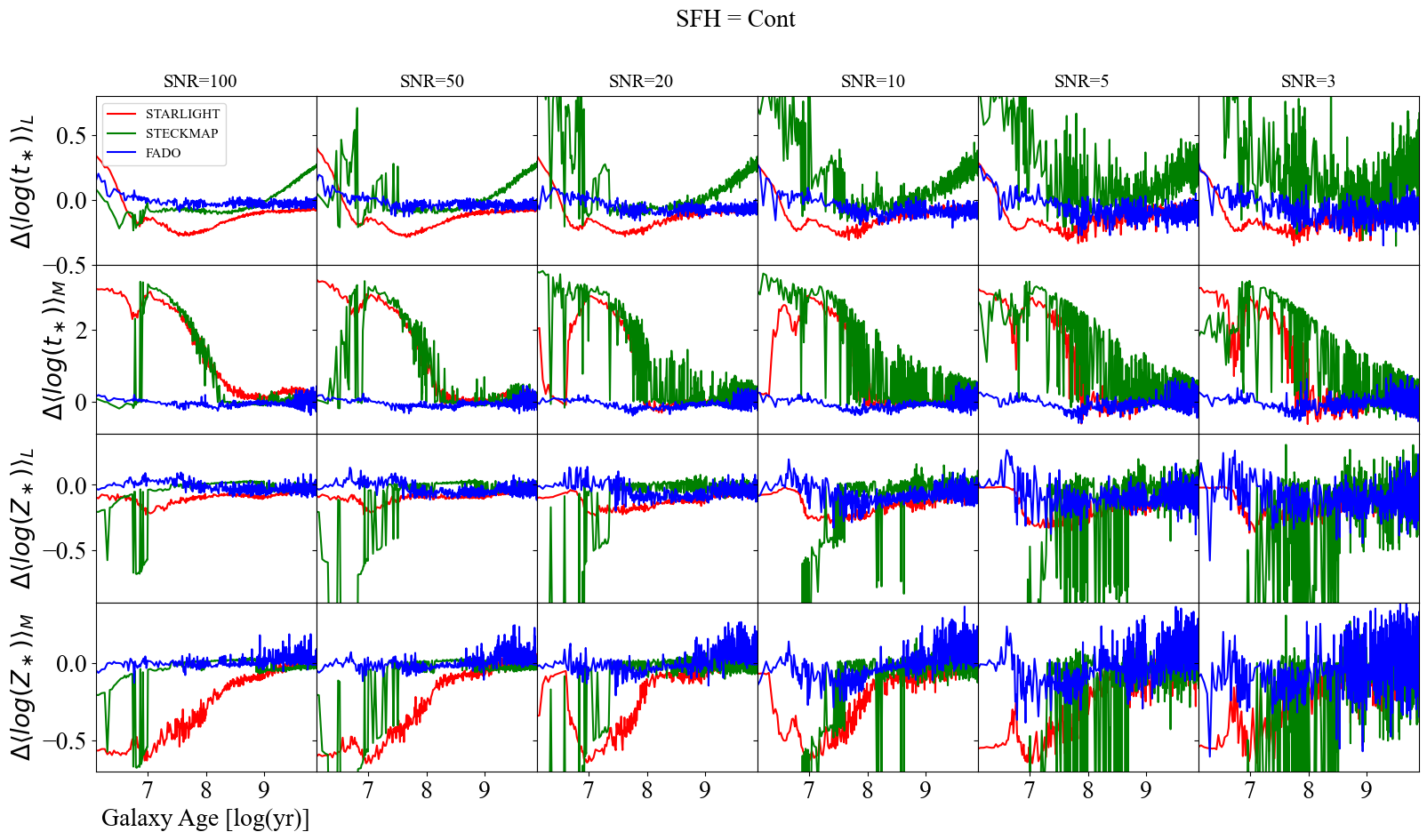}     
     \caption{Results of \Con\ model analysis. Luminosity- and mass-weighted mean stellar age (first and second row), and metallicity (third and fourth row) residuals for \fado\ (blue line), \starl\ (red line), \stec\ (green line) as a function of the galaxy age. Luminosity-weighted quantities for \fado\ and \starl\ are obtained considering the best-fitting populations vector at 4020 \AA. On top of each panel the S/N of the input \reb\ spectrum are reported.}
     \label{Cont-logtL}
 \end{figure*}

 \section{Results}
 \label{results}

  The synthetic spectra built with \reb\ in Section \ref{rebetikoSection} were analysed with the tools described in Section \ref{methods}, for six S/N values in the range $3 \le$ S/N $\le$ 100. For each SFH we estimated the mean stellar age and metallicity, weighted by mass or light, as defined in Eq. \ref{sspAge} and \ref{sspZ2}. Table \ref{table} lists the median of the logarithmic difference between the values recovered by each tool and the \reb\ input parameter. The results for \Con\ and \Tau\ models are discussed separately in the following subsections.

 \begin{table*}\centering
 \begin{tabular}{|c|c|c|c|c|c|c|c|c|c|c|c|}
 \hline
  tool & S/N & SFH & $\Delta t_L$ & $\Delta t_M$ & $\Delta Z_L$ & $\Delta Z_M$ & SFH & $\Delta t_L$ & $\Delta t_M$ & $\Delta Z_L$ & $\Delta Z_M$\\
 (1) & (2) &(3) &(4) &(5) &(6) & (7) & (8)&(9) &(10) & (11) & (12) \\
 \hline\hline
 \fado & 3 &  \Con  & 0.09  & 0.16  & 0.112  & 0.093  & \Tau  & 0.03  & 0.05  & 0.041  & 0.032 \\
 & 5   &&  0.10  & 0.13  & 0.103  & 0.058  && 0.02  & 0.03  & 0.031  & 0.020 \\
 & 10   &&  0.08  & 0.12  & 0.084  & 0.047  && 0.02  & 0.02  & 0.021  & 0.013 \\
 & 20   &&  0.06  & 0.11  & 0.072  & 0.035  && 0.01  & 0.02  & 0.017  & 0.007 \\
 & 50   &&  0.04  & 0.09  & 0.049  & 0.020  && 0.01  & 0.02  & 0.008  & 0.005 \\
 & 100   &&  0.03  & 0.07  & 0.037  & 0.013  && 0.01  & 0.01  & 0.007  & 0.004 \\
 \hline
 \starl & 3 &&  0.11  & 0.15  & 0.111  & 0.096  && 0.03  & 0.06  & 0.036  & 0.028 \\
 & 5   &&  0.10  & 0.12  & 0.110  & 0.079  && 0.02  & 0.04  & 0.022  & 0.018 \\
 & 10   &&  0.08  & 0.08  & 0.093  & 0.059  && 0.01  & 0.03  & 0.014  & 0.010 \\
 & 20   &&  0.08  & 0.14  & 0.082  & 0.043  && 0.01  & 0.03  & 0.009  & 0.008 \\
 & 50   &&  0.08  & 0.20  & 0.068  & 0.044  && 0.01  & 0.03  & 0.007  & 0.006 \\
 & 100   &&  0.08  & 0.24  & 0.070  & 0.042  && 0.01  & 0.03  & 0.007  & 0.005 \\
 \hline
 \stec & 3 &&  0.21  & 0.37  & 0.081  & 0.081  && 0.09  & 0.26  & 0.026  & 0.026 \\
 & 5   &&  0.19  & 0.33  & 0.055  & 0.055  && 0.08  & 0.20  & 0.021  & 0.021 \\
 & 10   &&  0.16  & 0.23  & 0.032  & 0.032  && 0.05  & 0.14  & 0.007  & 0.007 \\
 & 20   &&  0.15  & 0.19  & 0.026  & 0.026  && 0.04  & 0.11  & 0.003  & 0.003 \\
 & 50   &&  0.13  & 0.10  & 0.025  & 0.025  && 0.02  & 0.06  & $<10^{-4}$  & $<10^{-4}$ \\
 & 100   &&  0.11  & 0.07  & 0.023  & 0.024  && 0.01  & 0.04  & $<10^{-4}$  & $<10^{-4}$ \\
 \hline
 \end{tabular}
 \caption{Median of the logarithmic difference between the values of \logtL, \logtM, \logZL, and \logZM\ in dex recovered by each tool and the \reb\ input parameter. Col. 1: Analysis tool; col. 2: Signal-to-noise ratio of the input \reb\ spectra; cols. 3 and 8: SFH of the models; cols. 4-7: median $\Delta$\logtL, $\Delta$\logtM, $\Delta$\logZL, and $\Delta$\logZM\ obtained from \Con\ models; cols. 9-12: median $\Delta$\logtL, $\Delta$\logtM, $\Delta$\logZL, and $\Delta$\logZM\ obtained from \Tau\ models.}\label{table}
 \end{table*}

 \subsection{Continuous star formation: \Con\ models}
 
 Fig. \ref{Cont-logtL} shows the difference between \logtL, \logtM, \logZL, and \logZM\ for \Con\ models with respect to the input \reb\ value, whose median differences are reported in Table \ref{table}.
All tools show large differences of the fit at S/N = 3; \fado\ and \starl\ have a median $\Delta$\logtL$\sim$0.1 dex, while STECKMAP have a higher $\Delta$\logtL of 0.2. All tools have some problems at ages $\log(t/\mathrm{yr}) < 8$, where the results have a large spread as a consequence of the instability of the solutions. The variation in the solutions at those ages and S/N values have their origin in the large error bars of spectra, which enormously increases the number of models able to fit the data. This in turn produces almost flat $\chi^2$ maps, where spectral fitting algorithms could easily remain stuck in a local minimum. From Table \ref{table} we see that \fado\ and \starl\ underestimate \logtL\ by $\sim$ 0.1 dex, and in \fado\ this value is driven mainly by the large errors at $\log(t/\mathrm{yr}) < 8$, where results vary by $\sim$ 0.3-0.4 dex ($\sim$50\%). For older ages, the results become more stable and have a mild tendency to underestimate \logtL. For S/N = 3 \stec\ shows the highest discrepancies, have a median $\Delta$\logtL$\sim$ 0.2 dex, and large differences in the results.
 
 At S/N = 5 the solutions span a similar range in terms of $\Delta$\logtL, varying between 0.1 and 0.2 dex. The uncertainties of the fits at galaxy ages $\log(t/\mathrm{yr}) < 7.5$ are still present; \fado\ and \starl\ show some spread. However, the median $\Delta$\logtL for \fado\ stays within the 0.1 dex because of the higher reliability of the results at older galaxy ages. This trend is confirmed for spectra with S/N $>$ 20, where the uncertainties at $\log(t/\mathrm{yr}) < 8$ disappear, decreasing the $\Delta$\logtL$\sim$0.04 dex and keeping results differences under the 10\%.
 
  For S/N $> 20$ the median differences between the input \reb\ \logtL\ and the best fits from \fado\ are of the order of 0.03 dex ($\sim7$\%), a factor 3 and 4 lower than the 0.08 dex ($\sim$20\%), and 0.12 dex ($\sim$30\%) obtained from \starl\ and \stec, respectively. In contrast to \fado, \starl\ shows a systematic underestimation of the input \logtL\ by 0.08 dex, 20\%, for S/N $>$ 10. The fits show small variations, and the $\Delta$\logtL\ at ages above $\log(t/\mathrm{yr}) \sim 8$ decreases gradually, in correspondence with the predominance of the post-AGB stars. After this phase, we observe higher stability of the fit at $\log(t/\mathrm{yr}) > 9$. An interesting aspect to note are the systematic uncertainties in determining the mean stellar ages at $\log(t/\mathrm{yr}) \sim 7$, that is epochs with relevant nebular contribution to the overall continuum (see Fig. \ref{Ha}).
 
 \begin{figure}\centering
\includegraphics[clip=,width=0.49\textwidth]{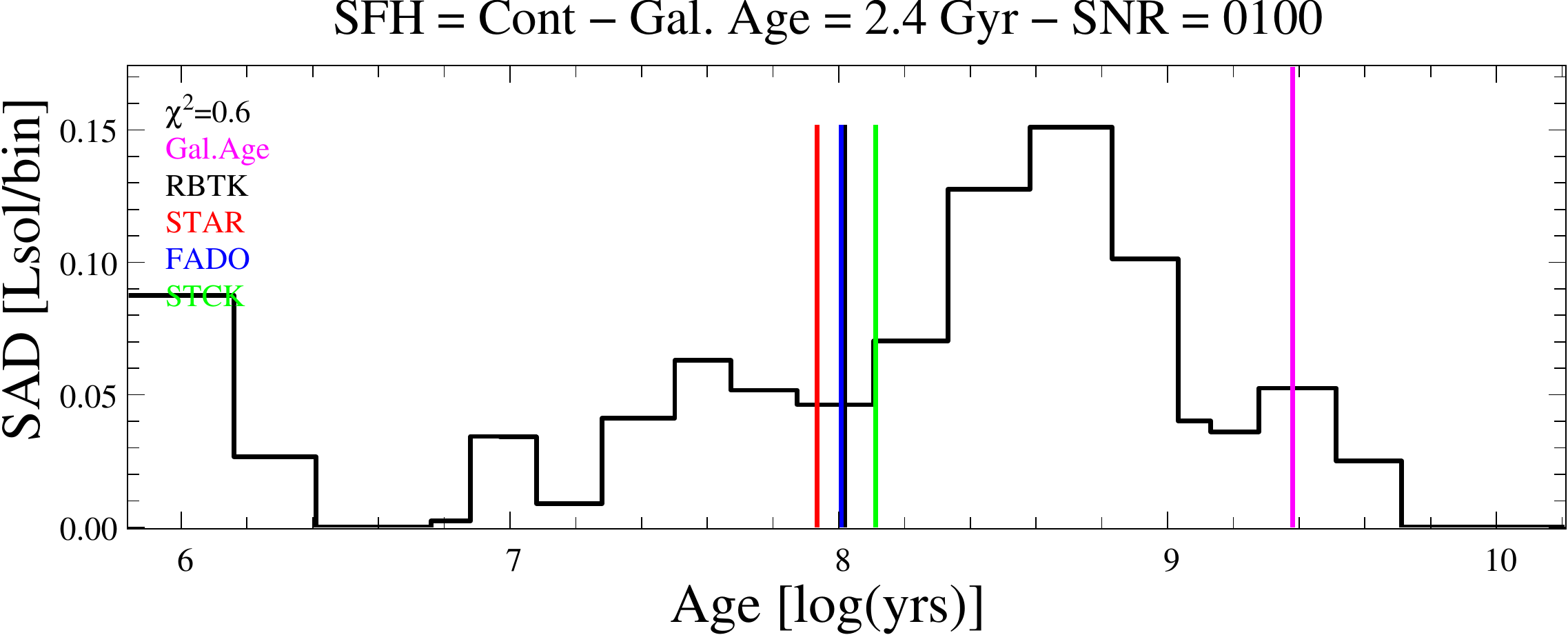} 
 \includegraphics[clip=,width=0.49\textwidth]{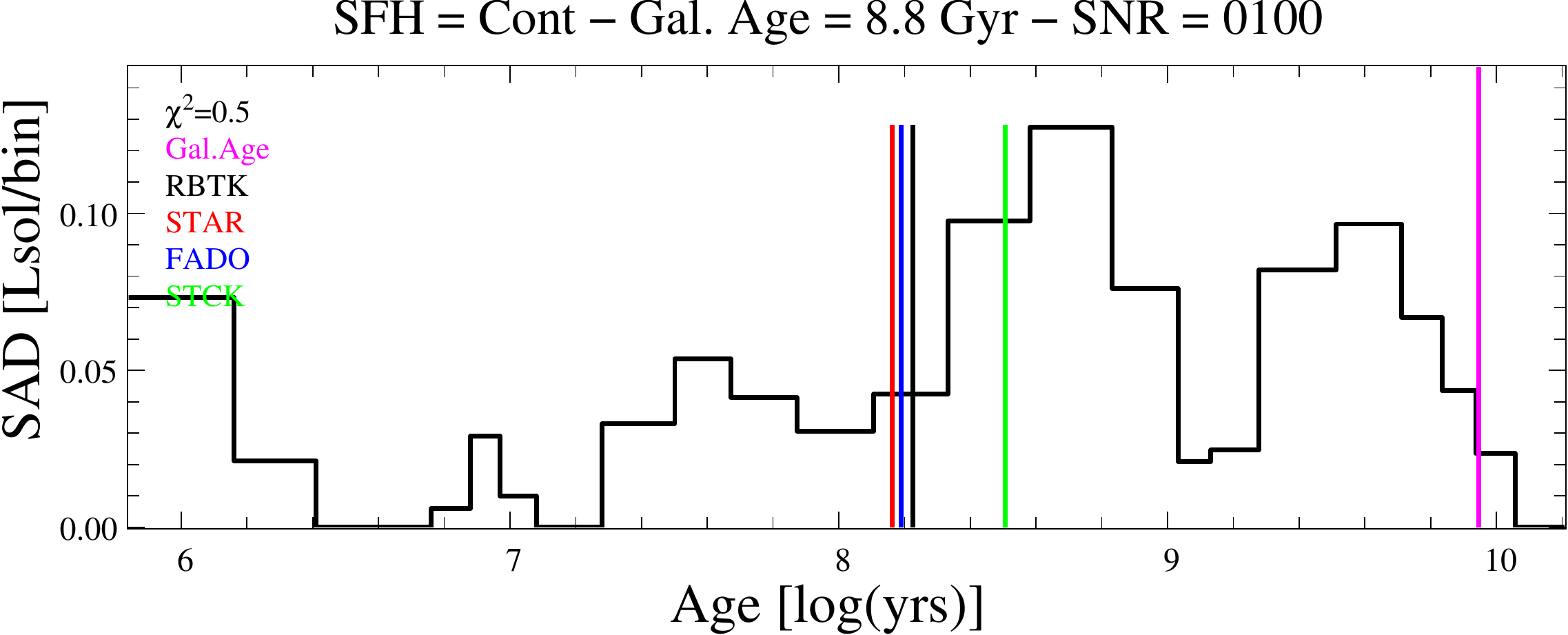}
 \includegraphics[clip=,width=0.49\textwidth]{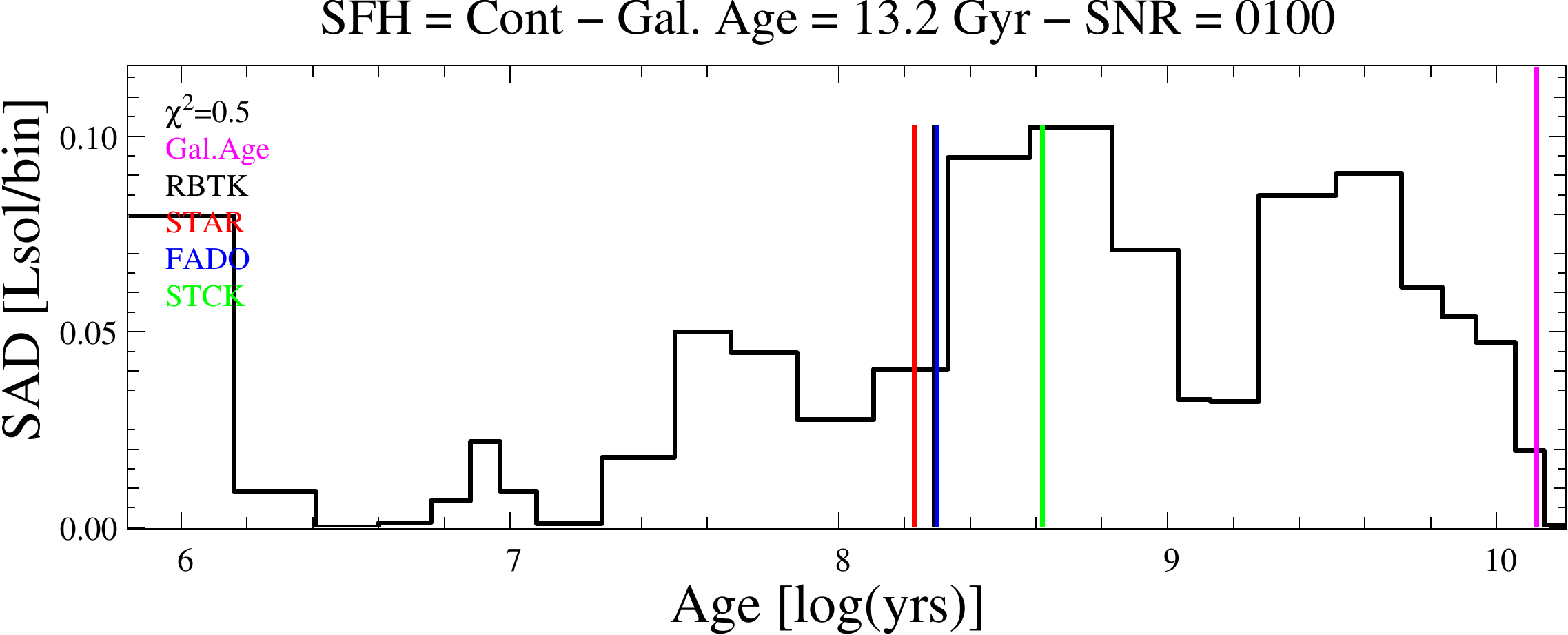}
 \caption{Luminosity-weighted SAD vs. SSPs age for \Con\ models and different \reb\ models: 2.4 (top), 8.8 (middle), and 13.2 (bottom) Gyr. The vertical lines represent the galaxy age from the model (magenta), the input \reb\ value of \logtL\ (black), and the values recovered with \stec\ (green), \fado\ (blue), and \starl\ (red). The top left of each panel indicate the $\chi^2$ of the \stec\ best-fit model.}\label{sparsecoverage}
 \end{figure}
 
 The \stec\ tool shows median differences of the order of 0.15 dex ($\sim$40\%) at all S/N, as shown in Table \ref{table}. On average the results are very stable between $7 < \log(t/\mathrm{yr}) < 9$ yr, with underestimation of \logtL$\sim$0.08 dex at $\log(t/\mathrm{yr}) \le 8$, similarly to that seen for \starl, and an overestimation of $\sim$20\% for ages $\log(t/\mathrm{yr}) > 9$, owing to an effect already mentioned in \cite{car}, and \cite{ocv3}. Both of these papers point towards the sparse coverage of the spectral library used for the analysis at ages $6<\log(t/\mathrm{yr})<7$, which prevents a proper fit for such epochs and produces, in turn, a decline of the luminosity-weighted stellar age distribution (SAD) at these time-steps. This lack of SSPs at those ages is compensated by an increased fraction of old stellar populations which, having a higher mass-to-light ratio, increases the value of the total mean \logtL. This is seen clearly in Fig. \ref{sparsecoverage}, in which we show the luminosity-weighted  SAD as a function of galaxy age for the three different models, at 2.4, 8.8, and 13.2 Gyr. The vertical lines report the value of \logtL\ recovered by the different tools and the input value from \reb. As we can see for each of the models shown, there is a prominent contribution in light from the first time bin at 1 Myr, followed by lower contributions from SSPs with ages between 6 $< \log(t/\mathrm{yr}) <$ 7. Then the shape of the SAD returns to the expected trend, with continuous star formation, but to compensate the missing light at 6 $< \log(t/\mathrm{yr}) <$ 7, the fitting algorithm during the minimisation increases the overall contribution of older SSPs. The final result of this process is the overestimation of \logtL\ for those galaxy ages (see Fig. \ref{sparsecoverage}).
 
 The third row panels of Fig. \ref{Cont-logtL} show the difference of luminosity-weighted mean stellar metallicity \logZL\ at different S/N with respect to the \reb\ input parameter. Broadly speaking, all tools converge towards the input solar metallicity, once input spectra have S/N above 10 and galaxy models have ages above $\log(t/\mathrm{yr}) > 8$. The results of \fado\ show small differences at S/N $>$ 5 on the order of 0.1 dex; there is a mild underestimation of the input value. On the other hand, \starl\ shows some discrepancies, $\Delta$\logZL$\sim 0.2$ dex, for ages below $\log(t/\mathrm{yr}) < 8$, with a similar pattern at S/N = 50 and 100. The \stec\ analysis tool shows the largest scatter in the results at galaxy ages $\log(t/\mathrm{yr})< 8$; there is a mild overestimation of the metallicity at higher S/N and old ages, in any case, lower than 0.01 dex.

 Mass-weighted quantities show fits that are more noisy with respect to light-weighted fits, as shown in Fig. \ref{Cont-logtL} and Table \ref{table}. The \fado\ uncertainties remain around $\sim$0.09 dex, 25\%, at S/N $\ge 10$, which is slightly larger than the 0.04 dex seen for \logtL. An interesting feature to note is the oscillation of the results for ages above 4 Gyr, which corresponds to the epochs where the contribution of nebular emission to the total continuum becomes negligible (see Fig. \ref{Ha}).
 
 The \starl\ results show median discrepancies between 0.12-0.24 dex and there is no real trend with increasing S/N. These values are driven by the large uncertainties shown at all S/N for ages younger than $\log(t/\mathrm{yr}) <$ 8 because of the combined effect of rapid variations of mass-to-light ratios at such ages, together with a large contribution of ionised gas to the observed continuum. This trend is confirmed in \stec, which shows similar $\Delta$\logtM\ for $\log(t/\mathrm{yr}) < 8$ at all S/N. With respect to \starl, these discrepancies decrease strongly with S/N, and uncertainties decrease from 0.23 to 0.1 dex for S/N = 10 and S/N = 50; this feature is discussed in detail in Section \ref{discussion}. We confirm an overestimation of the \logtM\ for ages above 3 Gyr, as seen already in \cite{car}. These variations are of the order of 0.2-0.3 dex for \starl, but decrease at 0.1 dex in \fado. For such ages, \stec\ shows more consistent results, where $\Delta$\logtM\ $< 0.1$ dex.
 
 Also mass-weighted mean stellar metallicities show for all codes more unstable results with respect to light-weighted quantities. In particular, all the tools have difficulties recovering the metallicity evolution at S/N$<$20. In \fado\ we see more stable fits at higher S/N, with mild differences at ages above $\log(t/\mathrm{yr}) > 9$, a feature seen also in \cite{car}. The \starl\ best fits systematically underestimate the metallicity by $\sim$0.01 dex, in particular at young ages, and this trend rises to 0.6 dex for $\log(t/\mathrm{yr}) < 8$, increasing with the S/N. In this context, \stec\ shows more stable results, with a mild underestimation of the metallicity at older ages, similarly to the light-weighted quantities. However, for this tool, we have to consider that the luminosity-weighted quantities are converted into mass-weighted quantities considering a conversion factor that takes into account the metallicity evolution obtained from the fit, normalised by construction. This implies that the differences in the metallicity are of the same order of magnitude and the effect of the mass-to-light ratio overall is negligible (see Eq. 3 in \citealt{ocv}).
 
 \subsection{Instantaneous burst: \Tau\ models}
 
 \begin{figure*}     \centering
     \includegraphics[clip=,width=0.99\textwidth]{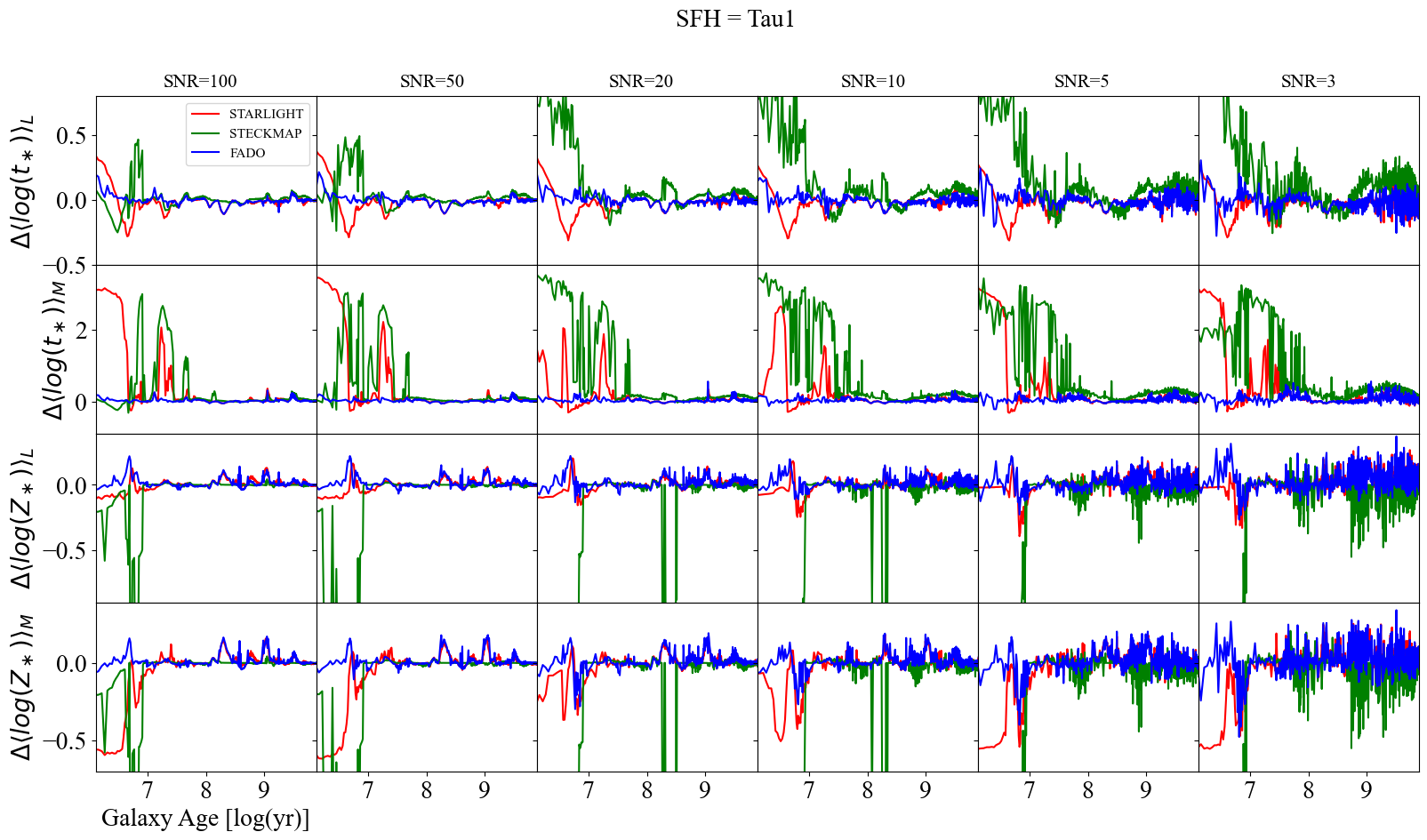}          
     \caption{Same caption of Fig. \ref{Cont-logtL} for \Tau\ SFH models.}\label{Tau1-logtL}
 \end{figure*}
 
 Figure \ref{Tau1-logtL} and Table \ref{table} report the same analysis as the previous subsection, considering SFH for the \Tau\ models. We have already seen in Sec. \ref{rebetikoSection} that the nebular contribution for this SFH models declines very rapidly, and for this reason, we expect higher reliabilities from pure stellar synthesis codes. This is confirmed in Fig. \ref{Tau1-logtL}, which shows an overall consistency of the results; the median $\Delta$\logtL\ is approximately 2-3\% for \fado\ and \starl\ already at S/N = 5. Slight oscillations are seen in \stec\ at ages $\sim$100-400 Myr, as a consequence of the post-AGB stars phase, which adds further degeneracy to the fit procedure, but overall the results are in good agreement with the input \reb\ parameters. A feature in common between \fado\ and \starl\ is a small bump at ages between 3-5 Gyr, presumably due to degeneracies between old SSPs in the adopted SSP library (see \citealt{car}). The results of \stec\ are dependent on the input S/N; a reasonable accuracy is reached at S/N $>$ 10. For lower S/N ratios the fits are more unstable and we observe an overestimation of \logtL\ at $\log(t/\mathrm{yr}) > 7.5$, whose origins are clarified in Section \ref{discussion}.
 
 The results for mean metallicities are more homogeneous than \Con\ models between the different tools. The solutions are more unstable at $\log(t/\mathrm{yr})\sim7$; \fado\ and \starl\ show a peak of $\sim$0.2 dex at $\sim$200 Myr in \logZL\ and \logZM\ resulting from the poor coverage of the spectral basis at those ages (see \citealt{car}). Curiously, \stec\ has similar problems when fitting mean metallicity at ages around $\log(t/\mathrm{yr})\sim 8$, but the sign of the deviation from the true value goes in the opposite direction with respect to \fado\ and \starl, where \stec\ underestimates the \reb\ values for low S/N and the other two codes showing an overestimation. In contrast to \fado\ and \starl, \stec\ fits show high instabilities at older ages above $\log(t/\mathrm{yr}) > 9$.
 
 Analogous to the results obtained for \Con\ models, mass-weighted quantities show a larger scatter, as shown in Fig. \ref{Tau1-logtL} and Table \ref{table}. Overall, the results of \fado\ fits have smaller differences, showing similar $\Delta$\logtL\ and $\Delta$\logtM, $\sim 0.1-0.2$ dex, while for \starl\ and \stec\ the average uncertainties increase by a factor of $\sim$3. Both methods show large uncertainties at ages below $\log(t/\mathrm{yr}) < 8$, where the nebular contribution is still significant. For \stec\ the scatter decreases at S/N $>$ 10, while for \fado\ and \starl\ the fits are more stable for S/N $\ge$ 5.
 
 \section{Discussion}
 \label{discussion}
 
 As we have seen in Sec. \ref{results}, the reliability of the results for each code is strongly dependent not only on the input SFH but also on other factors: the S/N; the contribution of the nebular continuum, particularly for pure stellar synthesis codes; and the age coverage of the spectral basis. Regarding stellar ages, the accuracy of the estimated \logtL\ and \logtM\ changes drastically according to the input SFH. For \Con\ models, Table \ref{table} shows that for \fado\ the uncertainties increase by a factor between 3-5 from the use of light- and mass-weighted mean stellar ages. For \Tau\ models there are no relevant differences between light- and mass-weighted stellar ages. In \starl\ the uncertainties are about a factor 2-4, for both \Con\ and \Tau\ models. The average accuracy of 0.15 dex for \logtM\ seen for \starl\ confirms the results of a previous analysis of \cite{cid}, and are typical of purely stellar codes \citep{ocv2,ocv,toj,kol,wil,car}. 
 
 Larger discrepancies are associated with evolutionary phases where the nebular contribution to the continuum is not negligible, which is particularly evident in \Con\ spectra. This confirms the importance of considering nebular emission for starburst systems, where this component can reach up to 50\% of the total optical and near-infrared emission \citep{kru,izo,papa,lei,sch}.  

 As discussed in Section \ref{rebetikoSection}, the parameters defined in Eq. \ref{sspAge} to quantify the assembly history of a galaxy must be used with caution. By definition, these are averaged quantities that miss important details of galaxy evolution, such as starburst episodes or sudden shutdown of the star formation, and they should be considered as the first moment of the distribution \citep{cid}. These details, however, can be investigated more thoroughly by analysing the luminosity-weighted SAD and the SFH recovered from the fitting process. Such analysis allows for a more complete insight into the fitting procedure, providing hints about the observed discrepancies and helping to interpret the results.
 
 For the mean stellar metallicities defined in Eq. \ref{sspZ2} the same point is valid, even if there is a further warning to consider. The synthetic spectra built in Section \ref{rebetikoSection} assumed a constant solar metallicity: since the SFHs chosen to model our galaxies are different, the constant metallicity imposed for \Con\ and \Tau\ models is a simplification. For \Con\ models, a constant SFR implies that the quantity of processed material within the galaxy increases with time, raising, in turn, the value of the average metallicity. To keep constant $Z$, a mechanism responsible for metals dilution must be introduced, such as a constant cosmic gas accretion from the circum-galactic medium. On the other hand, \Tau\ models require the opposite effect: a closed-box model, in which the galaxy does not produce further stellar populations, and the constant metallicity is guaranteed by the absence of any replenishment or inflow of low-metallicity gas from the circum-galactic medium \citep[see][for a more complete discussion]{tum} and the absence of new star formation episodes, responsible for ISM metal enrichment. Some additional cautionary remarks are relevant, for example the assumptions of a) no velocity broadening, b) constant spectral resolution, c) constant noise across the considered spectral range, d) constant electron density and temperature, e) no Lyman continuum photon leakage throughout the evolution of a galaxy, and f) no shocks are strong simplifications. Therefore the presented tests do not mean that application of the three codes to real spectra will closely match what is found in this work. However, it is reasonable to assume that in broad terms the biases found in this study, in particular the overestimation of \logtM\ in phases with strong nebular emission, should affect studies of real galaxy spectra.
 
\subsection{Uncertainties introduced by a different spectral basis or a change in the wavelength coverage}

An important and complex aspect to consider in our analysis is the impact on the results of a different spectral basis or a change in the wavelength coverage. These issues are beyond the goal of this paper, and it is part of ongoing works within our team (Gomes et al. in prep). However, in this section, a qualitative analysis gives some hints on possible systematics, reproducing in our tests existing instruments.

\begin{figure*}    \centering
    \includegraphics[clip=,width=0.9\textwidth]{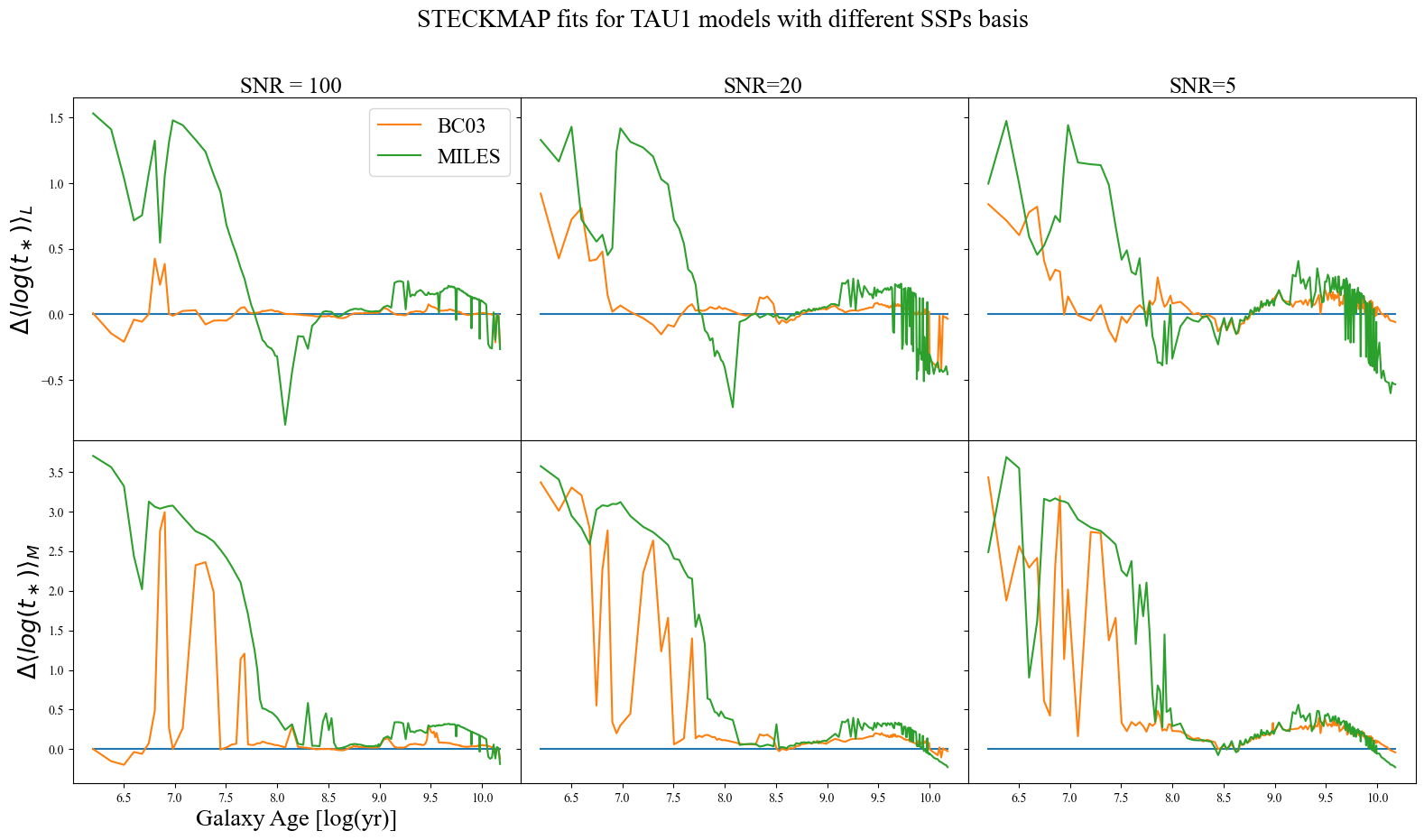}
    \caption{Luminosity- and mass-weighted mean stellar age (1st and 2nd row) residuals for \stec, obtained with {\tt BaseL} \citep[yellow line,][]{bc03} and MILES \citep[green line,][]{vaz} for an instantaneous burst SFH (\Tau\ models). The S/N of the input \reb \ spectra are indicated on the top of each panel.}\label{ssp}
\end{figure*}

A different SSP basis would certainly affect our results, reflecting at a first-order the systematic differences in the stellar basis used to build the SSPs, as shown for example in \cite{kol2}. The use of different stellar libraries has a strong effect on the near-infrared regime, while the effect in the optical is negligible, as seen for example in \cite{bal} (see their Fig. 7). In their work \cite{bal} show that the largest effect on the derived SFHs is related to the stellar spectral library chosen for the analysis, impacting more the results than the method considered. To better quantify this effect we compare in Fig. \ref{ssp} the residuals between the mean stellar ages estimated with \stec\ and the \reb\ input obtained with {\tt BaseL} as a function of the age of the galaxy for the \Tau\ models. The results were obtained using as a spectral basis the standard {\tt BaseL,} defined in Sec. \ref{methods} (yellow line), and MILES \citep[][green line]{vaz}.

The main differences in the light-weighted quantities are seen for ages below 10$^8$ yr, where the age coverage of MILES is sparser than {\tt BaseL}, resulting in larger uncertainties. For older stellar ages MILES fits overestimate the input value and have differences below 0.5 dexes. For mass-weighted stellar ages, the trends reported are qualitatively similar, and the overestimation starts at $\sim 10^8$ yr for both spectral bases. The overall differences are reported in Table \ref{sspT}, showing a general consistency of the results obtained with the two SSPs basis.

\begin{table}    \centering
\Tau\ Models
 \begin{tabular}{|c|c|c|c|c|c|c|c|c|c|c|c|}
 \hline
  tool & S/N & SSP & $\Delta t_L$ & $\Delta t_M$ & SSP & $\Delta t_L$ & $\Delta t_M$\\
 (1) & (2) &(3) &(4) &(5) &(6) & (7) & (8)\\
 \hline\hline
 \stec & 5 & {\tt BL} & 0.08 & 0.2  & {\tt ML} & 0.09  & 0.23\\
  & 20 &&  0.04  & 0.11  && 0.04 & 0.19 \\
 & 100   &&  0.01  & 0.04  && 0.12 & 0.24 \\
 \hline
 \end{tabular}
 \caption{Median of the logarithmic difference between the values of \logtL and \logtM\ in dex recovered by \stec\ and the \reb\ input parameter for \Tau\ models. Col. 1: Analysis tool; col. 2: Signal-to-noise ratio of the input \reb\ spectra; cols. 3, and 6: spectral basis used: {\tt BaseL} as defined in Sec. \ref{methods} ({\tt BL}), and MILES ({\tt ML}, \citealt{vaz}); cols. 4-5-7-8: median $\Delta$\logtL, $\Delta$\logtM\ for each SSP.}\label{sspT}
\end{table}

The problem of the spectral basis considered in a fit has been thoroughly addressed in \cite{car}, which showed how larger base libraries do not necessarily improve the results. The important part in building a good base is highly dependent on the type of sample in question (e.g. spectral coverage; passive or star-forming galaxies). For instance, a good base could require sufficient young stellar populations  $<$ 100 Myr in age to properly cover the star-forming regimes, while at the same time not having too many old stellar populations $>$ 1-10 Gyr in age because of the expected nature of the objects analysed.

\begin{figure*}    \centering
    \includegraphics[clip=,width=0.99\textwidth]{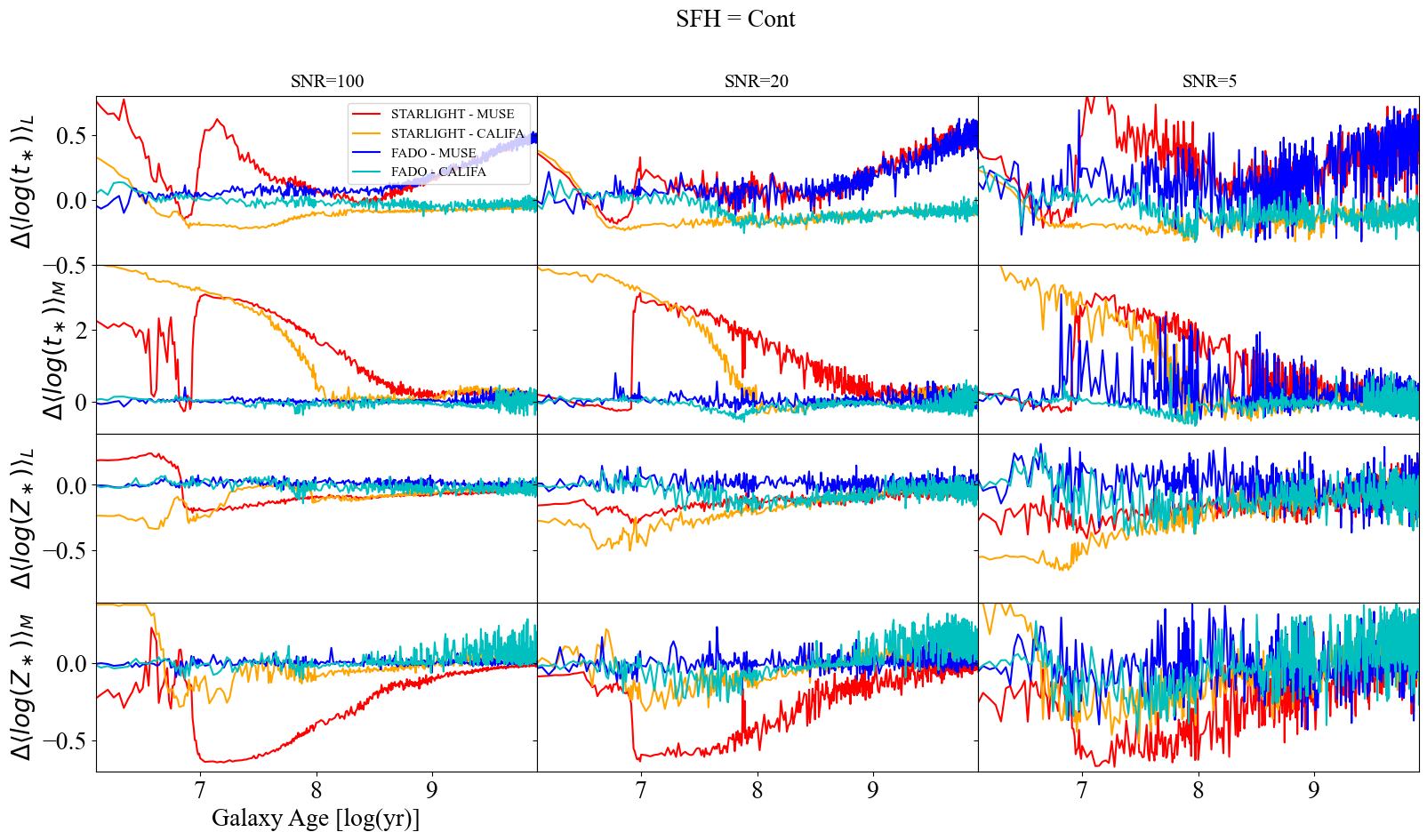}
    \caption{Luminosity- and mass-weighted mean stellar age (first and second row), and metallicity (third and fouth row) residuals for \fado \ (blue and cyan lines) and \starl \ (red and orange lines), obtained with MUSE (4800-9300 \AA) and CALIFA (3500-7000 \AA) spectral coverage (see labels) for a continuous SFH (\Con \ models). The S/N of the input \reb \ spectra are indicated on the top of each panel.}\label{spcCov1}
\end{figure*}
\begin{figure*}    \centering
    \includegraphics[clip=,width=0.99\textwidth]{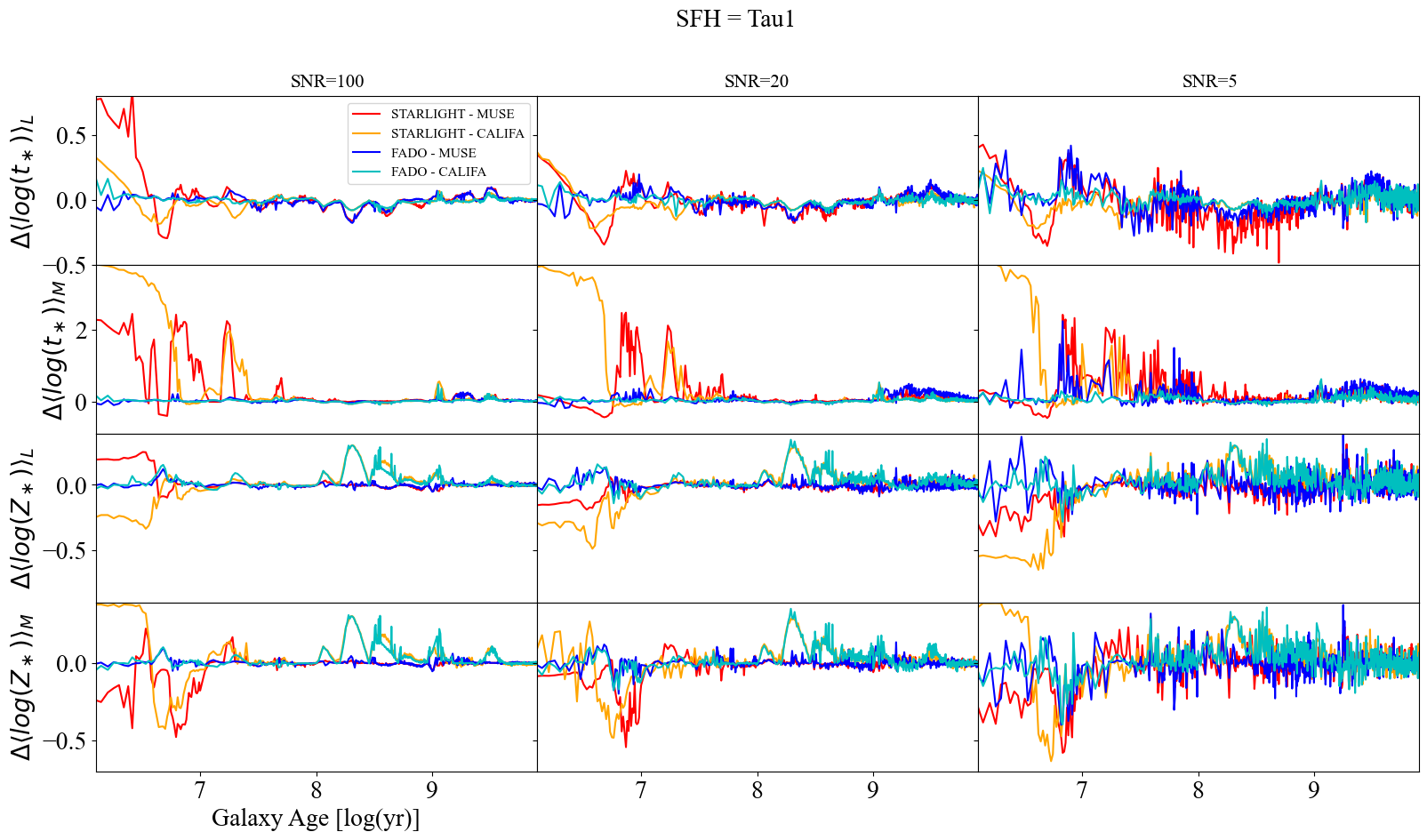}
    \caption{Same caption of Fig. \ref{spcCov1} for \Tau \ SFH models.}\label{spcCov2}
\end{figure*}

\bigskip

Concerning the wavelength coverage issue, Figures \ref{spcCov1} and \ref{spcCov2} show the difference between the estimated stellar properties (mean ages and metallicities) as a function of the age of the galaxy model for continuous (Fig. \ref{spcCov1}) and instantaneous (Fig. \ref{spcCov2}) SFHs. The results shown have been obtained by fitting the synthetic spectra from \reb \ in two spectral ranges: one simulating that of MUSE instrument \citep[][$\sim$4800-9300 \AA, including H$\alpha$+H$\beta$ as well as the Paschen jump]{bac}, and another in the range 3500-7000 \AA, simulating the combined V1200+V500 data from the Potsdam Multi-Aperture Spectrophotometer \citep[PMAS,][]{kel}, used in the CALIFA survey \citep{san}. The blue and red lines indicate the results obtained with MUSE coverage with \fado \ and \starl, while the cyan and orange lines show the best fits for the CALIFA coverage. The plots show results for S/N = 5, 20, and 100, but the trends seen are present in the other S/N values previously discussed, only varying in the amount of noise in the biases.

As a general comment, \fado \ fits show overall a better agreement with the input spectra compared to \starl in both coverage set-ups. However, each set-up yields new trends in both codes, justifying a more detailed interpretation in a future paper, as already mentioned at the beginning of this section. A bluer coverage in \fado, such as that in CALIFA, yields better light-weighted age estimates than its red counterpart (MUSE) for both SFHs. Interestingly, both \fado \ and \starl\ show an overestimation of stellar ages when MUSE coverage is considered for a continuous SFH with increasing galaxy age, linked to degeneracies between the old SSPs in the base (Fig. \ref{spcCov1}). This is rather intuitive since the continua of old SSPs ($>$ 1 Gyr) do not strongly differ in 6000-8900 \AA. A bluer coverage in \starl \ also yields better results than a redder coverage.

The results flip when looking at the light-weighted metallicity, and for such cases, a redder coverage gives better results than a bluer coverage in \fado \ and \starl\ for an instantaneous burst SFH (Fig. \ref{spcCov2}). The same holds for a continuous SFH  in \starl. When looking at the mass-weighted age and metallicity, \fado\ yields better results when assuming a red coverage; uncertainties in blue are probably linked to a poor age coverage of the adopted base library. The \starl\ tool also yields better mass-weighted age estimates when red coverage is available, but this effect is less prominent for an instantaneous burst. Finally, the blue coverage leads to a systematic metallicity overestimation in both codes (especially for older galaxy models), whereas a red coverage leads to a systematic metallicity underestimation in \starl\ for decreasing model age.

In summary, \fado\ seems to be more efficient than \starl\ in recovering the mean age and metallicity during strong star-forming evolutionary stages ($t >$ 0.1 Gyr) for MUSE and CALIFA coverages, regardless of S/N and SFH. These results show that the impact of the spectral coverage on this type of analysis is not negligible and this is something we will explore in a future paper (Gomes et al. in prep.). For example, an important caveat to this trend is the systematic overestimation of the light-weighted mean stellar age by \fado\ in the MUSE set-up after $t >$ 0.1 Gyr and for a continuous SFH, which is likely due to the absence of the Balmer jump at 3646 \AA\ needed to assure self-consistency between the stellar and nebular spectral contributions. This absence does not affect the instantaneous burst SFH (\Tau\ models) because after $t >$ 0.1 Gyr there is only a residual nebular spectral contribution coming from post-AGB stars.

\subsection{Star formation history from purely stellar codes}

 A systematic feature seen in \starl\ and \stec\ is the overestimation of \logtM\ for $\log(t/\mathrm{yr}) < 8$ (see Fig. \ref{Cont-logtL} and \ref{Tau1-logtL}). This feature affects both \Con\ and \Tau\ models, and there is a tendency to decrease with increasing S/N. The large errors, up to 3 dex, are associated with galaxy ages for which the contribution of nebular emission to the overall continuum is relevant (Fig. \ref{Ha}) and points towards the origin of these discrepancies.  
 
 \begin{figure*}\centering
     \includegraphics[clip=,width=0.69\textwidth]{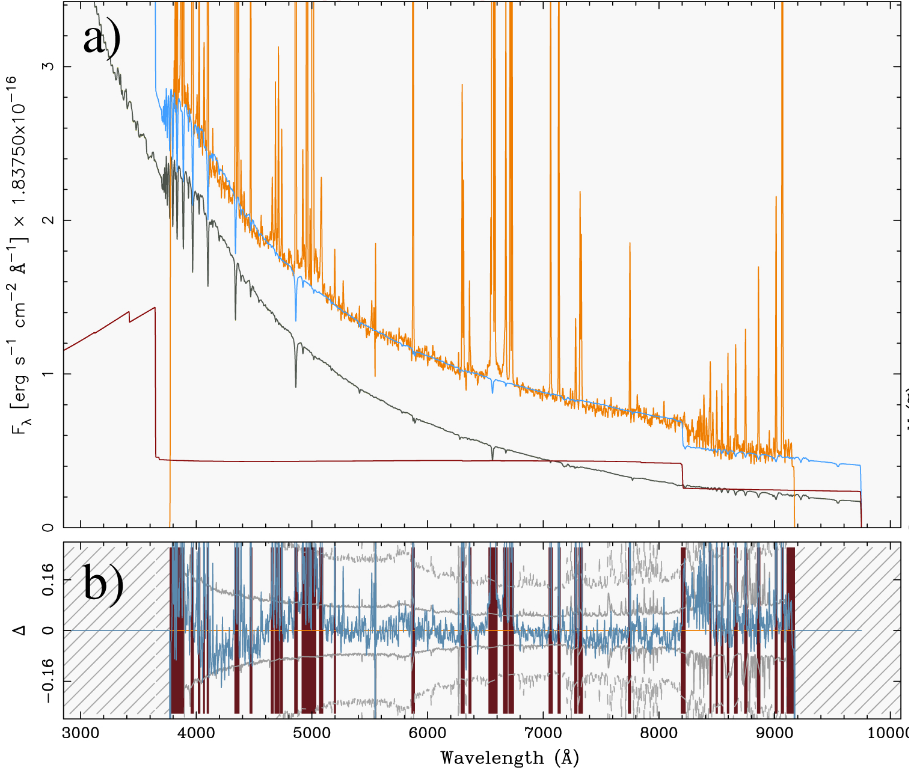}
     \caption{\fado\ spectral fit example, from \cite{gom}, for the  metal-poor blue compact galaxies (BCD CGCG 007-025, yellow line) observed in the SDSS (see also \citealt{gus}). Panel {\it a)}: Best-fitting synthetic SED with stellar and nebular continuum emission, shown in blue, grey and red, respectively. Panel {\it b)}: Residuals between fit and observed spectrum; the shaded area and the dashed curve delineate the $\pm1\sigma$ and $\pm3\sigma$ error spectrum, respectively.}\label{fadoexample}
 \end{figure*}
 
 Fig. \ref{fadoexample} shows an example of a best-fit spectrum obtained with \fado\ for a metal-poor blue compact galaxy (BCD CGCG 007-025). The contribution to the overall spectrum due to the nebular continuum is shown as a red line below the spectrum. The peak of this contribution is at wavelengths below 4000 \AA, coinciding with the Balmer jump at $\sim$3646 \AA. The contribution of the nebular continuum increases with increasing wavelength, becoming dominant in the region around the Paschen jump at $\sim$ 8207 \AA. In this part of the spectrum, both contributions are relatively small because of the decreased stellar continuum, which is typical of star-forming galaxies. In the redder part of the spectrum the nebular continuum is dominant, and this has profound implications for pure spectral synthesis codes: as discussed in \citealt{izo2} and \citealt{papa2} the reddish nebular continuum of starburst galaxies monotonically increases with wavelength and forces purely stellar SED fitting codes to artificially raise the fraction of old stars. 
 
 \begin{figure*}\centering
      \textbf{SFH = \Con}\par\medskip
 \includegraphics[clip=,width=0.49\textwidth]{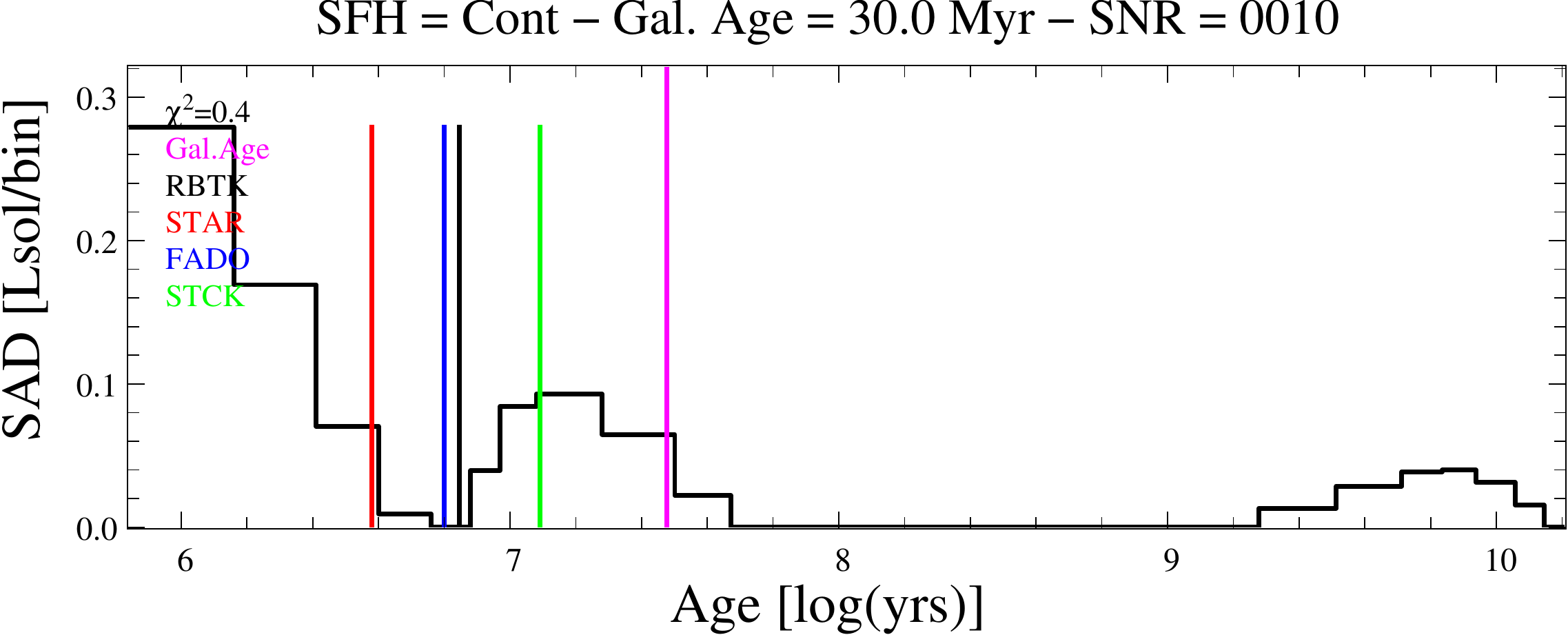}  
 \includegraphics[clip=,width=0.5\textwidth]{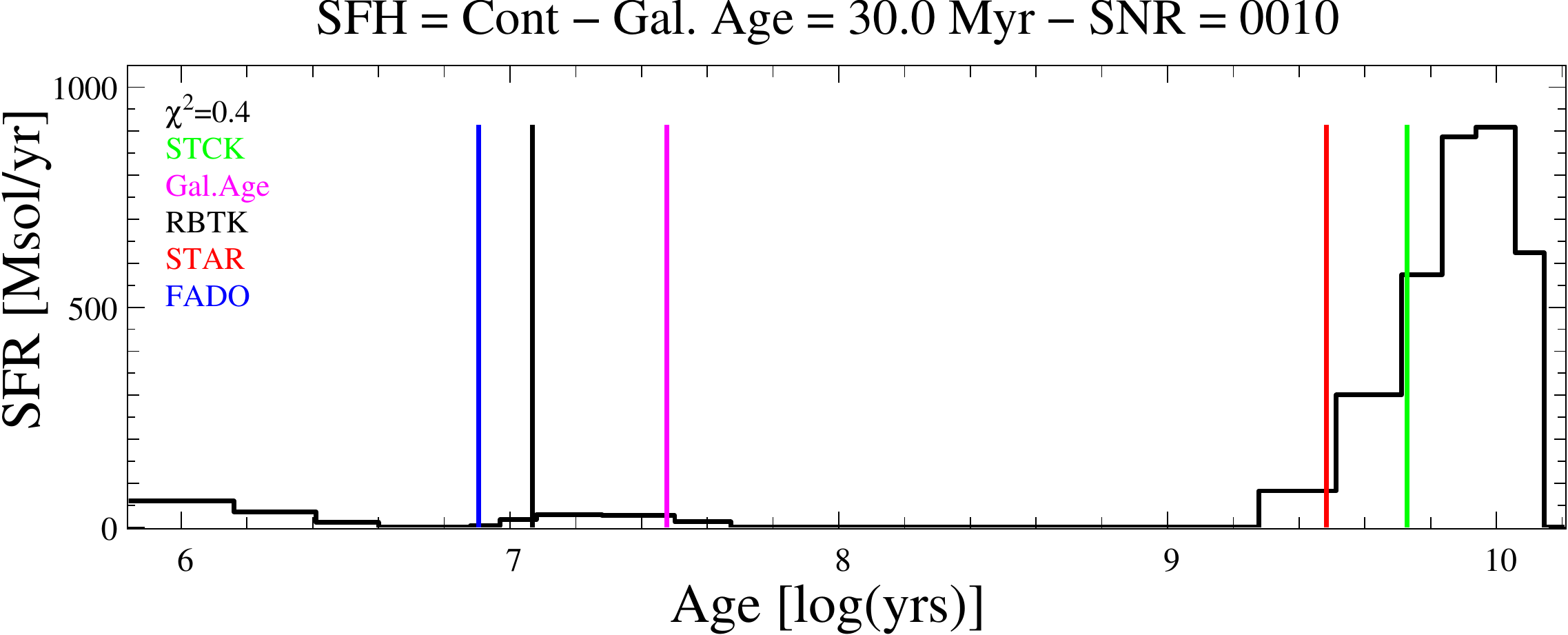} 
 
      \textbf{SFH = \Tau}\par\medskip
      
 \includegraphics[clip=,width=0.49\textwidth]{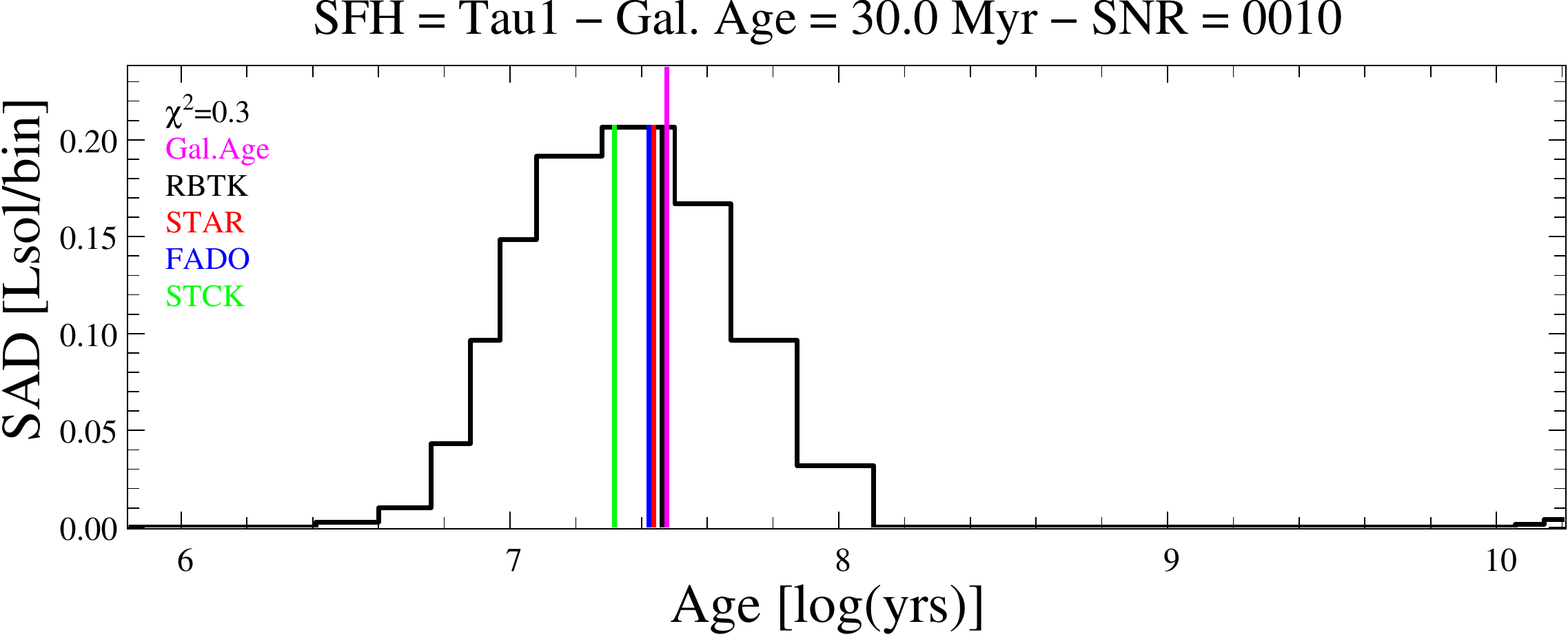}  
 \includegraphics[clip=,width=0.5\textwidth]{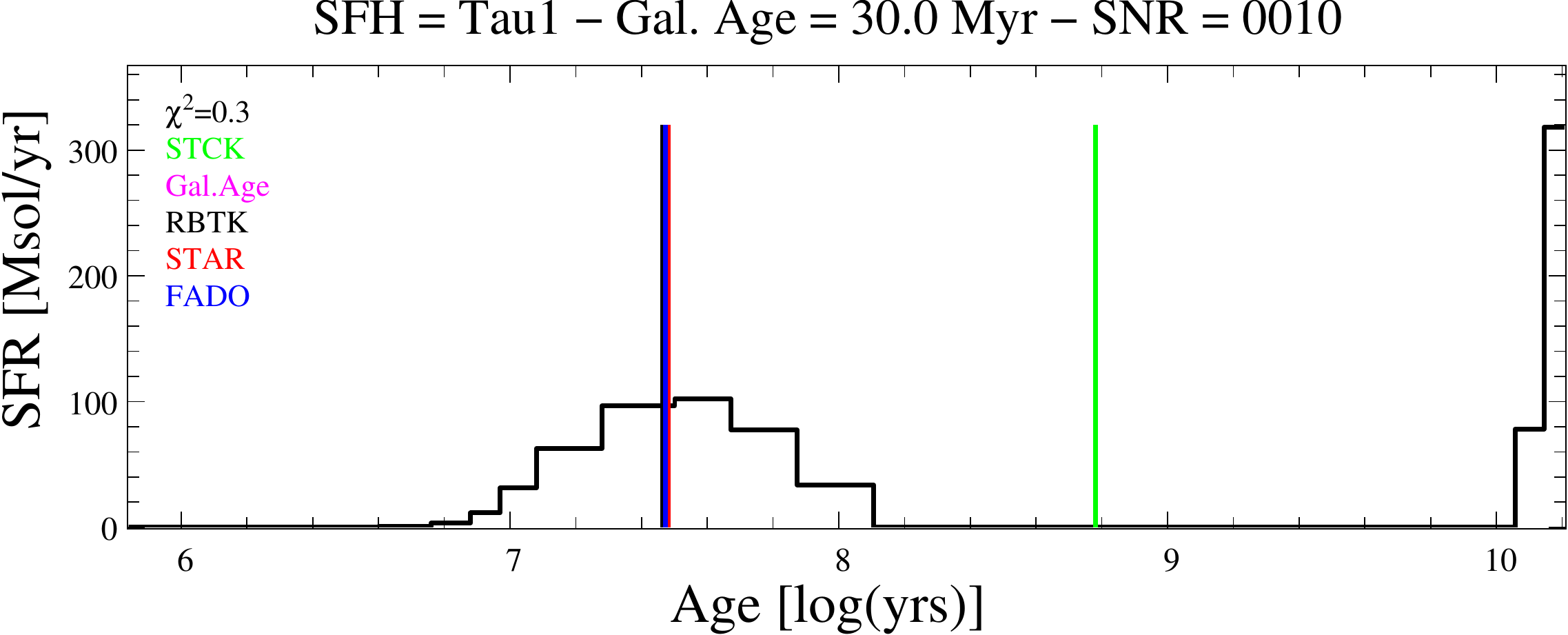}  
 \caption{Luminosity-weighted SAD (left panels) and SFR (right panels) vs. the SSPs ages obtained with \stec\ for \Con\ (top) and \Tau\ (bottom) models with S/N = 10 at a galaxy age of 30 Myr. The vertical lines are shown as in Fig. \ref{sparsecoverage}.}\label{redNebular}
 \end{figure*}

 For luminosity-weighted quantities this effect is less relevant, even in the presence of a strong nebular component. Fig. \ref{redNebular} shows the luminosity-weighted SAD and the SFH obtained for \Con\ and \Tau\ models with \stec\ at galaxy ages of 30 Myr. The \Con\ models show a small bump of old stellar populations around $\log(t/\mathrm{yr}) > 9$ (top left panel); but this bump, because of the high mass-to-light ratios of these stellar populations, does not affect the estimation of \logtL, which is recovered with an error of $\sim$0.1 dex. A similar effect is seen for \Tau\ models (bottom left panel), where \stec\ fits report a very small contribution from a stellar population with ages above 10 Gyr; this is barely visible from the figure.
 
\begin{figure*}\centering
      \textbf{SFH = \Con}\par\medskip
 \includegraphics[clip=,width=0.49\textwidth]{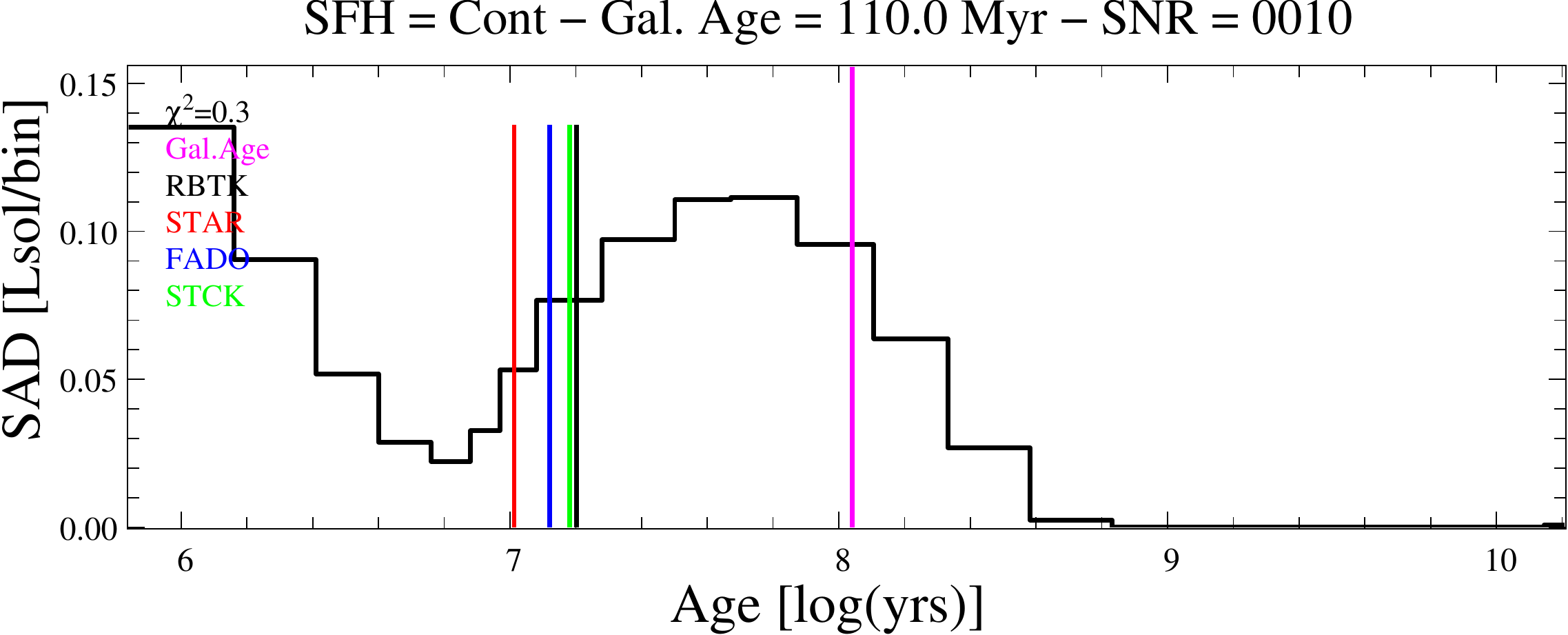}  
 \includegraphics[clip=,width=0.5\textwidth]{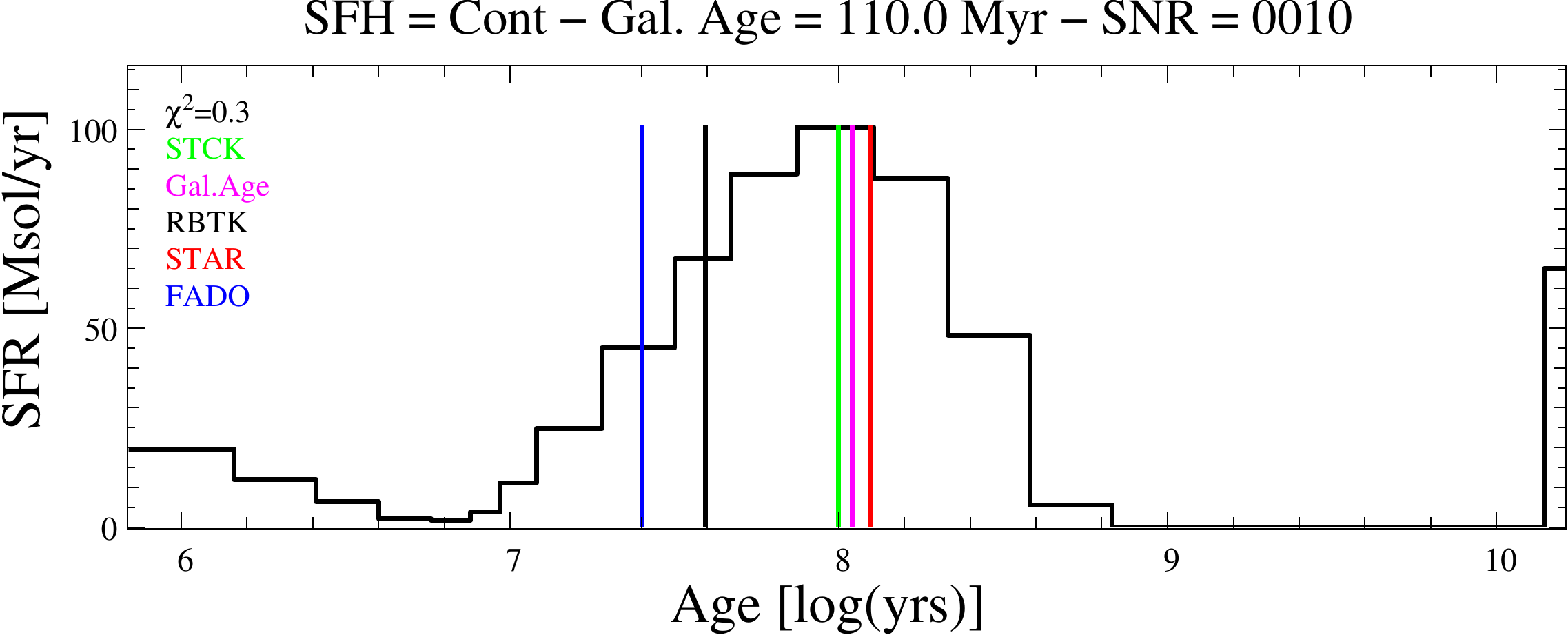}  
 
      \textbf{SFH = \Tau}\par\medskip

 \includegraphics[clip=,width=0.49\textwidth]{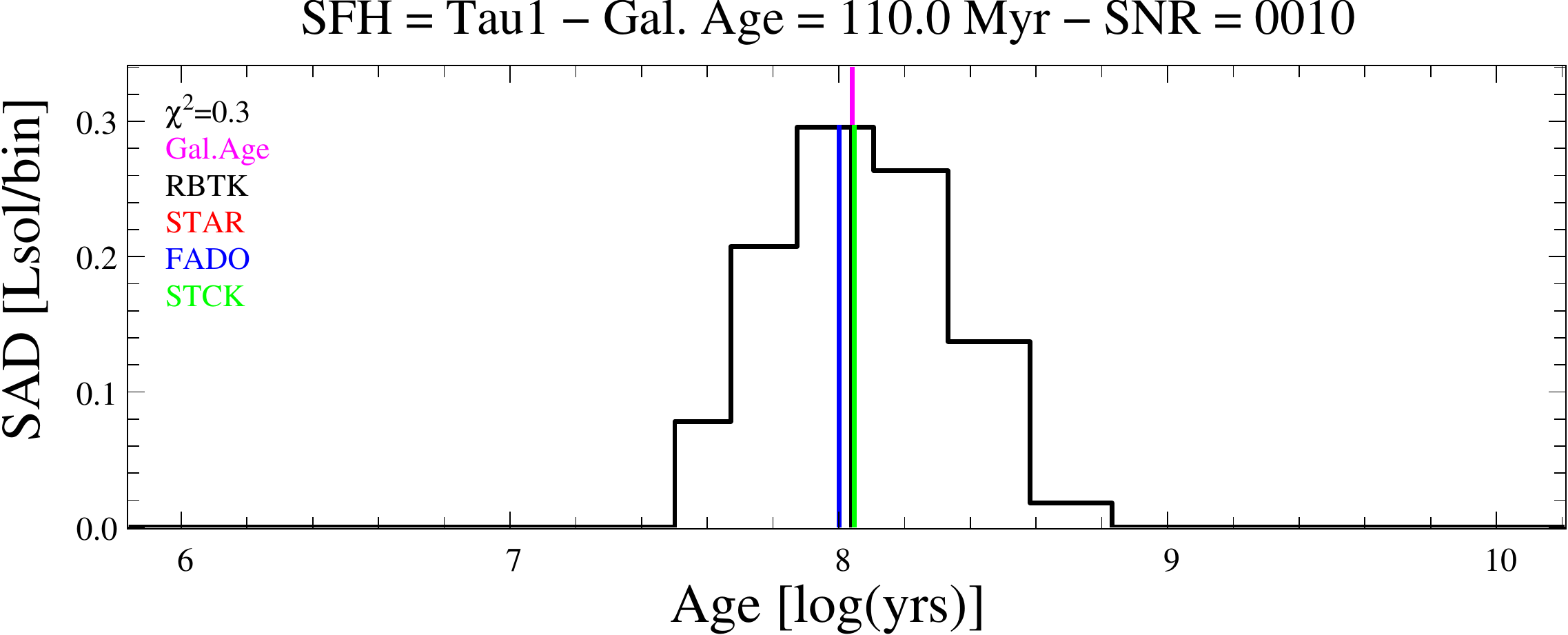}  
 \includegraphics[clip=,width=0.5\textwidth]{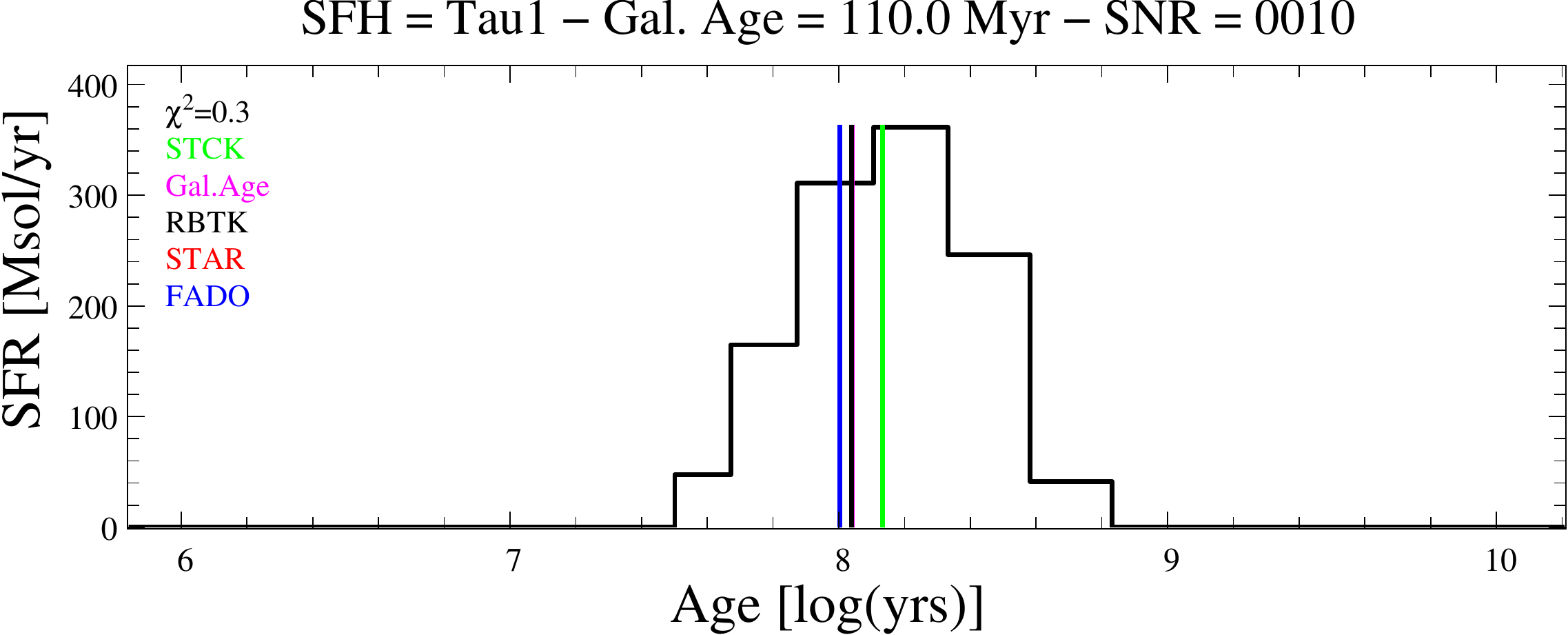}  
 \caption{Same caption of Fig. \ref{redNebular} for a galaxy age of 110 Myr.}\label{110Myr}
 \end{figure*}

 When we consider mass-weighted quantities, the small bumps seen in the SAD of Fig. \ref{redNebular} become the dominant stellar populations (right panels), producing an overall overestimation of \logtM, in agreement with the results of \cite[][see their Fig. B.1]{car}. These differences, as mentioned at the beginning of Section \ref{discussion}, depend strongly on the contribution of the nebular emission to the overall continuum. For this reason in the \Con\ and \Tau\ models at 110 Myr, where the nebular emission is lower, the fits consistently recover the main contributors in both \logtL\ and \logtM, as shown in Fig. \ref{110Myr}. In other words where the nebular contribution is relevant there is a systematic overestimation of \logtM, which has led to an overestimation of the stellar masses (see also \citealt{gom} and \citealt{car}).
 
  \subsection{The Balmer jump effect}
 
 Table \ref{table} and Figs. \ref{Cont-logtL}, and \ref{Tau1-logtL} show that the median $\Delta$\logtM\ measured with \stec\ span a relatively extended range of differences from \reb\ input parameters, for both \Con\ and \Tau\ models: from $\sim$ 0.4 dex at S/N = 3 up to 0.01 dex at S/N = 100. A relevant gap is seen between S/N = 50 and S/N = 10, where for example in \Con\ models $\Delta$\logtM\ decreases from 0.23 ($\sim$70\%) up to 0.1 ($\sim$25\%) dex. We explore the reasons for this decline.
 
 From the fourth row panels of Figs. \ref{Cont-logtL} and \ref{Tau1-logtL} it is clear that the improvement for \stec\ at higher S/N is mainly driven by better fits at $\log(t/\mathrm{yr}) < 8$. At those ages, \starl\ also shows a large bump of up to 3 dex, similar to that seen with \stec. In this context the \fado\ results are quite different, showing a small bump at $\log(t/\mathrm{yr}) < 7$ for S/N = 5, and overall a good agreement for higher S/N. This indicates that the contamination of nebular emission to the overall continuum is responsible for this scatter. However, since \stec\ fits improve with increasing S/N even for models in which the nebular contribution decreases relatively quickly, this effect must be related to specific features in the spectra involving other physical mechanisms.

 \begin{figure*}\centering
   \textbf{SFH = \Con}\par\medskip
   \includegraphics[clip=,width=0.49\textwidth]{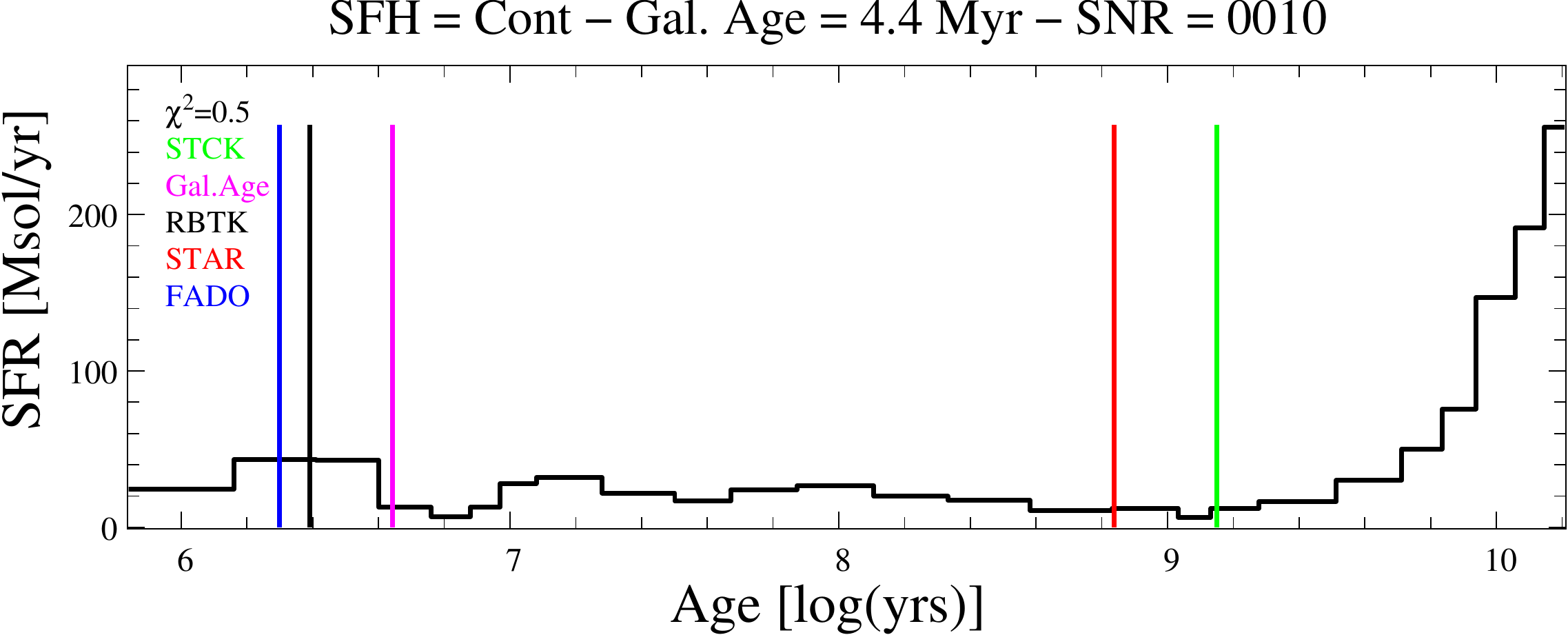}
   \includegraphics[clip=,width=0.5\textwidth]{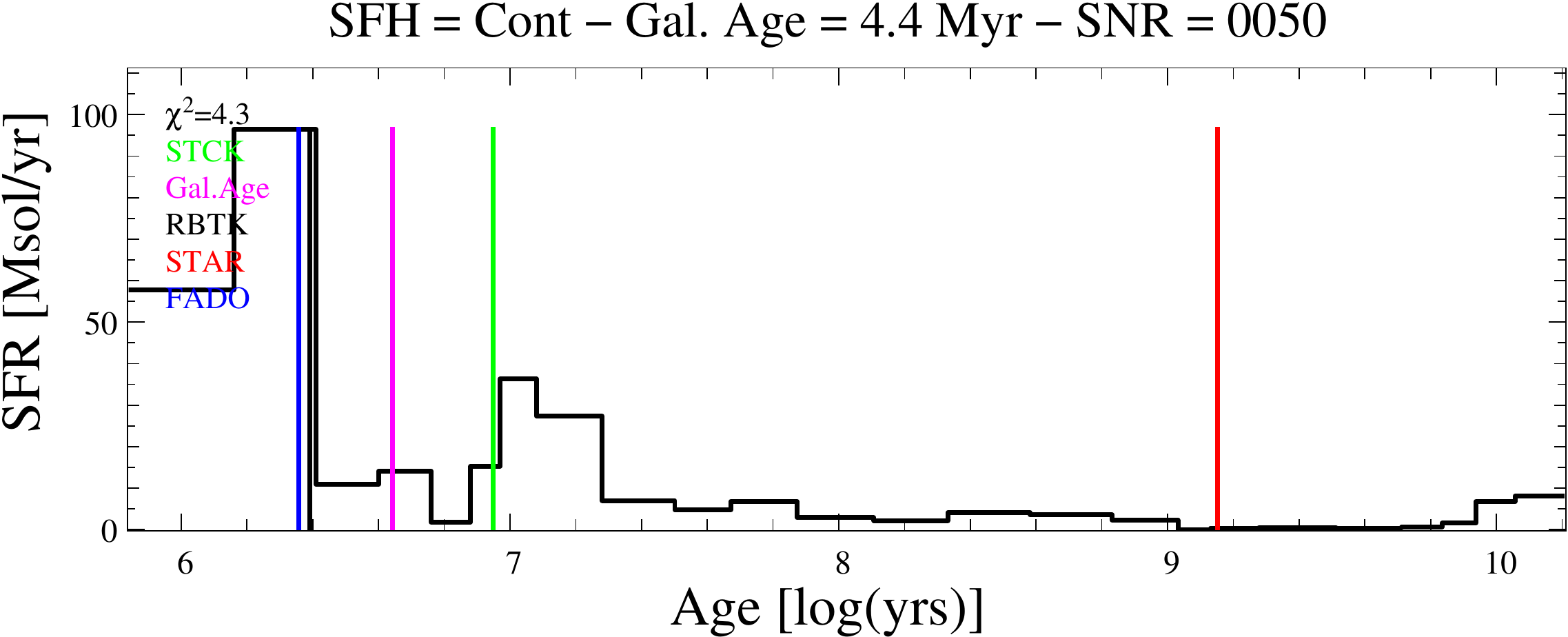}
   \includegraphics[clip=,width=0.49\textwidth]{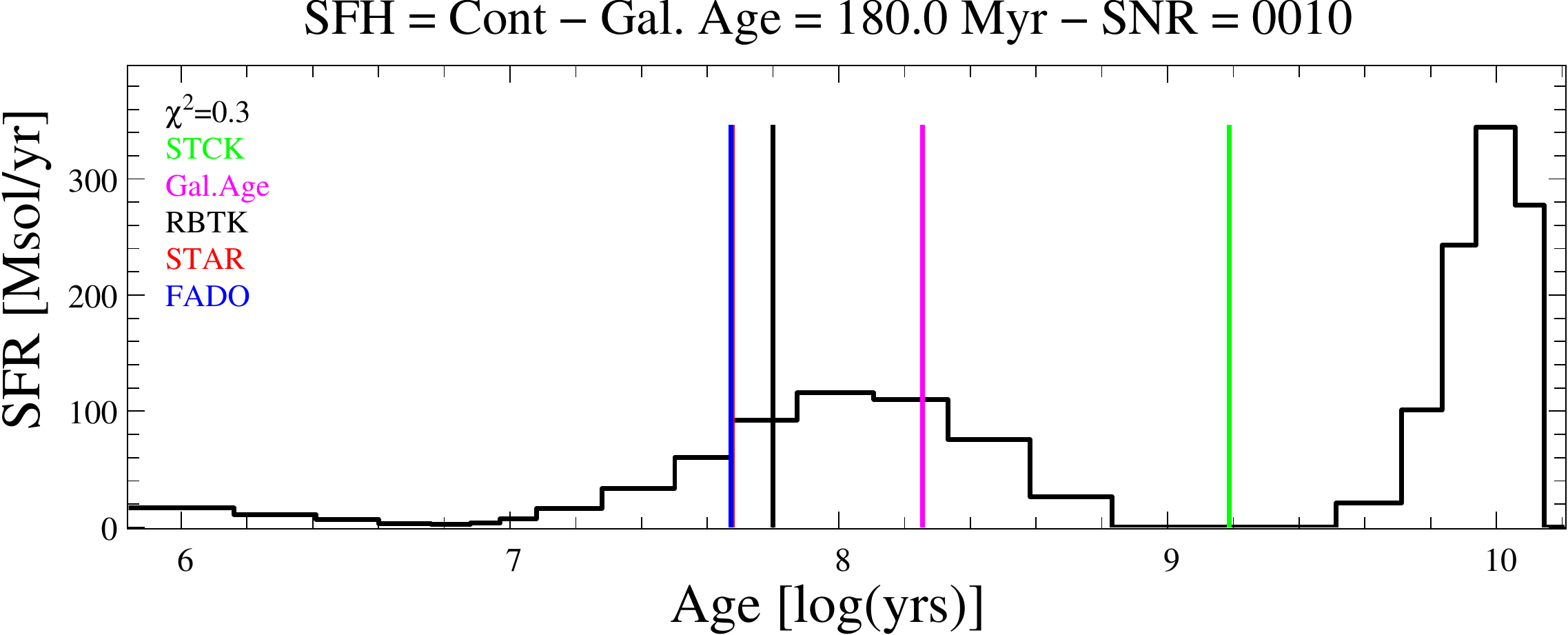}  
   \includegraphics[clip=,width=0.5\textwidth]{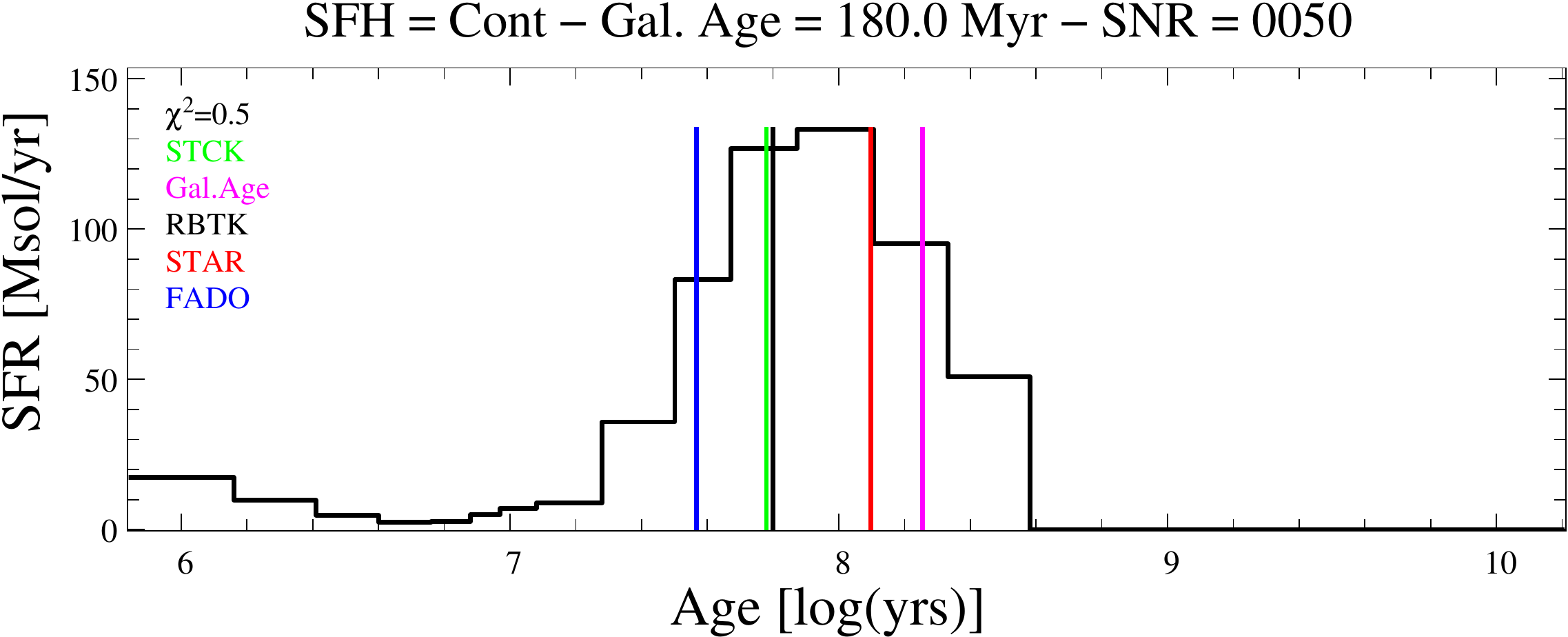}   
   \includegraphics[clip=,width=0.49\textwidth]{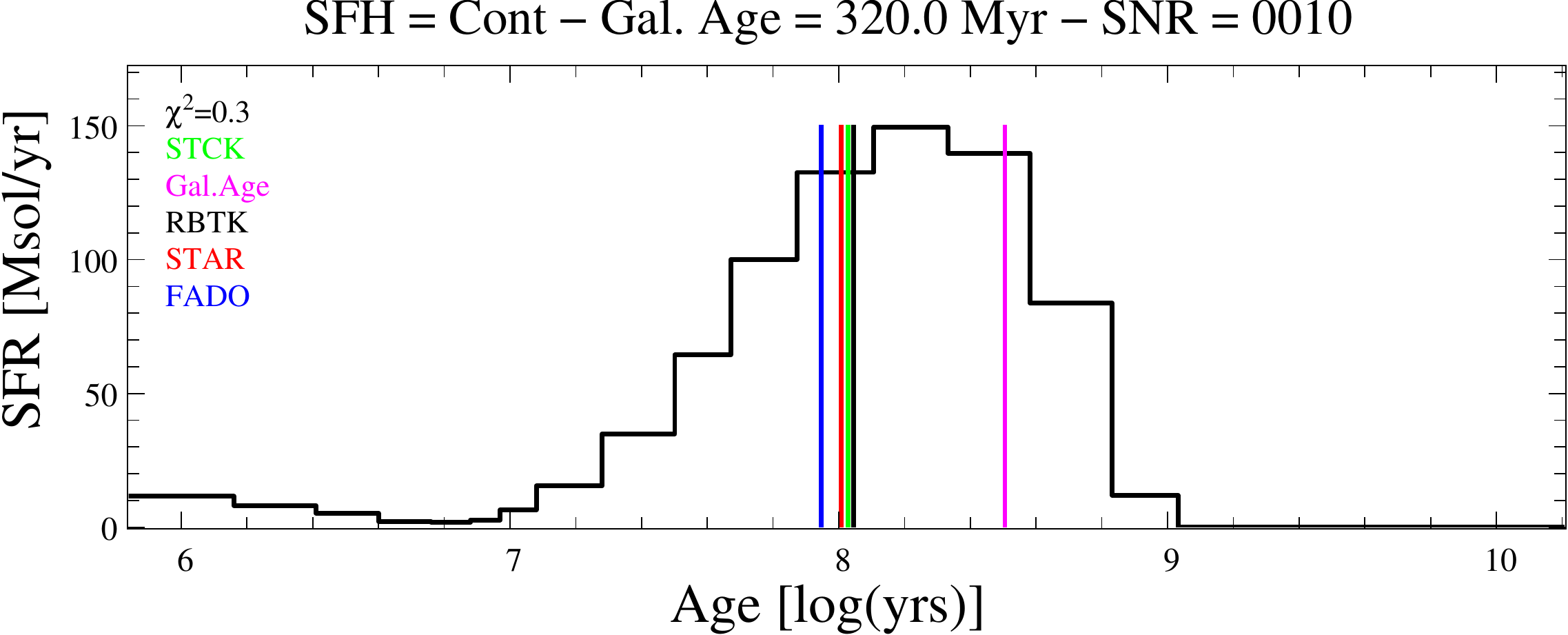}  
   \includegraphics[clip=,width=0.5\textwidth]{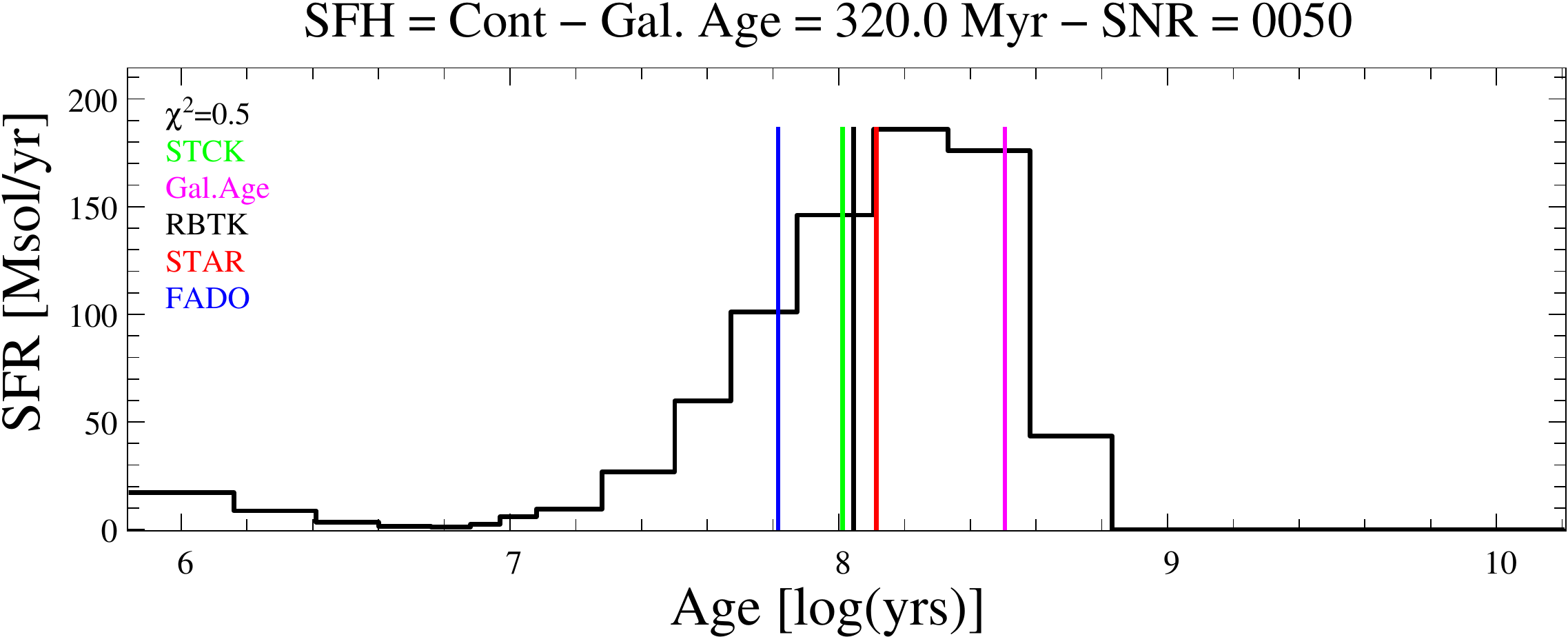}
   \textbf{SFH = \Tau}\par\medskip
   \includegraphics[clip=,width=0.49\textwidth]{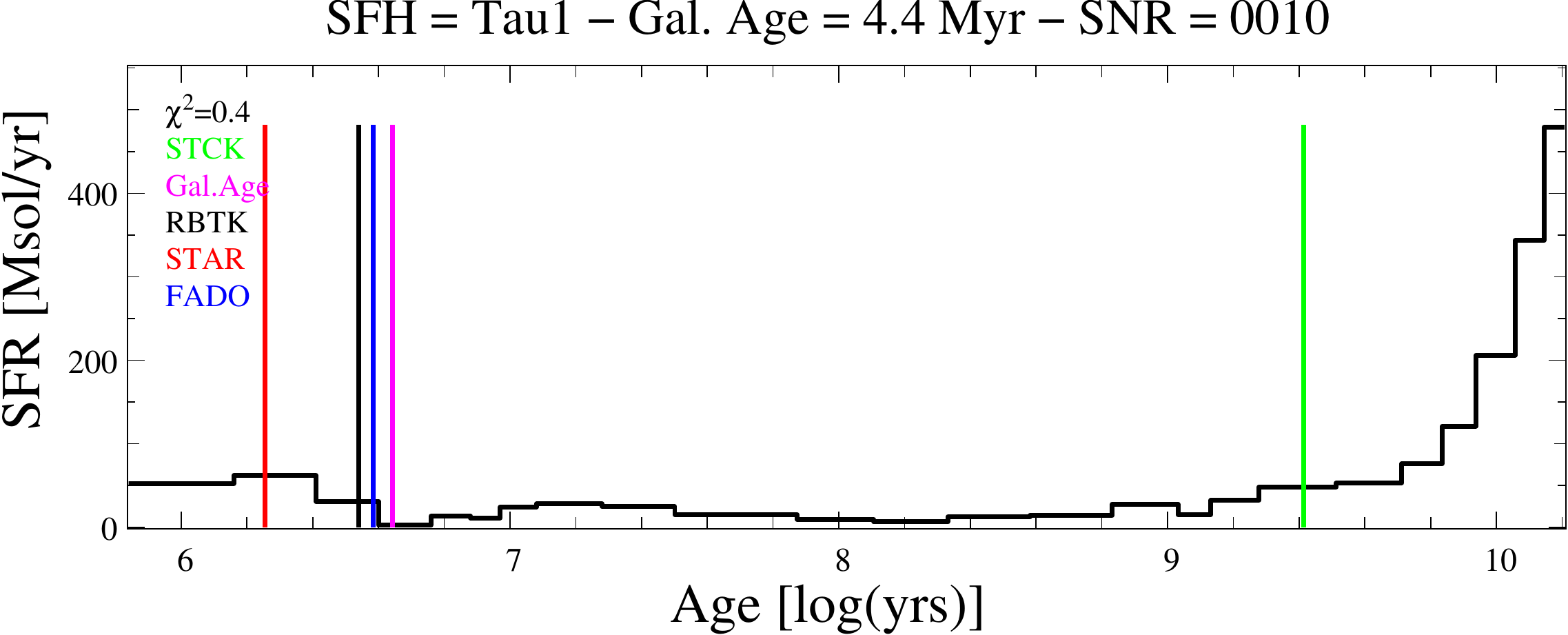}
   \includegraphics[clip=,width=0.5\textwidth]{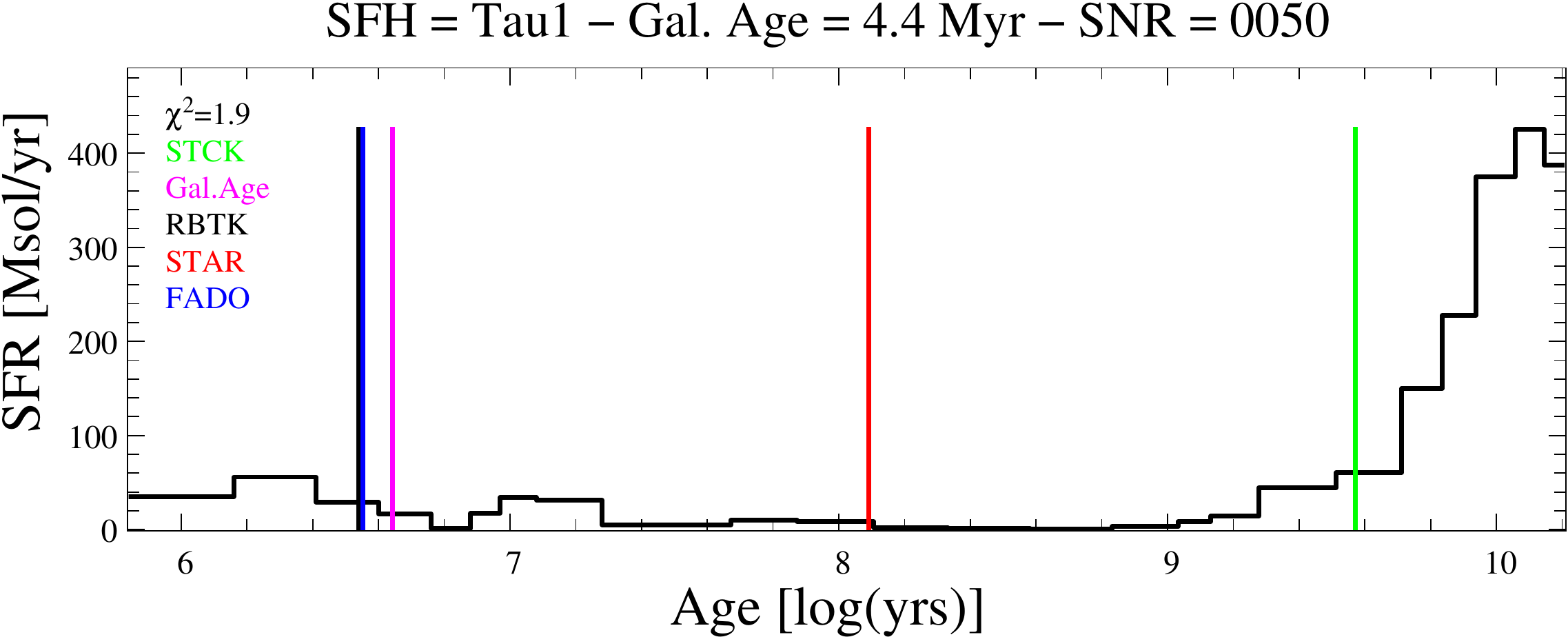}   
   \includegraphics[clip=,width=0.49\textwidth]{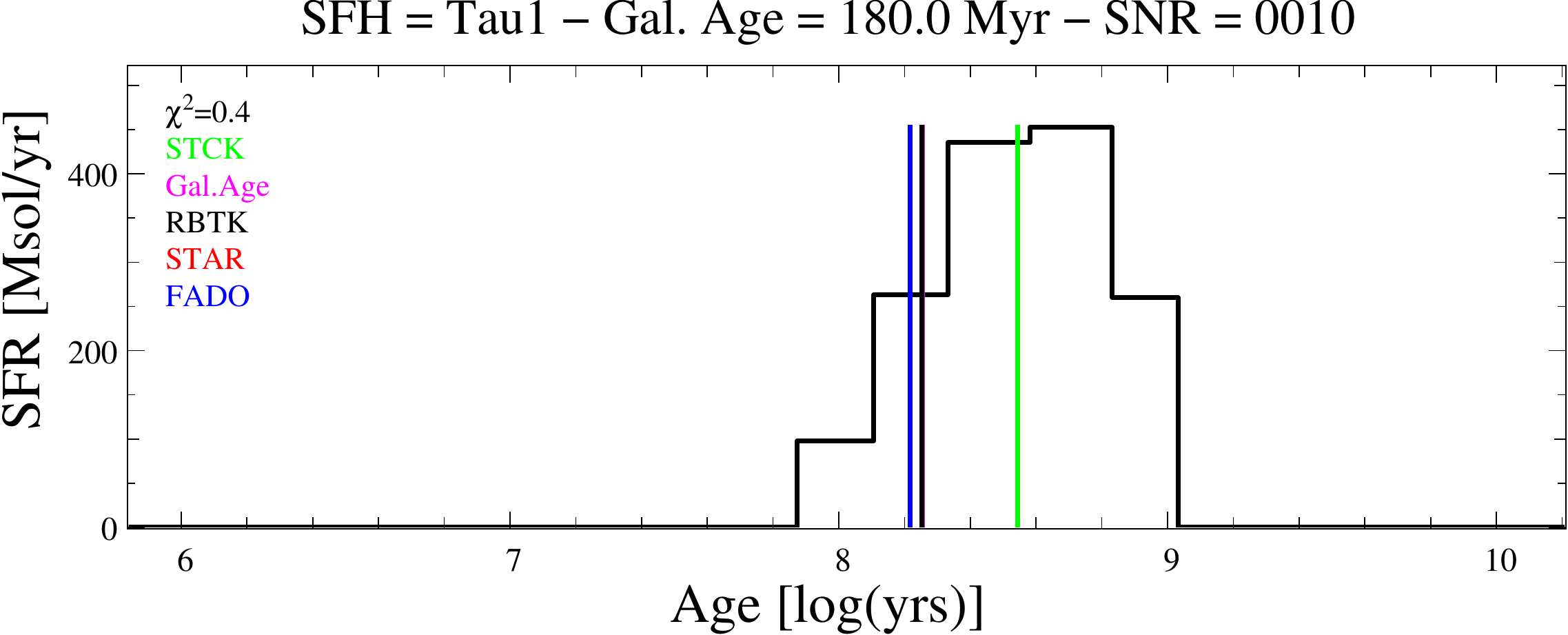}  
   \includegraphics[clip=,width=0.5\textwidth]{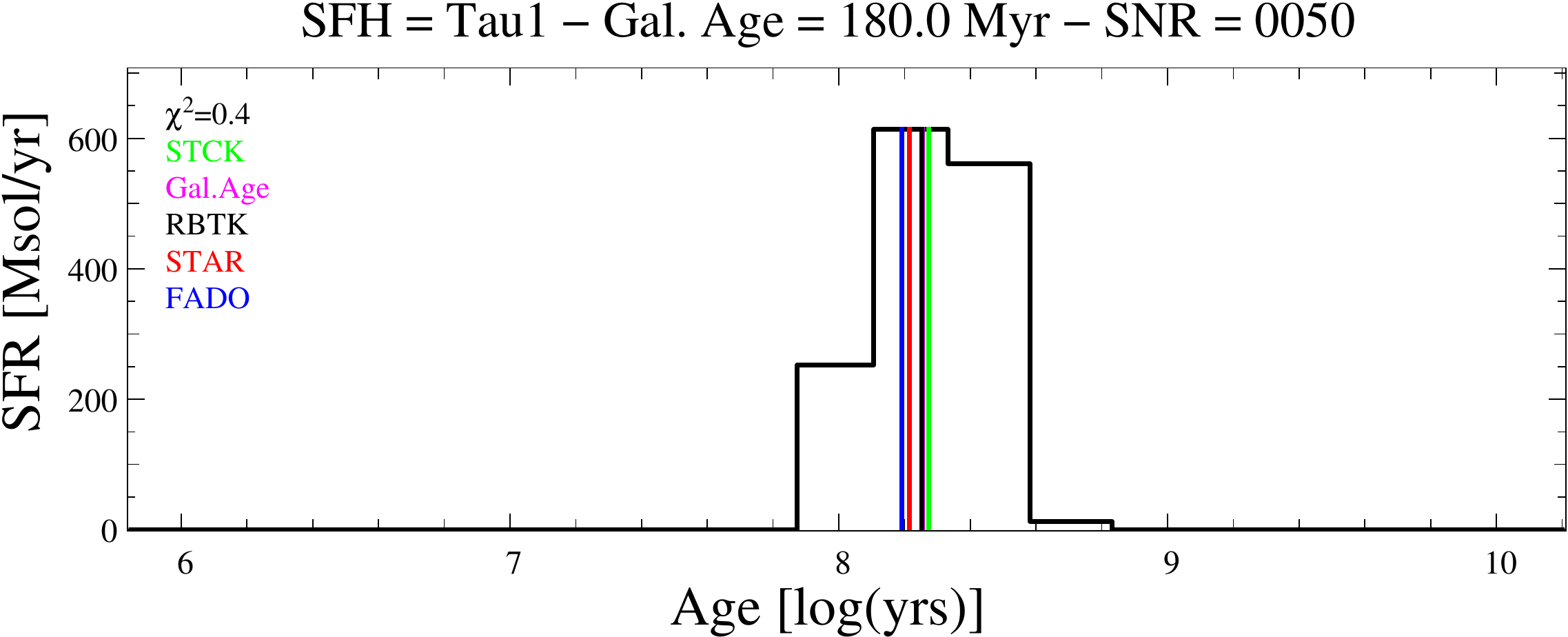}   
   \includegraphics[clip=,width=0.49\textwidth]{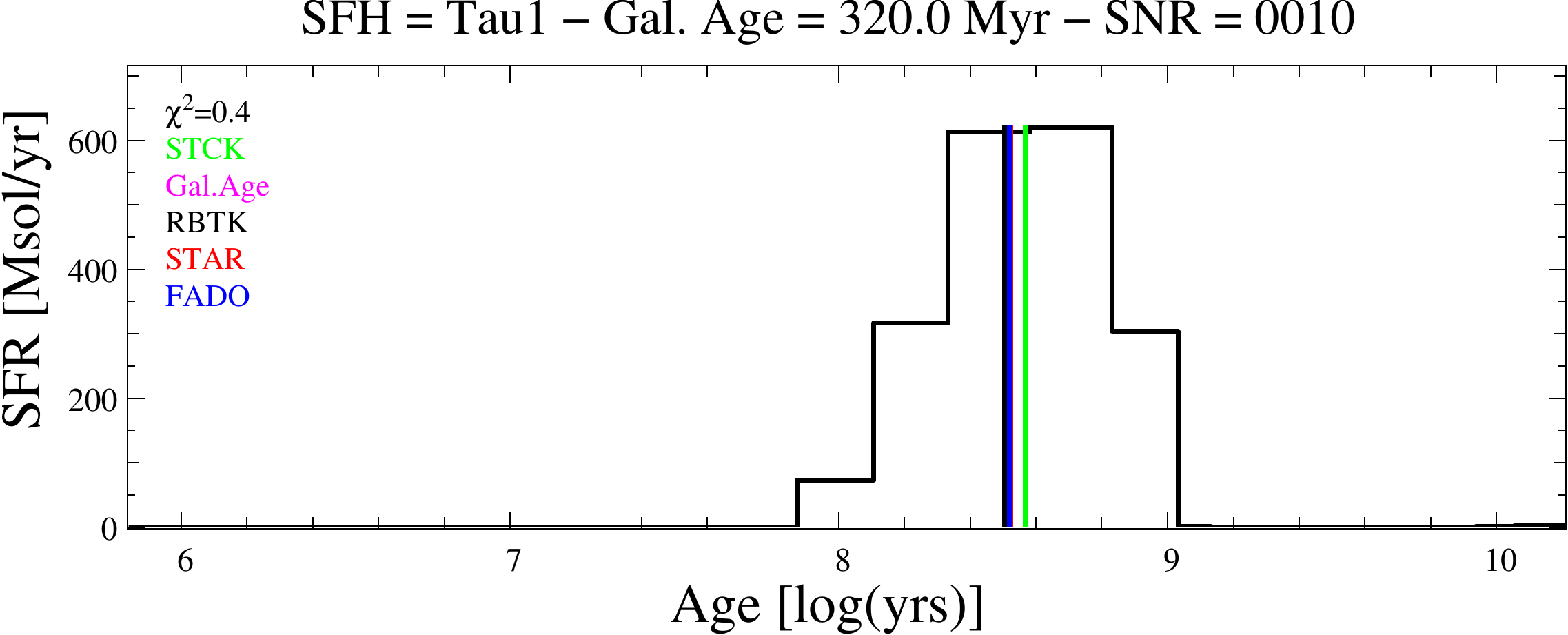}  
   \includegraphics[clip=,width=0.5\textwidth]{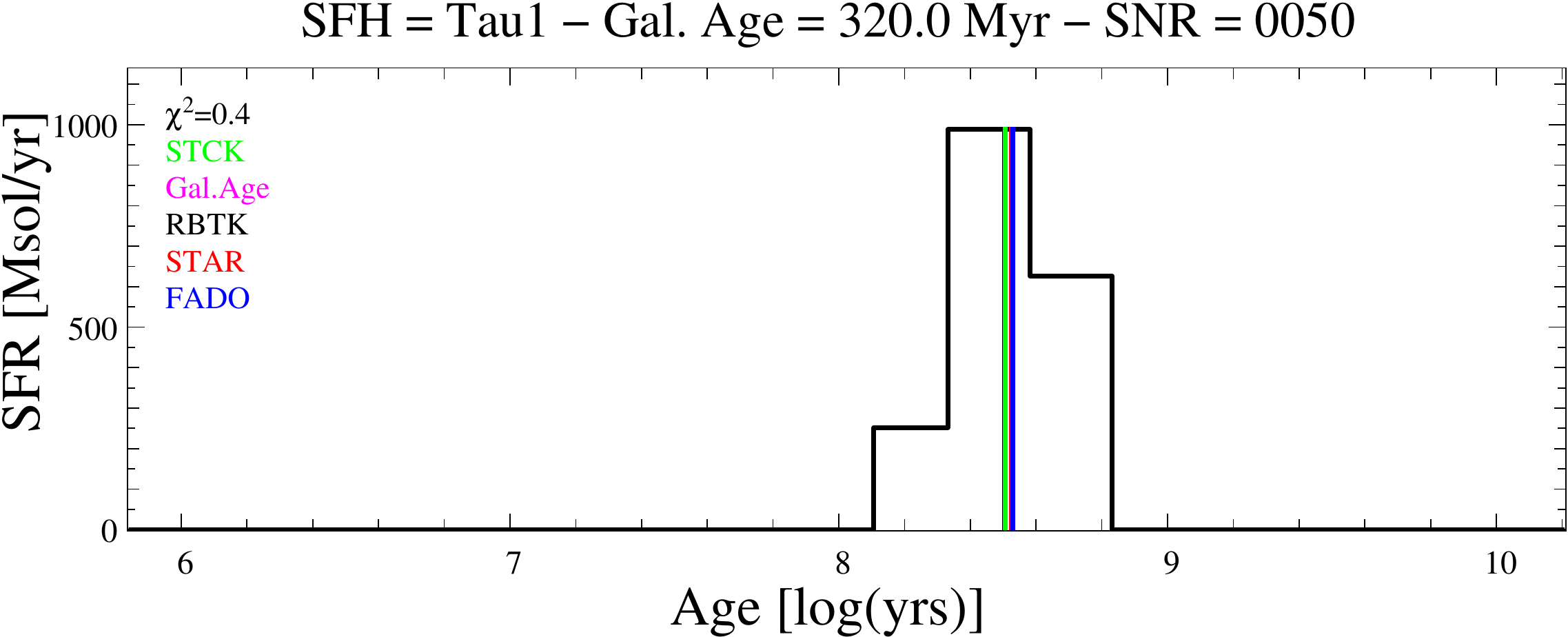}
   \caption{Star formation rate vs. the SSPs ages obtained with \stec\ for \Con\ (top panels) and \Tau\ (bottom panels) models with S/N = 10 (left column), and 50 (right column) at a galaxy age of 4.4 (top row), 180 (middle row), and 320 (bottom row) Myr. The vertical lines are indicated as in Fig. \ref{sparsecoverage}.}\label{balmerSFR}
   \end{figure*}   
 
 To investigate the origin of such S/N-linked overestimation of \logtM\ in detail, we analyse the SFH recovered from the \stec\ fits at different ages that are representative of galaxy evolutionary phases with prominent, intermediate, or negligible nebular contributions. Fig. \ref{balmerSFR} shows the SFH recovered with \stec\ for \Con\ and \Tau\ models at galaxy ages of 4.4, 180, and 320 Myr and S/N 10 and 50, respectively.
 
 For \Con\ models at S/N = 10 \stec\ fits have large discrepancies at 4.4 and 180 Myr. The best-fit spectra for the ages of these galaxies assume a large contribution from old stellar populations, with low mass-to-light ratios and a huge weight in term of total star formation. The fit improves substantially at 320 Myr, where the contribution of the different SSPs is well modelled, as Fig. \ref{Cont-logtL} confirms. When increasing the S/N to 50 the results at 320 Myr remains almost unchanged, but the \logtM\ estimated at 4.4 and 180 Myr improves notably.
 
 For \Tau\ models we observe a similar trend, but since the nebular contribution decreases more quickly for such SFHs, this effect is more evident only at 4.4 Myr, and \logtM\ is positively recovered at 180 and 320 Myr even for S/N = 10. For this reason for the rest of the analysis, we focus on the \Con\ models.
 
 \begin{figure}\centering
    \textbf{S/N = 10}\par\medskip
   \includegraphics[clip=,width=0.49\textwidth]{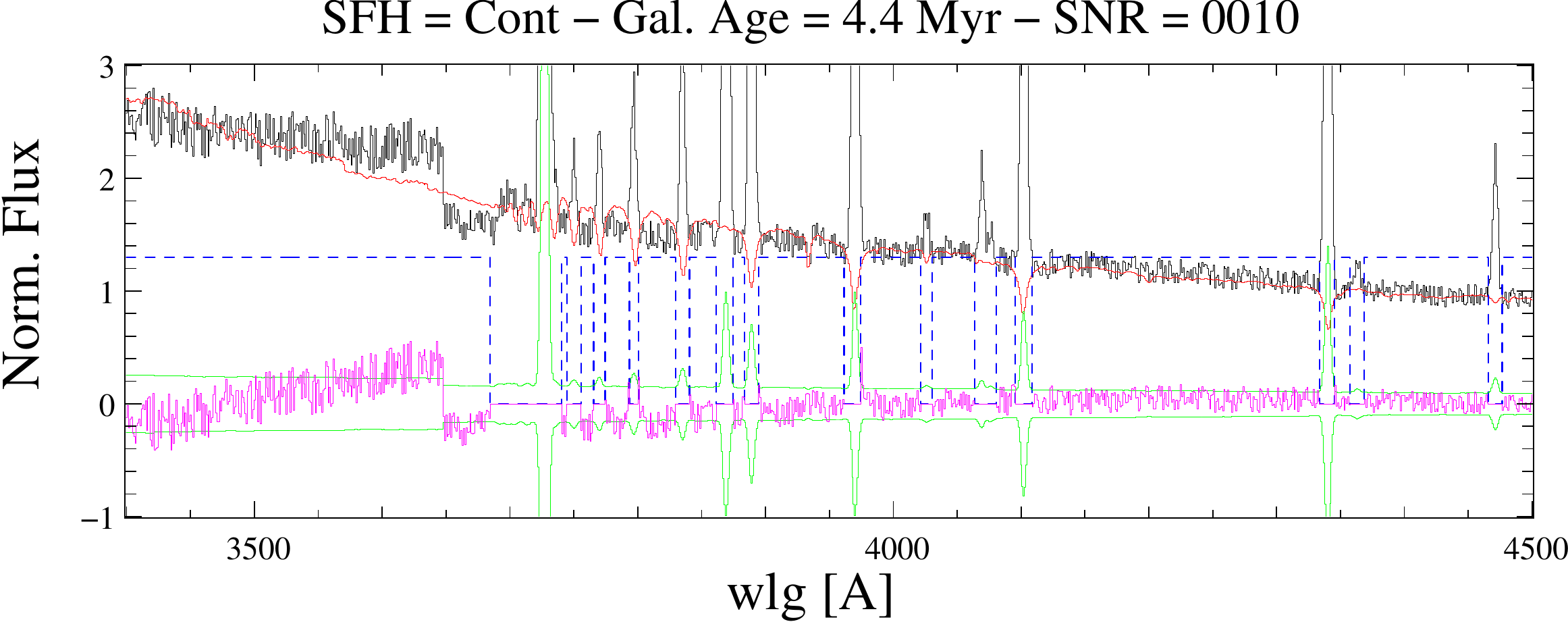}
   \includegraphics[clip=,width=0.49\textwidth]{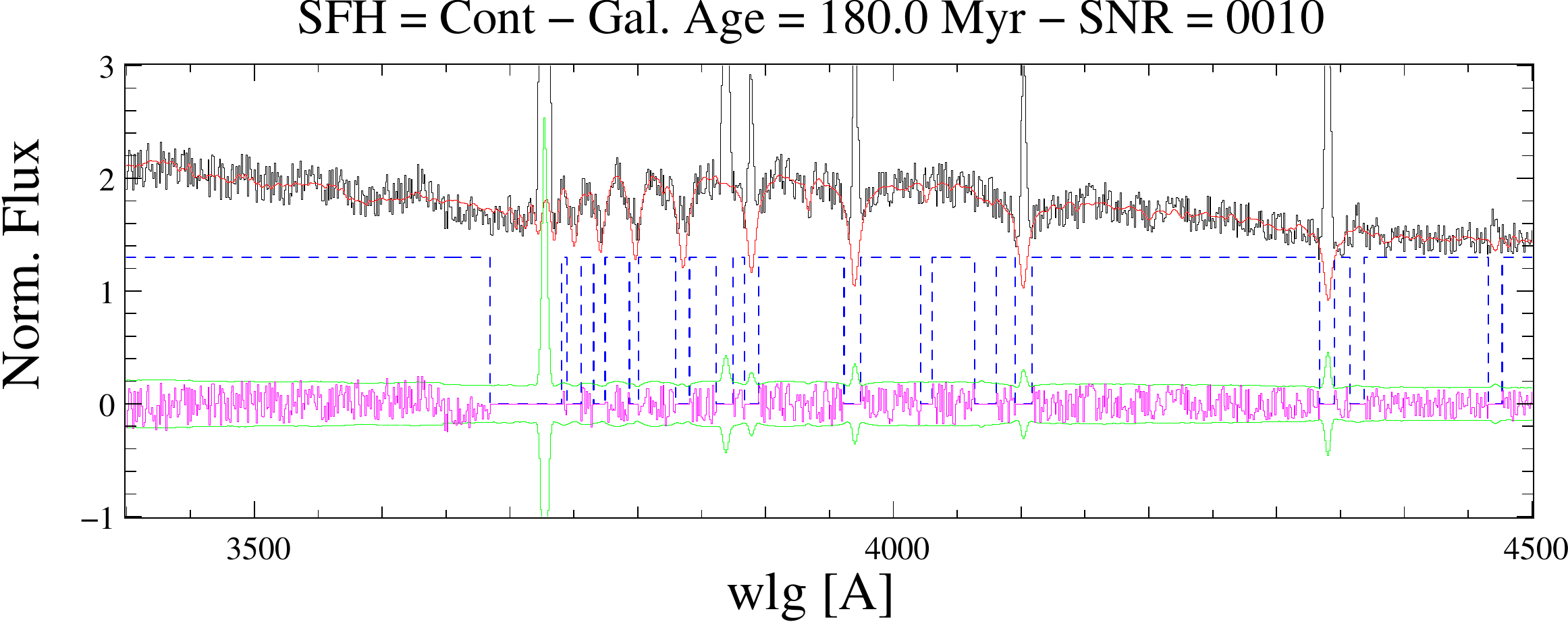}   
   \includegraphics[clip=,width=0.49\textwidth]{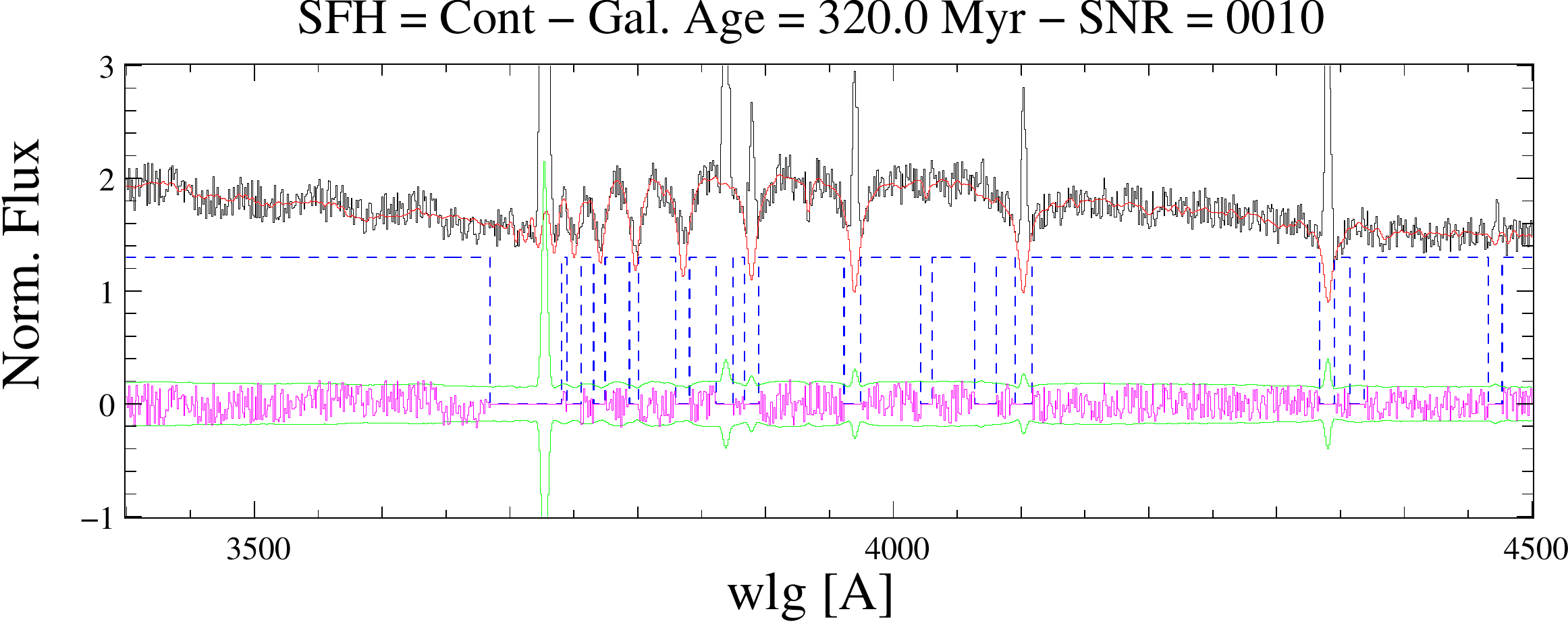}   
   
    \textbf{S/N = 50}\par\medskip
   \includegraphics[clip=,width=0.49\textwidth]{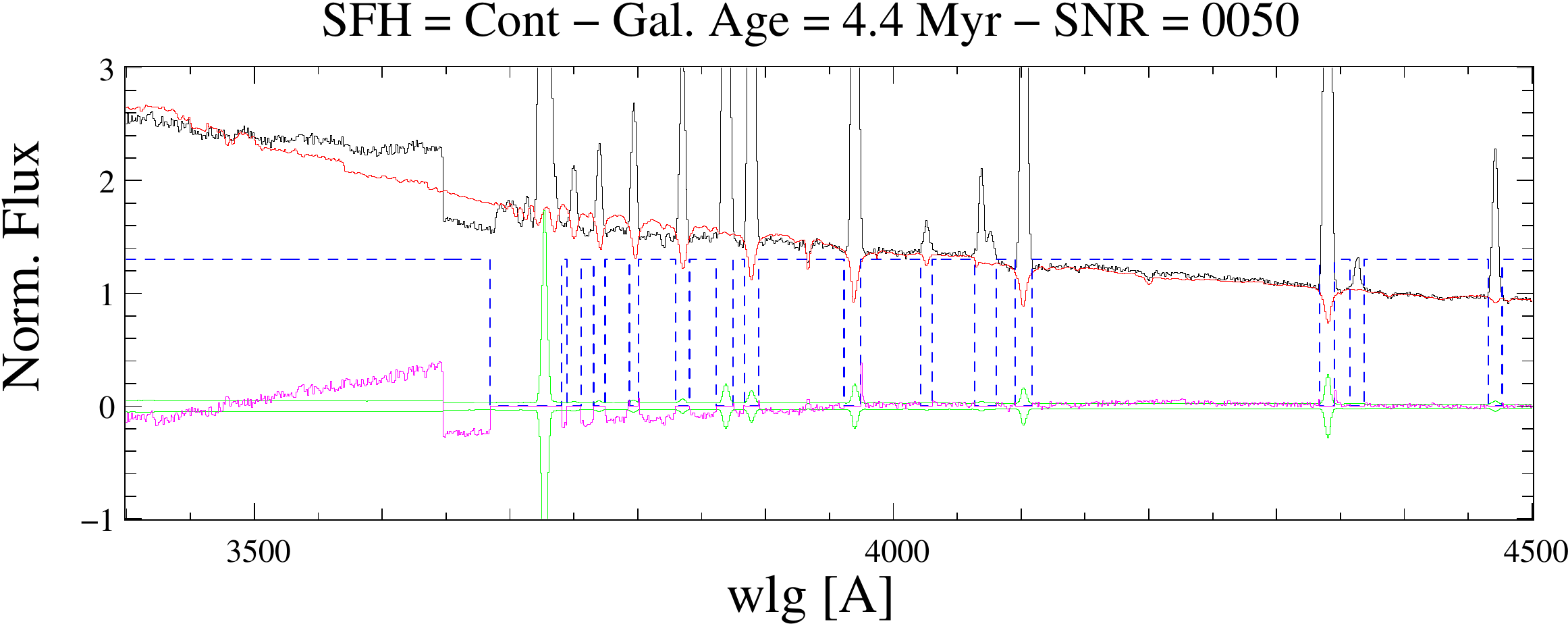}
   \includegraphics[clip=,width=0.49\textwidth]{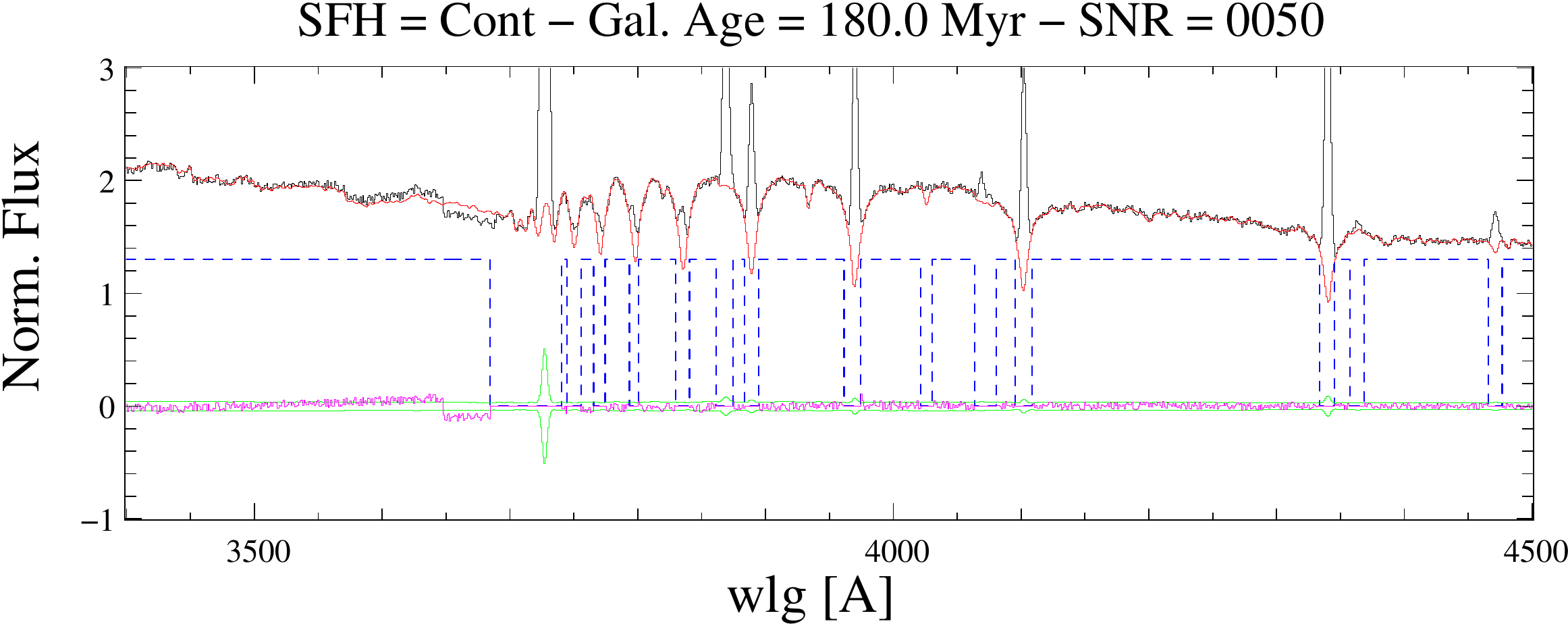}   
   \includegraphics[clip=,width=0.49\textwidth]{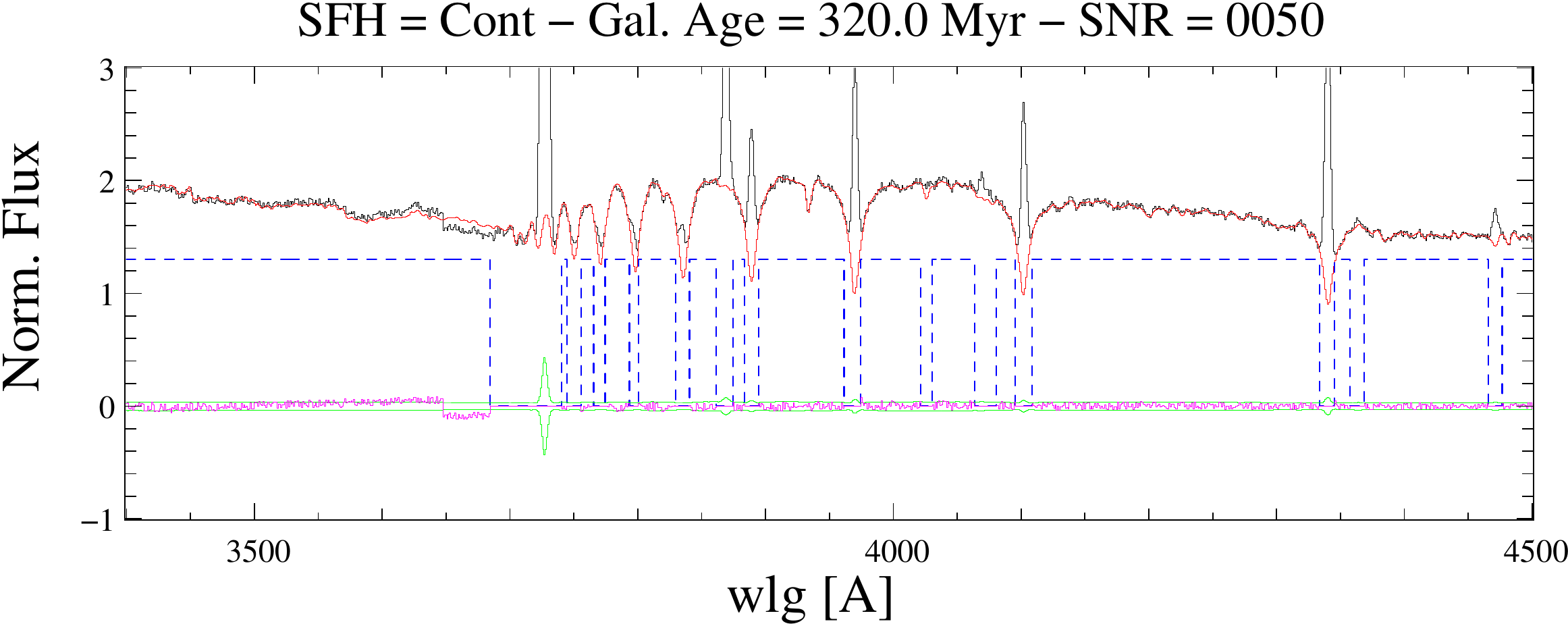}   
   \caption{\stec\ best-fit spectrum for \Con\ models at 4.4, 180, and 320 Myr and S/N = 10 (top panels), and 50 (bottom panels. The black and red lines show the input and the best-fit spectrum, respectively. The blue, magenta, and green lines show the mask, the residual and the errors, respectively.}\label{balmerSPC}
   \end{figure}   
 
  Increasing the S/N, \stec\ improves the quality of the fits, reproducing more the input spectrum more accurately. Fig. \ref{balmerSPC} shows the best-fit spectra obtained with \stec\ for the ages investigated in Fig. \ref{balmerSFR}, 4.4, 180, and 320 Myr, and focussing on the region of the spectrum between 3300-4500 \AA. This was done to underline an important effect visible regarding discontinuities, in particular around the Balmer jump, at 3646 \AA.
  
  As seen in Fig. \ref{fadoexample}, the nebular emission, according to the evolutionary stage of the galaxy, can contribute conspicuously to the shape and amplitude of the Balmer jump. The shape and the strengths of the Balmer jump are set by the evolution of the number of ionising photons, which evolves according to the stellar ages of the SSPs contributing to the observed spectrum. This is not only a theoretical expectation but it has been observed in galaxies with strong starburst activity \citep[e.g. ][]{gus}.
  
  Fig. \ref{balmerSPC} shows that one of the causes of the high \logtM\ recovered with purely stellar codes is related to the difficulty of such methods in reproducing the Balmer jump (see Fig. \ref{rebetikospectra}). To quantify this effect better, we defined a new parameter for each \reb\ synthetic spectrum built for \Con\ models called the Balmer jump strength (BJS), which is written as

  \begin{equation}
    \mathrm{BJS} = \frac{\int_{3580\AA}^{3600\AA} F(\lambda,t)d\lambda}{\int_{3660\AA}^{3680\AA} F(\lambda,t)d\lambda}
  \label{BJSeq}.\end{equation}
  
  \begin{figure}      \centering
      \includegraphics[clip=,width=0.49\textwidth]{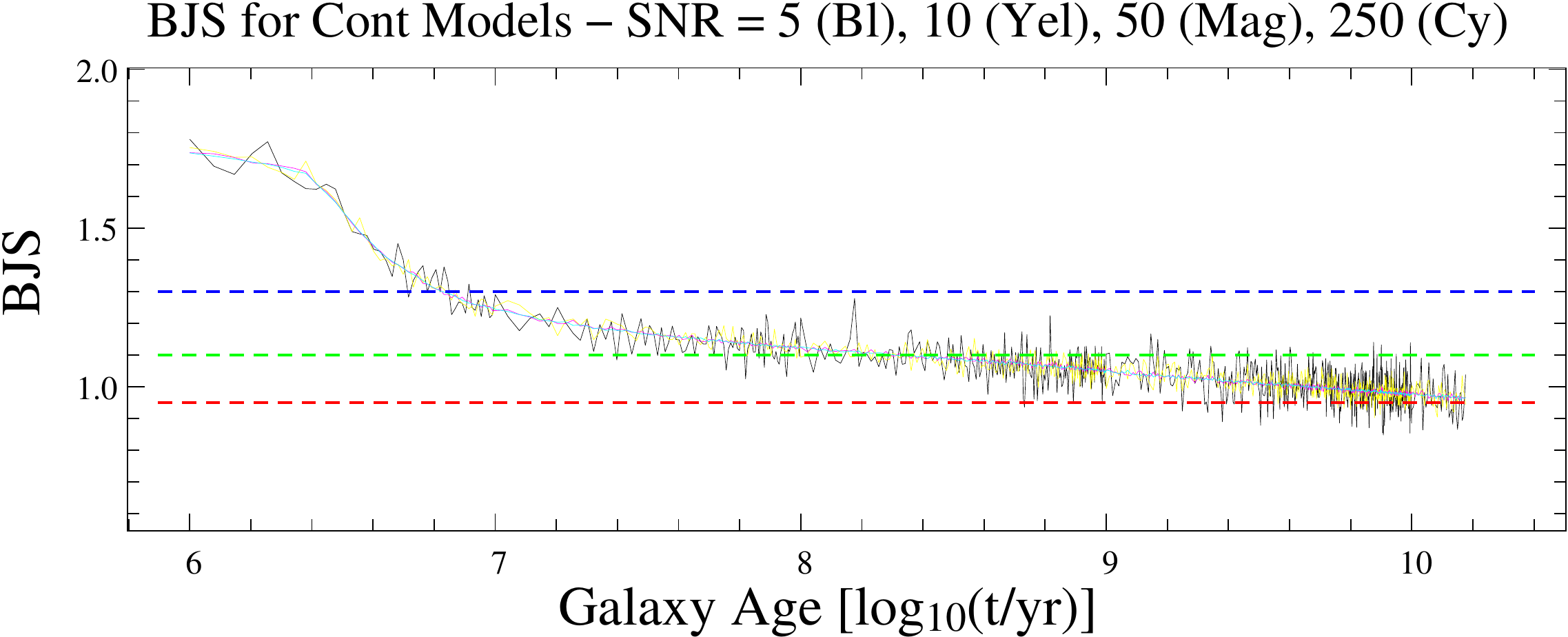}      
      \caption{Evolution of the BJS (see Eq. \ref{BJSeq}) as a function of galaxy ages for \Con\ models at S/N = 5 (black), 10 (yellow), 50 (magenta), and 250 (cyan). Horizontal lines show the subgroup identifying ages with weak (0.92 $<$ BJS $<$ 0.98, red line), intermediate (1.08 $<$ BJS $<$ 1.12, green line) and strong (1.2 $<$ BJS $<$ 2, blue line) BJS.}\label{BJS}
  \end{figure}
  
  The evolution of the BJS can be considered as a diagnostic of the `height' of the Balmer jump, and varies according to the evolutionary phase of the galaxy, as shown for different S/N and \Con\ models in Fig. \ref{BJS}. The effect of S/N is rather small, as expected, and its values span a range between 0.8 $<$ BJS $<$ 1.8, with an average of $\left<BJS\right>$ = 1.06$\pm$0.11. Within these \reb\ synthetic spectra we defined three subgroups (indicated by the horizontal lines in Fig. \ref{BJS}), according to the BJS value as follows:
  
  \begin{itemize}
      \item {\it High BJS}: 1.2 $<$ BJS $<$ 2 
      \item {\it Intermediate BJS}: 1.08 $<$ BJS $<$ 1.12
      \item {\it Low BJS}: 0.92 $<$ BJS $<$ 0.98
.  \end{itemize}
  
  These subsamples define families of spectra with different strengths in the Balmer jump, representatives of different evolutionary phases of a galaxy. The next step is to investigate the correlation of the $\Delta$\logtM\ analysed in Sec. \ref{results} with the BJS. 
  
  \begin{figure*}      \centering
      \includegraphics[clip=,width=0.89\textwidth]{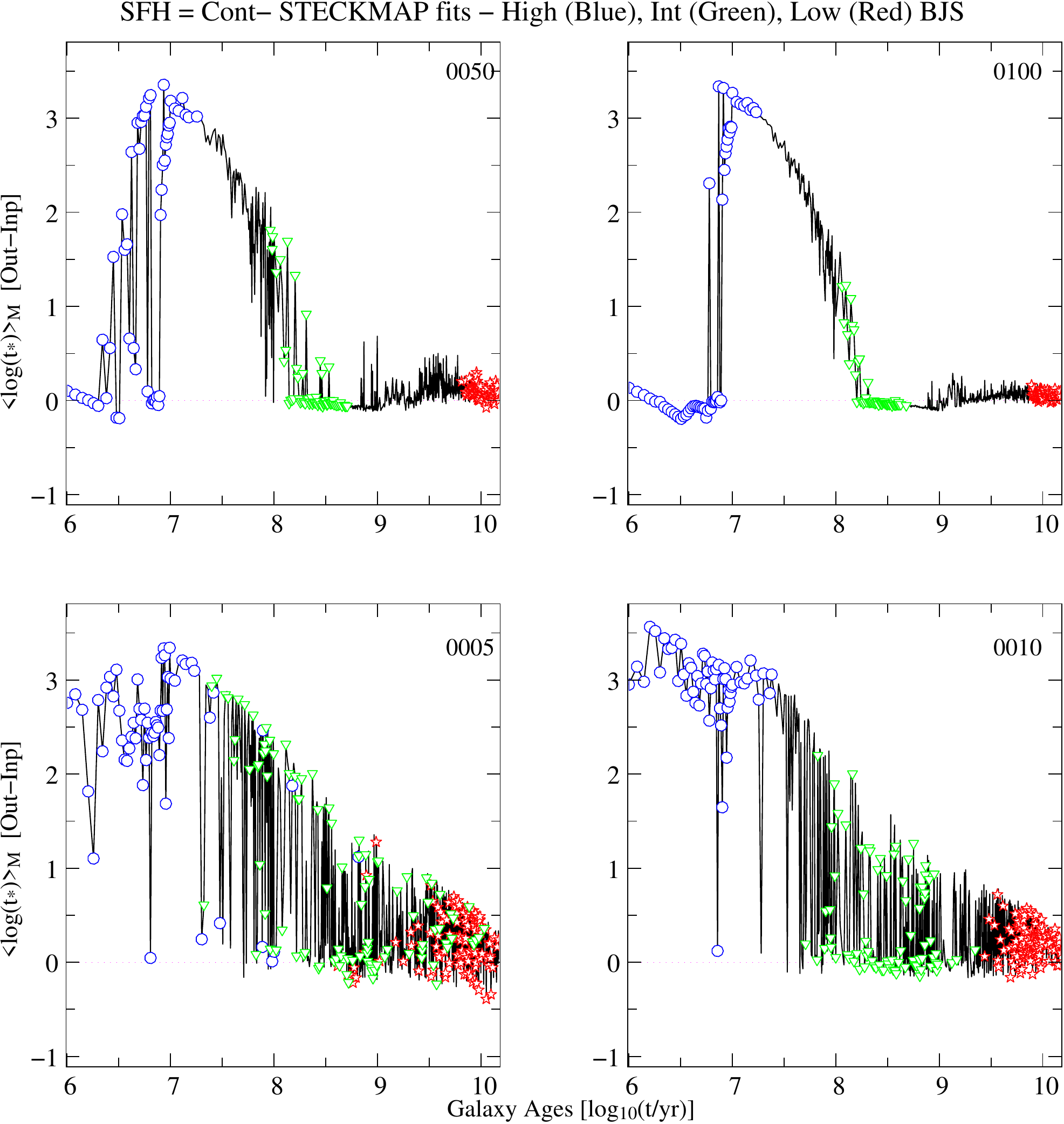}
      \caption{$\Delta$\logtM\ obtained from \stec\ vs. galaxy age at different S/N, reported on the top right of each panel. The blue circle, green triangles, and red stars point out the synthetic \reb\ spectrum with high, intermediate, and low BJS, respectively.}
      \label{BJSspc}
  \end{figure*}
  
  Fig. \ref{Cont-logtL} and Fig. \ref{BJSspc} show the residual of \logtM\ as a function of galaxy age, from analysis restricted to the \stec\ results and flagging the synthetic spectra with high (blue), intermediate (green), and low (red) BJS. According to the selection criteria defined above, we can identify three different cases:
  
  \begin{itemize}
      \item {\bf Case 1}: The spectrum has a high BJS because of the relative nebular continuum contribution to the overall SED. For these cases, purely stellar codes have immense difficulties in determining mean stellar ages independently of the quality of data, as shown by blue circles in Fig. \ref{BJSspc}. We see from the two lower panels of Figs. \ref{Cont-logtL} and \ref{Tau1-logtL} that the metallicity inferred by \stec\ is strongly sub-solar (off-scale in some panels). A reasonable explanation for this is that purely stellar codes, such as \stec\ and \starl,\ have no other option for fitting the sudden increase in flux at 3650 \AA\ than to assume a population of very hot, thus low-metallicity, massive stars.
      \item {\bf Case 2}: The spectra has an intermediate BJS, with a moderate contribution from nebular emission. In general, while the galaxy evolves, this component decreases and the fits converge to the input \logtM. However, for specific galaxy ages, purely stellar codes are not able to discriminate the correct stellar ages, unless the Balmer jump in these phases is small enough so that the uncertainties are dominated by the S/N and/or the intrinsic degeneracies between SSP, instead of the nebular continuum, as happens in the cases represented by the green triangles in Fig. \ref{BJSspc}.
      
      The SSPs at these ages change in short timescales, as shown in Fig. \ref{Ha}. The EW of H$\alpha$ in these spectra shows rapid jumps around 100 Myr because of the onset of the post-AGB phase, which produces further gas photo-ionisation \citep{bin,sta,cid2,gom2}. The $\Delta$\logtM\ at ages higher than $\log(t/\mathrm{yr}) \sim 8$ decreases gradually, depending on the importance of the post-AGB stars, adding further degeneracy to the fit procedure.
      \item {\bf Case 3}: The spectra has a low BJS, implying low contamination from nebular emission of the overall continuum or, alternatively, for evolved galaxies in which old stellar populations dominate the spectrum. For these evolutionary phases, purely stellar codes recover the \logtM\ with an accuracy better than 0.4 dex at S/N $\ge$ 5, as shown with red stars in Fig. \ref{BJSspc}. 
  \end{itemize}
  
  These considerations imply that when a galaxy enters into an evolutionary phase in which the observed spectrum has a strong BJS, the effect of neglecting the nebular emission in the fitting process has a strong impact on the estimation of the SFH at such ages. But even in the presence of mild contamination by nebula continuum, there is still a possibility to misinterpret the data, as a consequence of its poor quality. This can produce an overestimation of the stellar ages and a simultaneous underestimation of stellar metallicity, since the intermediate Balmer jump height cannot be fit by these purely stellar codes, introducing a higher contribution of old, low mass-to-light ratio SSPs to compensate for this effect. In our mock experiment, we identified this transition phase in the range 10 $<$ S/N $<$ 50, which is also the transition where purely stellar codes showed the largest differences in the results, as a consequence of unstable fits.

 \section{Conclusions}
 \label{conclusions}

 In this paper we investigated the limits and reliability of three spectral synthesis codes for the estimation of two specific parameters representative of the evolution of a galaxy: the mean stellar age and the mean stellar metallicity. The question addressed in the introduction was at which S/N values is it still possible to determine from a galaxy spectrum its star formation and chemical evolution history and how does this depend on the tool used to model the spectra.
 
 To address this question we built a set of synthetic spectra with both stellar and nebular emission, thereby reproducing the evolution of a galaxy with two different SFHs: a continuous star formation (\Con) and a single initial burst with an exponentially declining star formation law (\Tau). The synthetic spectra were modified for a range of S/N ratios and were analysed with the three spectral synthesis codes: \fado, \stec, and \starl. The main results from our analysis are as follows:
 
 \begin{itemize}
    \item At S/N = 3 all tools show large differences in the results; \fado\ and \starl\ have a median $\Delta$\logtL$\sim$0.1 dex and STECKMAP has a higher $\Delta$\logtL of 0.2. The codes have a large spread of \logtL\ at ages $\log(t/\mathrm{yr}) < 8$, as a consequence of the instability of the solutions.
     
    \item For S/N $> 20$ the median differences between the input \reb\ \logtL\ and the best fits from \fado\ are of the order of 0.03 dex ($\sim7$\%), which is a factor 3 and 4 lower than the 0.08 dex ($\sim$20\%) and 0.11 dex ($\sim$30\%) obtained from \starl\ and \stec, respectively. In contrast to \fado, \starl\ shows a systematic underestimation of the input \logtL = 0.08 dex, 20\%, for S/N $>$ 10.
    \item Mass-weighted quantities show results that are less reliable than light-weighted results (Fig. \ref{Cont-logtL}), with differences around 0.08 dex, 20\%, at S/N $\ge 10$ for \fado. The \starl\ results show median discrepancies between 0.13-0.19 dex; there is no real trend with increasing S/N.
    \item Since the nebular contribution for \Tau\ models declines very rapidly, the results for this SFH shows an overall consistency, with median $\Delta$\logtL\ around 5\% for \fado\ and \starl\ already at S/N = 5 (Fig. \ref{Tau1-logtL}).
    \end{itemize}
    
    The impact of a different spectral basis and/or wavelength coverage on the result is very complex and is postponed to a subsequent paper of this series (Gomes et al. in prep.). Our preliminary analysis has shown that a different spectral basis affects our results mainly as a consequence of the systematic differences in the stellar basis used to build the SSPs. Comparing the standard {\tt BaseL} basis defined in Sec. \ref{methods} with MILES \citep{vaz}, we see that the main differences in both light- and mass-weighted quantities arise for ages below 10$^8$ yr, where the age coverage of MILES is sparser than {\tt BaseL}. This confirms the importance of building an appropriate spectral basis that is calibrated for the type of galaxies investigated.
    
    The variation of wavelength coverage affects mean ages and metallicities in opposite ways: To estimate mean stellar ages a bluer coverage yields better estimates than its red counterpart, where \fado\ shows overall a better agreement compared to \starl. For the mean metallicity, the trend goes in the opposite direction, and a redder coverage gives better results than a bluer coverage with both \fado \ and \starl.
    
    Larger discrepancies in the results are associated with evolutionary phases in which the nebular contribution to the overall SED is significant, which is particularly evident in \Con\ spectra. Although less significant for its light-weighted counterparts, the presence of a strong nebular component leads to significant biases in the estimated mass-weighted quantities, mainly owing to a severe overestimation of old SSPs in pure-stellar fitting approaches (Fig. \ref{redNebular}). This implies that if the contribution from nebular continuum is relevant, there is a systematic overestimation of \logtM, which has led to an overestimation of the stellar masses up to 50\% for $t > 300$ Myr (see also \citealt{gom} and \citealt{car}).
    
    Detailed investigations of the best-fit spectrum in galaxies with overestimated \logtM allowed the determination of the origin of such discrepancies: the difficulties of such methods to reproduce the Balmer jump at 3646 \AA. Quantifying this effect through the definition of the BJS parameter (Eq. \ref{BJSeq}), we identified three possible cases, discriminated by the presence of high, intermediate, or weak BJS (Fig. \ref{BJSspc}):
    
    \begin{itemize}
      \item {\bf Case 1}: For galaxies with high BJS purely stellar codes have major difficulties in determining mean stellar ages, irrespective of the data quality.
      \item {\bf Case 2}: For galaxies with intermediate BJS pure stellar codes are not able to discriminate the correct stellar ages, unless the spectrum has a sufficiently high S/N, from our test at least S/N$>$10.
      \item {\bf Case 3}: For galaxies with weak BJS purely stellar codes recover the \logtM\ with an accuracy $<$0.4 dex, at S/N = 5, thereby providing consistent results.
  \end{itemize}
  
  These considerations imply that when a galaxy enters into an evolutionary phase in which the observed spectrum has a high BJS, the effect of neglecting the nebular emission in the fitting process has a strong impact on the estimation of the SFH of such galaxy. But even in the presence of a mild contribution from nebular continuum, there is still the possibility to misinterpret the data as a consequence of the poor quality of the observations. This can produce an overestimation of the stellar ages, for example Fig. \ref{Cont-logtL}, since intermediate Balmer jumps cannot be fit by a purely stellar code, which models the SFH of the spectrum adding an higher contribution from old SSPs with low mass-to-light ratios \citep{car}.
  
  Our work underlines the importance of using appropriate codes when considering the spectra from galaxies with high SFRs. The introduction of a self-consistent treatment of nebular emission in this context assumes a fundamental role, being the only viable route towards a reliable determination of the assembly history of galaxies in starburst phase, which are more common at higher redshifts.
  
\begin{acknowledgements}
    The authors would like to thank the anonymous referee for the constructive feedback on this manuscript, which helped to improve the content and the presentation of the article. This work was supported by Fundação para a Ci\^{e}ncia e a Tecnologia (FCT) through the research grants PTDC/FIS-AST/29245/2017, UID/FIS/04434/2019, UIDB/04434/2020 and UIDP/04434/2020.
    C. P. acknowledges support from DL 57/2016 (P2460) from the `Departamento de F\'{i}sica, Faculdade de Ci\^{e}ncias da Universidade de Lisboa' 
    L. S. M. C. acknowledges support by the project ``Enabling Green E-science for the SKA Research Infrastructure (ENGAGE SKA)" (reference POCI-01-0145-FEDER-022217), funded by COMPETE2020 and FCT. 
    P.P. acknowledges support by the project "Identifying the Earliest Supermassive Black Holes with ALMA (IdEaS with ALMA)" (PTDC/FIS-AST/29245/2017).
    T.S. acknowledges support from DL 57/2016/CP1364/CT0009 from the `Centro de Astrof\'{i}sica da Universidade do Porto, Portugal'.
    R.C. acknowledges support from the Fundação para a Ciência e a Tecnologia (FCT) through the Fellowship PD/BD/150455/2019 (PhD:SPACE Doctoral Network PD/00040/2012) and POCH/FSE (EC).
    A.P.A. acknowledge support by FCT under Project CRISP PTDC/FIS-AST-31546/2017 and UIDB/00099/2020.
    We would like to thank D. Munro for freely distributing his Yorick programming language (available at \texttt{http://www.maumae.net/yorick/doc/index.html}).
    C. P. acknowledges the Luisinho family, for the support offered during pandemic times in their house, and the countryside of Olho Marinho (Obidos), where most of this manuscript has been thought and written.
\end{acknowledgements}


\begin{thebibliography}{}
 \bibitem[Ahumada et al.(2020)]{ahu} Ahumada, R., Allende Prieto, C., Almeida, A., et al.\ 2020, \apjs, 249, 3
 \bibitem[Alsing et al.(2020)]{als} Alsing, J., Peiris, H., Leja, J., et al.\ 2020, \apjs, 249, 5. doi:10.3847/1538-4365/ab917f
 \bibitem[Alongi et al.(1993)]{alo} Alongi, M., Bertelli, G., Bressan, A., et al.\ 1993, \aaps, 97, 851
 \bibitem[Bacon et al.(2010)]{bac} Bacon, R., Accardo, M., Adjali, L., et al.\ 2010, \procspie, 7735, 773508. doi:10.1117/12.856027
 \bibitem[Baldwin et al.(2018)]{bal} Baldwin, C., McDermid, R.~M., Kuntschner, H., et al.\ 2018, \mnras, 473, 4698. doi:10.1093/mnras/stx2502
 \bibitem[Binette et al.(1994)]{bin} Binette, L., Magris, C.~G., Stasi{\'n}ska, G., et al.\ 1994, \aap, 292, 13
 \bibitem[Breda \& Papaderos(2018)]{bred} Breda, I., \& Papaderos, P.\ 2018, \aap, 614, A48
 \bibitem[Bressan et al.(1993)]{bres} Bressan, A., Fagotto, F., Bertelli, G., et al.\ 1993, \aaps, 100, 647
 \bibitem[Brinchmann et al.(2004)]{bri} Brinchmann, J., Charlot, S., White, S.~D.~M., et al.\ 2004, \mnras, 351, 1151. doi:10.1111/j.1365-2966.2004.07881.x
 \bibitem[Bruzual \& Charlot(2003)]{bc03} Bruzual, G., \& Charlot, S.\ 2003, \mnras, 344, 1000
 \bibitem[Bruzual A.(2010)]{bru} Bruzual A., G.\ 2010, Philosophical Transactions of the Royal Society of London Series A, 368, 783
 \bibitem[Cardoso, Gomes \& Papaderos (2019)]{car} Cardoso, L.~S.~M., Gomes, J.~M., \& Papaderos, P.\ 2019, \aap, 622, A56
 \bibitem[Cardoso et al.(2017)]{car2} Cardoso, L.~S.~M., Gomes, J.~M., \& Papaderos, P.\ 2017, \aap, 604, A99
 \bibitem[Chabrier(2003)]{cha} Chabrier, G.\ 2003, \apjl, 586, L133
 \bibitem[Chen et al.(2019)]{che} Chen, Y.-Y., Chen, Y.-M., Gu, Q.-S., et al.\ 2019, Research in Astronomy and Astrophysics, 19, 081. doi:10.1088/1674-4527/19/6/81
 \bibitem[Cicone et al.(2016)]{cic} Cicone, C., Maiolino, R., \& Marconi, A.\ 2016, \aap, 588, A41. doi:10.1051/0004-6361/201424514
 \bibitem[Cid Fernandes et al.(2005)]{cid} Cid Fernandes, R., Mateus, A., Sodr{\'e}, L., et al.\ 2005, \mnras, 358, 363
 \bibitem[Cid Fernandes et al.(2011)]{cid2} Cid Fernandes, R., Stasi{\'n}ska, G., Mateus, A., et al.\ 2011, \mnras, 413, 1687
 \bibitem[Cirasuolo \& MOONS Consortium(2016)]{moons} Cirasuolo, M., \& MOONS Consortium\ 2016, Multi-object Spectroscopy in the Next Decade: Big Questions, Large Surveys, and Wide Fields, 109
 \bibitem[Cirasuolo et al.(2020)]{moons2} Cirasuolo, M., Fairley, A., Rees, P., et al.\ 2020, The Messenger, 180, 10
 \bibitem[Coelho et al.(2020)]{coe} Coelho, P.~R.~T., Bruzual, G., \& Charlot, S.\ 2020, \mnras, 491, 2025
 \bibitem[Conroy(2013)]{con} Conroy, C.\ 2013, \araa, 51, 393
 \bibitem[Dalton(2016)]{weave} Dalton, G.\ 2016, Multi-object Spectroscopy in the Next Decade: Big Questions, Large Surveys, and Wide Fields, 97
 \bibitem[Doi et al.(2010)]{doi} Doi, M., Tanaka, M., Fukugita, M., et al.\ 2010, \aj, 139, 1628
 \bibitem[de Jong et al.(2019)]{4most} de Jong, R.~S., Agertz, O., Berbel, A.~A., et al.\ 2019, The Messenger, 175, 3
 \bibitem[Duckworth et al.(2020)]{duc} Duckworth, C., Tojeiro, R., \& Kraljic, K.\ 2020, \mnras, 492, 1869. doi:10.1093/mnras/stz3575
 \bibitem[Faber(1972)]{fab} Faber, S.~M.\ 1972, \aap, 20, 361 
 \bibitem[Fagotto et al.(1994a)]{fag} Fagotto, F., Bressan, A., Bertelli, G., et al.\ 1994, \aaps, 104, 365
 \bibitem[Fagotto et al.(1994b)]{fag2} Fagotto, F., Bressan, A., Bertelli, G., et al.\ 1994, \aaps, 105, 29 
 \bibitem[Florian et al.(2020)]{flo} Florian, J., Ziegler, B., Hirschmann, M., et al.\ 2020, \aap, 635, A41. doi:10.1051/0004-6361/201936441
 \bibitem[Girardi et al.(1996)]{gir} Girardi, L., Bressan, A., Chiosi, C., et al.\ 1996, \aaps, 117, 113 
 \bibitem[Gomes \& Papaderos(2016)]{gom3} Gomes, J.~M. \& Papaderos, P.\ 2016, \aap, 594, A49. doi:10.1051/0004-6361/201628316
 \bibitem[Gomes et al.(2016)]{gom2} Gomes, J.~M., Papaderos, P., Kehrig, C., et al.\ 2016, \aap, 588, A68
 \bibitem[Gomes \& Papaderos(2017)]{gom} Gomes, J.~M., \& Papaderos, P.\ 2017, \aap, 603, A63
 \bibitem[Guseva et al.(2007)]{gus} Guseva, N.~G., Izotov, Y.~I., Papaderos, P., et al.\ 2007, \aap, 464, 885
 \bibitem[Heavens et al.(2004)]{hea} Heavens, A., Panter, B., Jimenez, R., et al.\ 2004, \nat, 428, 625
 \bibitem[Iyer et al.(2019)]{iye} Iyer, K.~G., Gawiser, E., Faber, S.~M., et al.\ 2019, \apj, 879, 116
 \bibitem[Izotov et al.(1997)]{izo} Izotov, Y.~I., Lipovetsky, V.~A., Chaffee, F.~H., et al.\ 1997, \apj, 476, 698
 \bibitem[Izotov et al.(2011)]{izo2} Izotov, Y.~I., Guseva, N.~G., \& Thuan, T.~X.\ 2011, \apj, 728, 161
 \bibitem[Jin et al.(2016)]{jin} Jin, Y., Chen, Y., Shi, Y., et al.\ 2016, \mnras, 463, 913. doi:10.1093/mnras/stw2055
 \bibitem[Kelz et al.(2006)]{kel} Kelz, A., Verheijen, M.~A.~W., Roth, M.~M., et al.\ 2006, \pasp, 118, 129. doi:10.1086/497455
 \bibitem[Koleva et al.(2008)]{kol2} Koleva, M., Prugniel, P., Ocvirk, P., et al.\ 2008, \mnras, 385, 1998
 \bibitem[Koleva et al.(2009)]{kol} Koleva, M., de Rijcke, S., Prugniel, P., et al.\ 2009, \mnras, 396, 2133
 \bibitem[Krueger et al.(1995)]{kru} Krueger, H., Fritze-v. Alvensleben, U., \& Loose, H.-H.\ 1995, \aap, 303, 41
 \bibitem[Le Borgne et al.(2003)]{leb} Le Borgne, J.-F., Bruzual, G., Pell{\'o}, R., et al.\ 2003, \aap, 402, 433. doi:10.1051/0004-6361:20030243
 \bibitem[Leitherer et al.(1999)]{lei} Leitherer, C., Schaerer, D., Goldader, J.~D., et al.\ 1999, \apjs, 123, 3
 \bibitem[Leja et al.(2017)]{lej} Leja, J., Johnson, B.~D., Conroy, C., et al.\ 2017, \apj, 837, 170
 \bibitem[L{\'o}pez Fern{\'a}ndez et al.(2016)]{lop} L{\'o}pez Fern{\'a}ndez, R., Cid Fernandes, R., Gonz{\'a}lez Delgado, R.~M., et al.\ 2016, \mnras, 458, 184
 \bibitem[Maiolino et al.(2020)]{mai} Maiolino, R., Cirasuolo, M., Afonso, J., et al.\ 2020, The Messenger, 180, 24
 \bibitem[MacKay (2003)]{mac} MacKay D. J. C.,\ 2003, Information Theory, Inference and Learning Algorithms. Cambridge Univ. Press, Cambridge 
 \bibitem[Martinsson et al.(2013)]{mar} Martinsson, T.~P.~K., Verheijen, M.~A.~W., Westfall, K.~B., et al.\ 2013, \aap, 557, A130. doi:10.1051/0004-6361/201220515
 \bibitem[McGurk et al.(2010)]{mck} McGurk, R.~C., Kimball, A.~E., \& Ivezi{\'c}, {\v{Z}}.\ 2010, \aj, 139, 1261
 \bibitem[Ocvirk et al.(2006a)]{ocv2} Ocvirk, P., Pichon, C., Lan{\c{c}}on, A., et al.\ 2006, \mnras, 365, 46
 \bibitem[Ocvirk et al.(2006b)]{ocv} Ocvirk, P., Pichon, C., Lan{\c{c}}on, A., et al.\ 2006, \mnras, 365, 74
 \bibitem[Ocvirk(2010)]{ocv3} Ocvirk, P.\ 2010, \apj, 709, 88
 \bibitem[Osterbrock(1989)]{ost2} Osterbrock, D.~E.\ 1989, Astrophysics of Gaseous Nebulae and Active Galactic Nuclei
 \bibitem[Osterbrock \& Ferland(2006)]{ost} Osterbrock, D.~E., \& Ferland, G.~J.\ 2006, Astrophysics of gaseous nebulae and active galactic nuclei
 \bibitem[Pacifici et al.(2015)]{pac} Pacifici, C., da Cunha, E., Charlot, S., et al.\ 2015, \mnras, 447, 786
 \bibitem[Panter et al.(2003)]{pan1} Panter, B., Heavens, A.~F., \& Jimenez, R.\ 2003, \mnras, 343, 1145
 \bibitem[Panter et al.(2007)]{pan2} Panter, B., Jimenez, R., Heavens, A.~F., et al.\ 2007, \mnras, 378, 1550
 \bibitem[Papaderos et al.(1998)]{papa} Papaderos, P., Izotov, Y.~I., Fricke, K.~J., et al.\ 1998, \aap, 338, 43
 \bibitem[Papaderos \& {\"O}stlin(2012)]{papa2} Papaderos, P., \& {\"O}stlin, G.\ 2012, \aap, 537, A126
 \bibitem[Pappalardo et al.(2010)]{pap} Pappalardo, C., Lan{\c{c}}on, A., Vollmer, B., et al.\ 2010, \aap, 514, A33
 \bibitem[S{\'a}nchez et al.(2012)]{san} S{\'a}nchez, S.~F., Kennicutt, R.~C., Gil de Paz, A., et al.\ 2012, \aap, 538, A8. doi:10.1051/0004-6361/201117353
 \bibitem[Sarzi et al.(2007)]{sar} Sarzi, M., Bacon, R., Cappellari, M., et al.\ 2007, \nar, 51, 18. doi:10.1016/j.newar.2006.11.023
 \bibitem[Schaerer \& de Barros(2009)]{sch} Schaerer, D., \& de Barros, S.\ 2009, \aap, 502, 423 
 \bibitem[Stasi{\'n}ska et al.(2008)]{sta} Stasi{\'n}ska, G., Vale Asari, N., Cid Fernandes, R., et al.\ 2008, \mnras, 391, L29
 \bibitem[Storn \& Price (1996)]{sto} Storn, R., \& Price, K. V.
 , K. V.\ 1996, Minimizing the real function of the ICEC'96 contest by differential evolution, IEEE Conf. on Evolutionary Computation, 842, 844
 \bibitem[Storn \& Price (1997)]{sto2} Storn, R., \& Price, K. V.\ 1997, J. Global Optimization, 11, 34
 \bibitem[Struck-Marcell \& Tinsley(1978)]{stu} Struck-Marcell, C., \& Tinsley, B.~M.\ 1978, \apj, 221, 562
 \bibitem[Takada et al.(2014)]{pfs} Takada, M., Ellis, R.~S., Chiba, M., et al.\ 2014, \pasj, 66, R1
 \bibitem[Tinsley(1980)]{tin} Tinsley, B.~M.\ 1980, \fcp, 5, 287
 \bibitem[Tojeiro et al.(2007)]{toj} Tojeiro, R., Heavens, A.~F., Jimenez, R., et al.\ 2007, \mnras, 381, 1252
 \bibitem[Tumlinson et al.(2017)]{tum} Tumlinson, J., Peeples, M.~S., \& Werk, J.~K.\ 2017, \araa, 55, 389
 \bibitem[Vazdekis et al.(2010)]{vaz} Vazdekis, A., S{\'a}nchez-Bl{\'a}zquez, P., Falc{\'o}n-Barroso, J., et al.\ 2010, \mnras, 404, 1639. doi:10.1111/j.1365-2966.2010.16407.x
 \bibitem[Yuan et al.(2019)]{yua} Yuan, F.-T., Burgarella, D., Corre, D., et al.\ 2019, \aap, 631, A123
 \bibitem[Westera et al.(2004)]{wes} Westera, P., Cuisinier, F., Telles, E., et al.\ 2004, \aap, 423, 133
 \bibitem[Wilkinson et al.(2017)]{wil} Wilkinson, D.~M., Maraston, C., Goddard, D., et al.\ 2017, \mnras, 472, 4297
 \end{thebibliography}
\end{document}